\documentclass[a4paper,12pt]{article}

\usepackage[a4paper,text={16.8cm,22.8cm}]{geometry}
\usepackage{amsmath,amssymb,bm,graphicx,color}
\usepackage{extarrows}
\usepackage[utf8]{inputenc}
\usepackage{amsthm}
\usepackage{slashed}
\usepackage{tikz}
\usepackage{rotating}
\usepackage{hyperref}
\usepackage{array}
\usepackage{color}
\usepackage{multirow}
\usepackage{bbold}
\usepackage{cite}
\usepackage{pifont}
\usepackage{tabularx}
\usepackage{stackengine}
\usepackage{gensymb}

\allowdisplaybreaks 
\renewcommand{\arraystretch}{1.2}

\def\be{\begin{equation}}
\def\ee{\end{equation}}

\begin{document}

\begin{titlepage}

\vspace{1.2cm}
\begin{center}
\Large\bf
\boldmath
Revisiting the Universal Texture Zero of Flavour:  \\
a Markov Chain Monte Carlo Analysis
\unboldmath
\end{center}
\vspace{0.2cm}
\begin{center}
{\large{Jordan Bernigaud$^{a,b}$, Ivo de Medeiros Varzielas$^{c}$, Miguel Levy$^{c}$, Jim Talbert$^d$}}\\
\vspace{1.0cm}
{\small\sl 
${}^a$\,Institute for Astroparticle Physics (IAP), Karlsruhe Institute of Technology, Hermann-von-Helmholtz-Platz 1, D-76344 Eggenstein-Leopoldshafen, Germany,\\[0.2cm]
${}^b$\,Institute for Theoretical Particle Physics (TTP), Karlsruhe Institute of Technology, Engesserstrasse 7, D-76128 Karlsruhe, Germany\\[0.2cm]
${}^c$\,CFTP, Departamento de F\'{i}sica, Instituto Superior T\'{e}cnico, Universidade de Lisboa, Avenida Rovisco Pais 1, 1049 Lisboa, Portugal,\\[0.2cm]
${}^d$\,DAMTP, University of Cambridge, Wilberforce Rd., Cambridge, CB3 0WA, United Kingdom}\\[0.5cm]
{\bf{E-mail}}: ivo.de@udo.edu, miguelplevy@ist.utl.pt, rjt89@cam.ac.uk\\[1.0cm]
{\bf{\emph{This work is dedicated to the memory of Prof. Graham Garland Ross, FRS: 1944-2021.}}}
\end{center}

\vspace{0.5cm}
\begin{abstract}
\vspace{0.2cm}
\noindent 
We revisit the phenomenological predictions of the Universal Texture Zero (UTZ) model of flavour originally presented in \cite{UTZ}, and update them in light of both improved experimental constraints and numerical analysis techniques.  In particular, we have developed an in-house Markov Chain Monte Carlo (MCMC) algorithm to exhaustively explore the UTZ's viable parameter space, considering both leading- and next-to-leading contributions in the model's effective operator product expansion.  We also extract --- for the first time --- reliable UTZ predictions for the (poorly constrained) leptonic CP-violating phases, and ratio observables that characterize neutrino masses probed by (e.g.) oscillation, $\beta$-decay, and cosmological processes.  We therefore dramatically improve on the proof-in-principle phenomenological analysis originally presented in \cite{UTZ}, and ultimately show that the UTZ remains a minimal, viable, and appealing theory of flavour.  Our results also further demonstrate the potential of robustly examining multi-parameter flavour models with MCMC routines.
\end{abstract}
\vfil

\end{titlepage}


\tableofcontents
\noindent \makebox[\linewidth]{\rule{16.8cm}{.4pt}}


\section{Introduction}
\label{sec:INTRO}
The bulk of the free, unexplained parameters in the Standard Model (SM) of particle physics originate in its flavour sector, thanks to the replication of SM fermion generations with distinct masses and quantum mixings.  These parameters are technically natural, in that sending them to zero recovers a global U(3)$^5$ flavour symmetry of the Lagrangian \cite{Gerard:1982mm,Chivukula:1987py}.  However, the Yukawa couplings of SM fermions to the Higgs boson break this symmetry in a deeply flavour-non-universal manner, with a mass ratio of $\sim\mathcal{O}(10^{12})$ between (e.g.) neutrinos and the top quark.  Furthermore the Cabibbo–Kobayashi–Maskawa (CKM) quark mixing matrix  exhibits a hierarchical, approximately unit structure \cite{Workman:2022ynf}, while the Pontecorvo–Maki–Nakagawa–Sakata (PMNS) leptonic mixing matrix is extremely non-hierarchical, with large mixings amongst generations \cite{Esteban:2020cvm}.  These highly disparate patterns of fermionic mass and mixing strongly hint that the origins of flavour in the SM may be dynamical, as opposed to a random, soft deviation from an accidental symmetric limit. 

The \emph{flavour puzzle} therefore remains a compelling motivation to search for physics Beyond-the-SM (BSM), as it can be solved  dynamically via  the breakdown of an ultraviolet (UV), BSM symmetry in specific directions of flavour space.  This symmetry breaking typically occurs when exotic scalar \emph{familons} develop special vacuum expectation values (vev) as determined by a family-symmetric scalar potential, although other alignment mechanisms are conceivable.  When familons couple to SM fermions and the Higgs boson, their flavoured vevs shape the otherwise free Yukawa matrices of the SM, and therefore control their associated mass eigenvalues and mixing angles after electroweak symmetry breaking.  These predictions can be compared to global flavour data sets to falsify the model, serving as an indirect probe of the new physics proposed.    

While the predictions of flavour models --- derived from either top-down or bottom-up considerations --- are rich, they are also becoming increasingly difficult to falsify, given that experiment is rapidly resolving all SM flavour parameters to a high degree of precision, such that the models' predictions should actually be considered \emph{postdictions}.  Indeed, virtually all quark masses and CKM mixings are measured with exceptional accuracy, while only the PMNS angle $\theta_{23}^l$,\footnote{In what follows we use the label $l$ for leptons, $q$ for quarks, and $u, d, e, \nu$ for individual families of either. We also include neutrino mass and mixing when we reference `SM' flavour parameters in the text, despite these being fundamentally BSM objects.} the Dirac CP-violating phase $\delta^l$, absolute neutrino mass eigenvalues, and Majorana CP-violating phases (if relevant) are poorly constrained in the leptonic sector.  While physical observables that depend on non-trivial combinations of these parameters, e.g. neutrinoless-double-$\beta$ decay rates ($0\nu\beta\beta$), single $\beta$-decay rates, and the sum of neutrino mass eigenvalues (as constrained by cosmology), offer additional independent probes of flavour models, it is conceivable that a believable BSM theory will also make falsifiable predictions for a subset of the aforementioned, unresolved constituent flavour parameters. Complicating matters further, many (most) BSM flavour models introduce a number of UV theory parameters that are difficult to numerically sample in a fully generic manner, and so extracting concrete predictions from said models is challenging in its own right.

In light of this experimental situation, and in response to the need for more robust analysis routines for exploring viable model parameter spaces, we will re-examine the Universal Texture Zero (UTZ) Model originally presented in \cite{UTZ}. The UTZ is an effective theory (EFT) valid at mass scales above those characteristic of the SM, but below those of hypothetical (and potentially unfalsifiable), renormalizable UV completions, e.g. those incorporating ultra-heavy fermionic messenger fields $\mathcal{V}$:  $\Lambda_{\mathcal{V}} > \Lambda_{\text{UTZ}} > \Lambda_{\text{SM}}$.  Its Yukawa sector is therefore generated only at the non-renormalizable level, with EFT expansion parameters in inverse powers of the messenger masses $M_i$.  The UTZ Lagrangian is symmetric under a $\Delta(27) \simeq \left(\mathbb{Z}_3 \times \mathbb{Z}_3 \right) \rtimes \mathbb{Z}_3$ \cite{deMedeirosVarzielas:2006fc,Ma:2006ip,Luhn:2007uq,deMedeirosVarzielas:2015amz,Ishimori:2010au} non-Abelian discrete family symmetry and a further $\mathbb{Z}_N$ discrete shaping symmetry, and is consistent with an underlying stage of SO(10) grand unification as all fermions and their conjugates --- including right-handed (RH) gauge-singlet neutrinos --- are assigned as triplets ${\bf{3}}$ under $\Delta(27)$. Critically, the additional scalars introduced are charged such that a $\Delta(27)$-invariant scalar potential exists that drives family-symmetry breaking as mentioned above, yielding symmetric mass matrices with a characteristic texture zero in the (1,1) position for \emph{all} family sectors.  As shown in \cite{UTZ}, this UTZ structure is capable of explaining quark and lepton flavour data with as few as nine infrared (IR) theory parameters, and therefore amounts to an appealing and predictive theory for the origin of SM flavour patterns. The UTZ stands as a continuation of similar solutions employing texture zeroes, explored already in e.g. \cite{Ramond:1993kv, Ross:2007az}. 

However, the numerical exploration of the UTZ parameter space presented in \cite{UTZ} only achieved a `proof-in-principle' fit demonstrating the model's phenomenological viability.  It did not exhaustively explore the predictions of the UTZ Lagrangian at leading order (LO) in its EFT expansion parameters, nor did it consider the complete set of corrections generated by operators present at next-to-leading order (NLO) in $1/M_i$.  Most importantly, the analysis in \cite{UTZ} did not present robust \emph{predictions} for the aforementioned unresolved leptonic flavour parameters nor any other observables (e.g. $\beta$-decay rates) that depend on them, and hence it did not provide a reliable means of falsifying the UTZ model space as data continues to improve.  In this paper we aim to remedy these shortcomings by applying a Markov Chain Monte Carlo (MCMC) fitting algorithm to the UTZ.  Inspired by similar analyses \cite{Bernigaud:2021kpw,Sarazin:2021nwo} (also see \cite{Alcaide:2018vni,Jurciukonis:2019bsq,AlcaidedeWandeleer:2020flx} for advanced statistical analyses of flavour-texture models), the MCMC technology we employ allows for a robust exploration of multi-parameter models and their associated likelihoods. It also allows one to simultaneously extract predictions for poorly-constrained observables which are controlled (in part and in different combinations) by the same parameters that control exceptionally well-constrained observables, thereby accounting for the intricate correlations between UTZ theory parameters and their associated phenomenology.  In this way we are capable of presenting predictions in experimentally-preferred regions of the UTZ parameter space, for both the LO and NLO UTZ Lagrangian. As we will show, the theory is phenomenologically viable at \emph{both} orders in its operator product expansion, with the latter NLO terms yielding only minor corrections to the dominant LO predictions.  The UTZ is therefore a stable, predictive, and minimal theory of flavour.

The remainder of the paper develops as follows:  in Section \ref{sec:REVIEW} we review the UTZ model as conceived in \cite{UTZ}, including the field and symmetry content composing the (N)LO contributions to its operator product expansion, as well as its qualitative predictions in the quark and lepton sectors.  Then in Section \ref{sec:CONSTRAINTS} we review the most up-to-date experiment that constrains its predictions in the Yukawa sector, and also discuss the uncertainties associated to renormalization group evolution (RGE) from the UV to the IR.  In Section \ref{sec:MCMC} we discuss the MCMC algorithm we have developed to explore the UTZ parameter space, and also present the results and analysis following from our scans. We conclude in Section \ref{sec:CONCLUDE}. 

\section{The Universal Texture Zero Model}
\label{sec:REVIEW}

\renewcommand{\arraystretch}{1.5}
\begin{table}[t]
\centering
\begin{tabular}{|c|c|c|c|c|c|c|c|c|c|c|}
\hline
$\text{Fields}$ & $\psi_{q,e,\nu}$  & $\psi^{c}_{q,e,\nu}$ & $H$ & $\Sigma$ & $S$ & $\theta_{3}$ &$\theta_{23}$ &$\theta_{123}$ & $\theta$ & $\theta_{X}$ \\
\hline
\hline
$\Delta(27)$ & 3 & 3 & $1_{00}$ & $1_{00}$ & $1_{00}$ & $\bar{3}$ & $\bar{3}$ & $\bar{3}$ & $\bar{3}$ & $3$  \\
\hline
$\mathbb{Z}_{N}$ & 0 & 0 & 0 & 2 & -1 & 0 & -1 & 2 & 0 & $x$  \\
\hline
\end{tabular}
\caption{The fields and $\Delta(27) \times \mathbb{Z}_N$ family symmetry content of the UTZ flavour model.  Note that $\theta_{X}$ only appears in the scalar potential, and hence the only restriction on its $\mathbb{Z}_{N}$ charge is that it does not contribute significantly to the fermionic mass matrices.  Its $\mathbb{Z}_{N}$ charge can therefore be left generic, as shown.}
\label{tab:Zcharges}
\end{table}

The field and $\Delta(27) \times \mathbb{Z}_N$ symmetry content of the UTZ \cite{UTZ} is given in Table \ref{tab:Zcharges}.  There one observes that all SM fermions $\psi_a$ are assigned as triplets ${\bf{3}}$ under the family symmetry, as are additional gauge singlet `sterile' neutrinos that participate in a seesaw mechansim.   Besides the fermionic content, we also have a set of BSM scalar familons $\theta_i$, charged as $\Delta(27)$ anti-triplets ${\bf{\bar{3}}}$, a lepton-number-violating (LNV) anti-triplet familon $\theta$ necessary for describing the Majorana neutrino mass sector, and finally a triplet familon $\theta_X$ necessary for successful vacuum alignment.  All such familons are SM gauge singlets.  There is also a $\Delta(27)$ singlet sector composed of the $\Sigma$ and Higgs $H$ scalars, both associated to an underlying stage of grand unification consistent with the following symmetry-breaking chain: 
\begin{equation}
\label{eq:GUTsymmbreak}
    \text{SO(10)} \rightarrow \text{SU(4)} \times \text{SU(2)}_L \times \text{SU(2)}_R \rightarrow \text{SU(3)} \times \text{SU(2)} \times \text{U(1)}\,,
\end{equation}
where the SO(10) breaking proceeds via an $H$ vev and where $\langle \Sigma \rangle \propto B - L + \kappa \, T_3^R$ is associated to Pati-Salam breaking. As seen below, this latter $\Sigma$ field selects unique Dirac textures for distinct fermion families out of an otherwise universal mass matrix structure. Finally, the $\Delta(27)$ singlet scalar $S$ is a shaping field that, along with the $\mathbb{Z}_N$ shaping symmetry, restricts the class of operators that appear in the UTZ EFT.  Its main role is to indirectly forbid terms $\propto \theta_{123}\theta_{123}$ in the UV Majorana Lagrangian presented in \eqref{eq:MMAJ}, which would destroy the desirable UTZ texture.  We note that this field and symmetry content exhibits explicit discrete gauge anomaly freedom at the relevant scale of our EFT --- see (e.g.) \cite{Ibanez:1991hv,Ibanez:1991wt,Banks:1991xj,Araki:2006sqx,Araki:2008ek,Ishimori:2010au,Talbert:2018nkq,Gripaios:2022vvc,Davighi:2022icj}.

Besides the Yukawa Lagrangian to be discussed in upcoming Sections, the familons $\theta_i$, $\theta_X$ and $\theta$ also compose an associated scalar potential $V = V_A + V_B$, 
\begin{align}
\nonumber
V_A &= \sum\limits_{i = 3,123} {\left( {{V_1}({\theta _i}) + {V_2}({\theta _i})} \right)}  + {V_3} + {V_4} + {V_5}\,, \\
\label{eq:scalarsum}
V_B &= V_1(\theta) + V_2(\theta) + V_6\,.
\end{align}
While we leave the complete description of the vacuum alignment mechanism to \cite{UTZ}, we recall that the individual components $V_i$ of $V$ are given by
\begin{align}
\nonumber
V_1(\theta_i) &= m_i^2|\theta_i|^2\,,\,\,\,V_2(\theta_i) = h_i {\left( {{\theta _i}} \right)^2}{\left( {{\theta ^{\dag i}}} \right)^2}\,, \,\,\,\, V_3 = k_1{\theta _{X,i}}\theta _{123}^{\dag i}{\theta _{123,j}}\theta_X ^{\dag j}\,\,\,\,\,\,(k_1>0)\,,\,\,\,V_4 = k_2 m_0 \theta_X^1\theta _X^2\theta _X^3\\
\label{eq:potentialterms}
V_5 &= k_3\theta _{23,i}\theta _X^i\theta _{23}^{\dag j}\theta_X^{\dag j}+
k_4{\theta _{23,i}}\theta _{3}^{\dag i}{\theta _{3,i}}\theta_{23} ^{\dag i}\,\,\,\,\,\,\, (k_{3}>0\,\, \text{and}\,\, k_{4}<0)\,\,, \,\,\,\,\,V_6 = k_5 \theta_{3, i} \theta^{\dagger i} \theta_i \theta_{3}^{\dagger i}\,\,\,\,\,(k_5 < 0)\,.
\end{align}
The first term $V_1$ sets the scale of the scalar familon fields, and is sufficient to break the family symmetry spontaneously upon $m_i^2$ being driven to negative values, perhaps via radiative corrections in the manner of \cite{Casas:1999ac}.  Then the second term $V_2$ aligns the $\theta_{3,123}$ vevs in flavour space as a function of the sign of $h_i$; $h_{123} \equiv h_i > 0$ while $h_3 \equiv h_i < 0$.  The terms $V_{3,4,5,6}$ account for the final alignment of the $\theta_{23,X}$ and $\theta$ vevs, with $V_3$ sourcing the dominant coupling of $\theta_X$, $V_4$ selecting $\langle \theta_X \rangle \propto \left(2,-1,1\right)$ 
out of the two degenerate vacua $V_3$ allows, and $V_5$ and $V_6$ respectively driving the final $\theta_{23}$ and $\theta$ orientations upon minimization.\footnote{Observe that in \eqref{eq:scalarsum} we have only included terms that are consistent with a spontaneously broken, supersymmetric (SUSY) underlying theory with triplet mediators.  Additional quartic terms may appear, but must be suppressed in order to preserve \eqref{eq:finalvac}.  All other aspects of the tree-level phenomenology of the UTZ model can be studied without reference to hypothetical UV completions, and we adopt this agnosticism to be as generic as possible in what follows.} 

All subtleties considered, the potential in \eqref{eq:potentialterms} aligns the scalar familon fields in special directions in flavour-space,
\begin{equation}
\nonumber
\left\langle {{\theta _{(3)}}} \right\rangle  = {{\rm{v}}_{\theta (3)}}\left( {\begin{array}{*{20}{c}}
0\\
0\\
1
\end{array}} \right),\quad \left\langle {{\theta _{123}}} \right\rangle  = \frac{{{\rm{v}}_{123}}}{{\sqrt 3 }}\left( {\begin{array}{*{20}{c}}
e^{i\beta}\\
e^{i\alpha}\\
{ - 1}
\end{array}} \right), \quad
 \left\langle {{\theta _{23}}} \right\rangle  = \frac{{{\rm{v}}_{23}}}{{\sqrt 2 }}\left( {\begin{array}{*{20}{c}}
0\\
e^{i\alpha}\\
1
\end{array}} \right),\quad \left\langle {{\theta ^{\dag}_X}} \right\rangle  = \frac{{{\rm{v}}_X}}{{\sqrt 6 }}\left( {\begin{array}{*{20}{c}}
2e^{i\beta}\\
 - e^{i\alpha}\\
1
\end{array}} \right)\,,
\label{eq:finalvac}
\end{equation}
where the parentheses on the first term indicate that both $\theta_3$ and $\theta$ are aligned in the third-family direction, and where we have included the generic phases $\alpha$, $\beta$ for completeness, although we will eventually set these to zero following the discussion in \cite{Roberts:2001zy}.  We note that, of the above potential terms, many are not invariant under SU(3)$_\mathcal{F}$, and so the use of $\Delta(27)$, a non-Abelian discrete subgroup of SU(3)$_\mathcal{F}$, was instrumental in the above discussion.

\subsection{The Leading-Order Effective Yukawa Lagrangian}
\label{sec:LOLAGRANGIAN}

Upon demonstrating that a successful vacuum alignment is plausible upon family-symmetry breaking, a meaningful BSM Yukawa sector can be subsequently formed from the field and symmetry content of Table \ref{tab:Zcharges}.  This leads to the following LO UTZ effective Lagrangian in the Dirac sector of the theory:
\be 
\mathcal{L}^{LO}_{D,f}=\psi_i\left ({c^{(6)}_{3}\over\, M_{3,f}^2}\theta_3^i  \theta_3^j  +{c^{(7)}_{23}\over M_{23,f}^3}\,\theta_{23}^i\theta_{23}^j\Sigma+{c^{(7)}_{123}\over M_{123,f}^3}\,(\theta_{123}^i\theta_{23}^j+\theta_{23}^i\theta_{123}^j)  S    \right)\psi_j^{c}H  \,,
\label{eq:DiracUTZLO}
\ee
where $f\in\lbrace u,\;d,\;e, \; \nu \rbrace$. Here $c^{(n)}_{i}$ are free Wilson coefficients whose superscript denotes the mass dimension $n$ of the operator, while $M_{i,f}$ represent the mass scales associated to heavy messenger fields that have been integrated out of the spectrum in forming the EFT, a l\'a the Froggatt-Nielsen mechanism \cite{Froggatt:1978nt}.  These messenger fields are associated to distinct UV completions and are typically taken to be vector-like fermions, although we do not wish to commit ourselves to any particular scenario.  In what follows we will simply point out the implications and constraints on said UV messengers coming from the (falsifiable) IR spectrum associated to \eqref{eq:DiracUTZLO}. 

To that end, one quickly notices that a natural hierarchy for the third-family fermions is realized, thanks to the power suppression (assuming only mild hierarchies amongst messenger masses) of the second and third terms with respect to the first, which only contributes to the (3,3) entry of the Dirac mass matrices.  While this helps realize an approximate SU(2)$_\mathcal{F}$ symmetry of the quark mass matrices and associated CKM mixing matrix, it also implies that the ratio $\theta_3/M_{3,f}$ is large \cite{deMedeirosVarzielas:2005ax}, at least in the up sector. This is acceptable if $\theta_3$ is the dominant contributor to the messenger mass, which we assume for all charged fermion sectors.  For an alternative solution to this issue involving Higgs mediators, see \cite{deMedeirosVarzielas:2012esq}.     

Besides \eqref{eq:DiracUTZLO}, the field and symmetry content of Table \ref{tab:Zcharges} also permits a Majorana mass Lagrangian, which at leading order in the OPE is of the following form:
\be
\mathcal{L}^{\nu}_{\mathcal{M}}=\psi_i^c \left ({c^{(5)}_M\over M}\,\theta^i  \theta^j  +{1\over M^4} [c^{(8)}_{M,1}\,\theta_{23}^i\theta_{23}^j(\theta^k\theta^k\theta_{123}^k)+c^{(8)}_{M,2}\,(\theta_{23}^i\theta_{123}^j+\theta_{123}^i\theta_{23}^j)(\theta^k\theta^k\theta_{23}^k)]\right )\psi^c_j\,.
\label{eq:MajoranaUTZLO}
\ee
Here one notices that there are two insertions of the LNV scalar $\theta$ in each operator, as is consistent with our underlying SO(10) $\rightarrow$ SU(4)$\times$SU(2)$_L \times$SU(2)$_R$ GUT embedding,
and also that the leading contributions in this effective Lagrangian are at dimension five and eight in the $1/M$ expansion of the EFT, as opposed to six and seven in the case of the Dirac Lagrangian given in \eqref{eq:DiracUTZLO}.  This results in an extremely dominant third-family hierarchy that has important phenomenological implications in the neutrino sector upon applying the seesaw, as mentioned below.  Further discussion regarding the relative power suppression between Dirac and Majorana sectors will be given in Section \ref{sec:HO}. 

\subsubsection*{Qualitative Charged Fermion Masses and Sum Rules}
While the SM's quark and charged lepton flavour sector is exceptionally well-measured and therefore offers little opportunity for novel predictions, we do note that the UTZ Lagrangian above has been designed to realize successful charged fermion mass ratios, as well as two long-standing and successful phenomenological ans\"atze: the Georgi-Jarlskog mechanism and the Gatto-Sartori-Tonin sum rule. This is due to the UTZ structure of the Dirac mass matrices, given qualitatively by 
\be
\renewcommand{\arraystretch}{1.3}
M_f^D\approx m_3 \left( {\begin{array}{*{20}{c}}
0&{\varepsilon _f^3}&{  \varepsilon _f^3}\\
{\varepsilon _f^3}&{{r_f}\varepsilon _f^2}&{  {r_f}\varepsilon _f^2}\\
{  \varepsilon _f^3}&{  {r_f}\varepsilon _f^2}&1
\end{array}} \right),\quad {r_{u,d}}=1/3,\quad {r_e} =   -1\,,
\label{eq:mm1}
\ee
with $f$ again indicating the family sector, $f \in \lbrace u, d, e \rbrace$ and $\epsilon_f$ associated small parameters. Phenomenologically viable values are given by $\epsilon_u \approx 0.05$ and $\epsilon_{d,e} \approx 0.15$. This family splitting is accommodated via the UTZ relations
\be
\label{eq:epsilons}
r_f\,\epsilon_f^2 \equiv \frac{\langle \theta_{23}\rangle^2 \langle \Sigma \rangle}{M_{23,f}^3}\cdot \frac{M_{3,f}^2}{\langle \theta_3 \rangle^2}\,, \,\,\,\,\,\,\,\,\,\,\,\,\,\epsilon_f^3 \equiv \frac{\langle \theta_{23}\rangle\langle \theta_{123} \rangle \langle S \rangle}{M_{123,f}^3}\cdot \frac{M_{3,f}^2}{\langle \theta_3 \rangle^2}\,,
\ee
which hold up to $\mathcal{O}(1)$ coefficients and signs.  Here one sees that $\epsilon_u < \epsilon_d$ is realized if $M_{3,u}/M_{23,u} < M_{3,d}/M_{23,d}$, and of course $\epsilon_e$ and $\epsilon_d$ can be equal given the symmetry breaking in \eqref{eq:GUTsymmbreak} and the fact that both are $T_{R,3} = -1/2$ states which (in SUSY models) acquire their mass from the same Higgs boson ($H_d$).  Note that we assume the messenger masses carry both lepton and quark quantum numbers, and so it is important that RH messengers associated to SU(2)$_R$ breaking in \eqref{eq:GUTsymmbreak} dominate for $\epsilon_u < \epsilon_d$, as opposed to the LH messengers associated to SU(2)$_L$, whose up and down masses are of course equal due to SU(2)$_L$ invariance.

Also associated to this pattern of family suppression are the $r_f$ coefficients in \eqref{eq:mm1}, which are sourced from the $\Sigma$ vev and therefore implement the Georgi-Jarlskog mechanism \cite{Georgi:1979df}, resulting in 
\be
\label{eq:GJR}
m_\tau = m_b \,, \,\,\,\,\,\,\,\, m_\mu = 3\,m_s \,, \,\,\,\,\,\,\,\,\,\,\,\,\, m_e = \frac{1}{3} \,m_d
\ee
at the scale of grand unification, as is consistent with RGE and threshold corrections \cite{Ross:2007az}.

In addition to these characteristic mass relations, the quark sector realizations of \eqref{eq:mm1} also implement the Gatto-Sartori-Tonin mixing sum rule \cite{Gatto:1968ss},
\be 
\sin{\theta _c} = \left\vert \sqrt {\frac{{{m_d}}}{{{m_s}}}}  - {e^{i\delta }}\sqrt {\frac{{{m_u}}}{{{m_c}}}} \right\vert\,,
\label{eq:GST}
\ee
relating the Cabibbo angle $\theta_c \simeq \theta_{12}^q$ to mass ratios from the first- and second-generation quarks of both the up and down families.  Setting $\delta \approx \pi/2$ and again accounting for RGE and threshold corrections \cite{Ross:2007az} (cf. Section \ref{sec:RGE}), both \eqref{eq:GJR} and \eqref{eq:GST} remain successful predictions that are maintained in our UTZ framework.\footnote{These predictions are a consequence of the texture zeroes in the charged fermion structures and are therefore not unique to the UTZ, see e.g. \cite{Ramond:1993kv, King:2003rf, deMedeirosVarzielas:2005ax}.}
\subsubsection*{Qualitative Neutrino Masses and Sum Rules}
As discussed above, the bulk of the parameters left to be constrained experimentally are in the neutrino sector, and so it is worthwhile to discuss the qualitative, analytic predictions of the UTZ construction in this area.  As with other flavour models, we employ the Type-I seesaw mechanism \cite{Minkowski:1977sc,Gell-Mann:1979vob,Mohapatra:1979ia,Yanagida:1980xy}.  In this framework the right-handed (RH) Majorana mass terms generated by \eqref{eq:MajoranaUTZLO} are parametrically heavier than the Dirac neutrino masses coming from \eqref{eq:DiracUTZLO}.  Integrating the RH neutrino fields out of the spectrum generates a left-handed (LH) Majorana neutrino mass term,
\be
\label{eq:seesaw}
M_\nu = \frac{1}{2}\,M^{D}_\nu \cdot M_M^{-1} \cdot M^{D,T}_\nu\,,
\ee
which is of course naturally light due to the heavy $M_M$ suppression. In the presence of a sequentially dominant \cite{King:1998jw,King:1999cm,King:1999mb,King:2002nf} RH neutrino spectrum
\be
\label{eq:sequentialdominance}
M_{M,3} \gg M_{M,2} \ge M_{M,1}\,,
\ee
which is naturally realized thanks to the hierarchical suppression of the second term in \eqref{eq:MMAJ} with respect to the dominant (third-family)  first term, the see-saw contribution coming from $\nu_3^c$ exchange is negligible.  This results in the lightest active neutrino having a parametrically smaller mass compared to the two heaviest active neutrinos.  This spectrum is described by an effective 2$\times$2 neutrino mass structure in the IR, which can be analyzed analytically. 
In particular, after application of the Type-I seesaw mechanism, one can extract sum rules for the PMNS mixing angles as a function of neutrino mass eigenvalues,
\begin{equation}
 \label{eq:PMNSsumrules}
 \sin \theta^{\nu}_{13} \approx \sqrt{{m_2\over 3m_3}}\,, \quad \quad
 \sin\theta^{\nu}_{23} \approx \vert {1\over \sqrt{2}}-e^{i\eta}\sin\theta^{\nu}_{13} \vert \,, \quad \quad
 \sin\theta^{\nu}_{12} \approx \frac{1}{\sqrt{3}}\,,
 \end{equation}
where the phase $\eta$ is defined from the predicted ratio of the heavy neutrino masses $m_2/m_3$ (see the discussion in \cite{UTZ}).
One notes that the relationships in \eqref{eq:PMNSsumrules} are similar to the renowned `Tri-Bimaximal' (TBM) texture \cite{Harrison:2002er} ($\sin \theta_{13}^{TBM} = 0$, $\sin \theta_{23}^{TBM} = 1/\sqrt{2}$, $\sin \theta_{12}^{TBM} = 1/\sqrt{3}$) that is often a starting point for neutrino mass model building. However, the salient difference with respect to prior models of this type --- see e.g.
\cite{King:2003rf, deMedeirosVarzielas:2005ax} or the more recent
\cite{Bjorkeroth:2015ora,Bjorkeroth:2015uou,Bjorkeroth:2017ybg} --- is that the (1,1) texture zero of the mass matrix remains \emph{after} application of the seesaw, such that our UTZ setup leads to a non-negligible departure from the TBM texture and naturally allows for a large(r) reactor mixing angle $\theta_{13}^l$ (which also receives corrections from the charged-lepton sector), in accord with data.  
 Finally, we note that our $\Delta(27)$ family-symmetry breaking realizes the $\mathbb{Z}_2 \times \mathbb{Z}_2$ \cite{Lam:2007qc} residual symmetry of the IR neutrino mass term only `indirectly,'\footnote{...following the classification system of \cite{King:2013eh}. For a pedagogical algorithm and exhaustive discussion regarding the reconstruction of effective Lagrangians analogous to \eqref{eq:DiracUTZLO}-\eqref{eq:MajoranaUTZLO} in the alternative `direct' symmetry-breaking scenario, see \cite{Bernigaud:2020wvn}.} in that it is not a subgroup of $\Delta(27)$ and appears only accidentally thanks to \eqref{eq:finalvac}.  

\subsection{Higher-Order Contributions}
\label{sec:HO}
The Lagrangians in \eqref{eq:DiracUTZLO}-\eqref{eq:MajoranaUTZLO} represent the LO contributions to the UTZ operator product expansion.  Higher-order terms in this series are suppressed by further powers of the relevant mediator masses, and should therefore represent small corrections to the qualitative structures and predictions discussed above.  However, these corrections can a priori be non-negligible as noted in \cite{UTZ}, and so it is important that we consider them robustly as we revisit the UTZ.

In the Dirac sector, the NLO $\Delta(27) \times \mathbb{Z}_N$ invariant terms composed of the same field content as in Table \ref{tab:Zcharges} arise at mass-dimension eight, i.e. with four powers of mediator mass suppression,
\be
 \label{eq:DiracUTZHO}
\mathcal{L}^{HO}_{D,f} = \psi_i\left ({c^{(8)}_{23}\over M_{23,f}^4}(\theta_{23}^i\theta_{3}^j + \theta_{3}^i\theta_{23}^j)\Sigma S+{c^{(8)}_{123}\over M_{123,f}^4}(\theta_{123}^i\theta_{3}^j+\theta_{3}^i\theta_{123}^j)  S^2    \right)\psi_j^{c}H\,.
\ee
While these terms contribute at the same order in the EFT's power counting, we have already identified in the discussion below \eqref{eq:epsilons} that the LO Dirac mass contribution $\propto \langle \Sigma \rangle$ is parametrically larger than that $\propto S$.  If one assumes roughly universal messenger masses, one finds that
\be
\frac{\langle \theta_{23} \rangle \langle \theta_{23} \rangle \langle \Sigma \rangle}{M_{23}^{3}} \sim \mathcal{O}(\epsilon^{2}), \,\,\,\,\,\,\,\frac{\langle \theta_{23} \rangle \langle \theta_{123} \rangle \langle S \rangle}{M_{123}^{3}} \sim \mathcal{O}(\epsilon^{3}) \,\,\,\Longrightarrow \,\,\, \frac{\langle \theta_{23} \rangle \langle \Sigma \rangle}{\langle \theta_{123} \rangle \langle S \rangle} \sim \mathcal{O}\left( \frac{1}{\epsilon} \right)\,,
\ee  
from which once can readily conclude that the HO contributions $\propto S^{2}$ in \eqref{eq:DiracUTZHO} are also parametrically smaller than those $\propto \Sigma S$:
\be
\label{eq:HOcomp}
\frac{\langle \theta_{3} \rangle \langle \theta_{23} \rangle \langle \Sigma \rangle \langle S \rangle}{M^{4}} \sim \frac{1}{\epsilon} \frac{\langle \theta_{3} \rangle \langle \theta_{123} \rangle \langle S \rangle^{2}}{M^{4}} \,.
\ee
In \cite{UTZ} we used \eqref{eq:HOcomp} to justify ignoring the $S^2$ contribution to the Dirac mass matrix entirely.  However, we will now include both terms in \eqref{eq:DiracUTZHO} for completeness.

The UTZ's operator product expansion is of course infinite-dimensional in the absence of an explicit UV completion.  Hence further, next-to-next-to-leading order contributions can also be written down.  However, these operators will have at least three additional insertions of $\Delta(27)$ triplets, and are therefore highly suppressed. We neglect their contributions as a result. We also note that the NLO contributions to the Dirac Lagrangian given in \eqref{eq:DiracUTZHO} enter at the same mass-dimension as LO contributions to the Majorana Lagrangian given in \eqref{eq:MajoranaUTZLO},
\be
\mathcal{O}^{\text{HO}}_{D} \sim \mathcal{O}^{\text{LO}}_{\mathcal{M}} \sim \mathcal{O}(1/M^4)\,,
\ee 
and so we do not consider any corrections to \eqref{eq:MajoranaUTZLO} to be consistent in our power counting.

\subsection{Complete Effective Mass Matrices in the Ultraviolet}

The discussions in the Subsections above lead to the LO and NLO Lagrangians of \eqref{eq:DiracUTZLO}, \eqref{eq:MajoranaUTZLO} and \eqref{eq:DiracUTZHO}. After family- and electroweak-symmetry breaking, these Lagrangians generate the following Dirac and Majorana fermion UTZ mass matrices: 
\begin{equation}
\label{eq:MDIR}
\mathcal{M}_{f}^{\mathcal{D}} \simeq
\left(
\begin{array}{ccc}
0 & a\,e^{i(\alpha+\beta+\gamma)}  & a\,e^{i(\beta+\gamma)}+c\,e^{i(\beta+\zeta)}  \\
a\,e^{i(\alpha+\beta+\gamma)}  & (b\,e^{-i\gamma}+2a\,e^{-i\delta})\,e^{i(2\alpha+\gamma + \delta)} & b\,e^{i(\alpha + \delta)}+c\,e^{i(\alpha + \zeta)}+d\,e^{i(\alpha + \psi)}\\
a\,e^{i(\beta + \gamma)} + c\,e^{i(\beta + \zeta)} & b\,e^{i(\alpha + \delta)}+c\,e^{i(\alpha + \zeta)}+d\,e^{i(\alpha + \psi)} & 1 - 2 a\,e^{i\gamma} + b\,e^{i\delta} -2 c\,e^{i\zeta} + 2 d\,e^{i\psi}
\end{array}
\right) \,,
\end{equation}
\begin{equation}
\label{eq:MMAJ}
\mathcal{M}^{\mathcal{M}} \simeq
\left(
\begin{array}{ccc}
0 & y\,e^{i(\alpha+\beta+\rho)}  & y\,e^{i(\beta+\rho)}  \\
y\,e^{i(\alpha+\beta+\rho)}  & (x\,e^{-i\rho}+2 y\,e^{-i\phi})\,e^{i(2\alpha+\rho + \phi)} & x\,e^{i(\alpha + \phi)}\\
y\,e^{i(\beta + \rho)} & x\,e^{i(\alpha + \phi)} & 1 -  2 y\,e^{i\rho} + x\,e^{i\phi}
\end{array}
\right)  \,.
\end{equation}
Here the matrices have been normalized such that $\mathcal{M}_{f}^{\mathcal{D}} \equiv M_{f}^{\mathcal{D}}/s_f$,  $\mathcal{M}^{\mathcal{M}} \equiv M^{\mathcal{M}}/M_\theta$, where $M_\theta$ and $s$ are the overall scale-setting parameters of \eqref{eq:MDIR}-\eqref{eq:MMAJ} which, along with the relative-scale-setting parameters $\lbrace a,b,c,d,x,y \rbrace$, are defined in terms of scalar vevs and other coefficients:
\begin{align}
\nonumber
{a^{\prime}_f} &= \frac{c_{123}^{(7)}{{{\rm{v}}_{123}}{{\rm{v}}_{23}}\left\langle S \right\rangle }}{{\sqrt 6 M_{123,f}^3}},\quad {b^{\prime}_f} = \frac{c_{23}^{(7)}{{r_f}{\rm{v}}_{23}^2\left\langle \Sigma  \right\rangle }}{{2M_{23,f}^3}},\quad {c^{\prime}_f} = \frac{c_{123}^{(8)} \rm{v}_{123}\rm{v}_{3}\langle S \rangle^2}{\sqrt{3} M_{123,f}^4}, \quad {d^{\prime}_f} = \frac{c_{23}^{(8)} r_f \rm{v}_{23} \rm{v}_3 \langle \Sigma \rangle \langle S \rangle}{\sqrt{2}M_{23,f}^4}, \\ 
\label{eq:paramprime}
\quad {s_f} &= \frac{c_{3}^{(6)}{{\rm{v}}_3^2}}{{M_{3,f}^2}},\quad \quad \quad  x^\prime = \frac{c_{M,1}^{(8)} \rm{v}_{23}^2 \langle \Theta_{123} \rangle }{2M^4},\quad y^\prime = \frac{c_{M,2}^{(8)}  \rm{v}_{23}\rm{v}_{123} \langle \Theta_{23} \rangle}{\sqrt{6} M^4},\quad M_\theta = \frac{c_{M}^{(5)}\rm{v}_\theta^2}{M}\,,
\end{align}
with $r_{u,d,e,\nu}=(1,1,-3,-3)/3$ and $\langle \Theta_{23,123} \rangle \equiv \langle \theta^k \theta^k \theta^k_{23,123}\rangle$, i.e. the vev of the singlet contractions with $k$ superscript in \eqref{eq:MajoranaUTZLO}.
The relationship between primed and unprimed parameters, along with associated complex phases, is then given by
\begin{align}
\nonumber
\frac{a^{\prime}}{s} &= \bigg\vert \frac{a^{\prime}}{s} \bigg\vert \, e^{i \gamma} \equiv a \, e^{i \gamma}, \,\,\,\,\,\,\,\,\,\frac{b^{\prime}}{s} = \bigg\vert \frac{b^{\prime}}{s} \bigg\vert \, e^{i \delta} \equiv b \, e^{i \delta}, \,\,\,\,\,\,\,\,\, \frac{c^{\prime}}{s} = \bigg\vert \frac{c^{\prime}}{s} \bigg\vert \, e^{i \zeta} \equiv c \, e^{i \zeta}, \,\,\,\,\,\,\,\,\,\frac{d^{\prime}}{s} = \bigg\vert \frac{d^{\prime}}{s} \bigg\vert \, e^{i \psi} \equiv d \, e^{i \psi}\,,\\
\label{eq:params}
\frac{x^{\prime}}{M_\theta} &= \bigg\vert \frac{x^{\prime}}{M_\theta} \bigg\vert \, e^{i \phi} \equiv x \, e^{i \phi}, \,\,\,\,\,\,\,\,\,\frac{y^{\prime}}{M_\theta} = \bigg\vert \frac{y^{\prime}}{M_\theta} \bigg\vert \, e^{i \rho} \equiv y \, e^{i \rho}\,,
\end{align}
and it is clear that $c$ and $d$ account for the HO Dirac corrections in \eqref{eq:DiracUTZHO}.  

Given \eqref{eq:MDIR}-\eqref{eq:MMAJ}, the values of the `physical' fermionic mass, mixing, and CP-violating parameters can be extracted numerically as described in \cite{UTZ} or analytically, using flavour-invariant theory as described in (e.g.) \cite{Jenkins:2009dy,Talbert:2021iqn,Wang:2021wdq}.  Then, given \eqref{eq:paramprime}-\eqref{eq:params}, one can compare the number of IR theory parameters vs. IR physical parameters as a measure of the predictivity of the UTZ.  At LO, there are a priori two coefficients (a,b) and two phases ($\gamma$, $\delta$) for each charged fermion sector, plus the additional two family-universal phases ($\alpha$, $\beta$) from vacuum alignment.  However, following \cite{Roberts:2001zy}, we can set all but two of these phases to zero without loss of generality.  Assuming the GUT embedding discussed above to relate the down quarks to charged leptons taking into account the Georgi-Jarlskog factors, one then has $\left( 2 + 2 \right) \cdot 3 + 2 - 4 - 4 = 6$ UTZ model parameters (including two phases) to describe three CKM mixing angles, one CKM Dirac phase, four quark mass ratios and two charged lepton mass ratios, totalling 10 physical parameters.  The neutrino sector's predictivity is even more striking, in the sequentially dominant limit of \eqref{eq:PMNSsumrules}.  There, only three parameters, including a phase and an overall mass scale, are necessary to reproduce the neutrino mass differences, which when combined with the aforementioned charged lepton parameters also generate PMNS angles and phases.  In total, we see that only nine theory parameters are required to reproduce 18 physical parameters at LO in the UTZ OPE. This is to be compared to the SM where, before allowing for weak basis transformations or rephasing freedoms, the same physical parameters are controlled by three, $3 \times 3$ complex matrices (charged fermion Yukawas) and unspecified neutrino mass operator(s) (taking the Weinberg operator \cite{Weinberg:1979sa} as an infrared limit of the seesaw mechanism, at least one additional complex symmetric $3 \times 3$ matrix must be introduced).

\section{Experimental Constraints}
\label{sec:CONSTRAINTS}

The core experimental constraints on the UTZ model presented in Section \ref{sec:REVIEW} are of course the fermionic mass eigenvalues and CKM/PMNS mixings extracted from a host of low- and high-energy flavour experiments. Regarding the charged fermion sector, this information is regularly collated in the PDG review \cite{Workman:2022ynf}, which reports bounds on  fermion masses and mixing angles.  We have reported these IR bounds for the mass sector in Table \ref{tab:massfits}, translating the 
uncertainties on individual masses into uncertainties on mass ratios, given that the UTZ only predicts the charged fermion mass spectrum up to a common scale.  On the other hand, uncertainties on mixing angles and the Dirac CP phase can be extracted from global fits to the CKM matrix and Jarlskog invariant given by \cite{Workman:2022ynf}
\renewcommand\arraystretch{1.8}
\begin{equation}
\label{eq:PDGCKM}
\vert V_{\rm CKM} \vert \equiv \vert U_u^\dagger U_d \vert \in
\left(
\begin{array}{ccc}
\left(\stackanchor{0.97419}{0.97451}\right)  & \left(\stackanchor{0.22433}{0.22567}\right) & \left(\stackanchor{0.00358}{0.00388}\right) \\
\left(\stackanchor{0.22419}{0.22553}\right)   & \left(\stackanchor{0.97333}{0.97365}\right) & \left(\stackanchor{0.04108}{0.04267}\right)\\
\left(\stackanchor{0.00839}{0.00877}\right)  & \left(\stackanchor{0.04038}{0.04193}\right) & \left(\stackanchor{0.999082}{0.999149}\right)
\end{array}
\right)\,,\,\,\,\,\, \mathcal{J}^{\rm CKM} \in \left(\stackanchor{3.23}{2.95}\right) \cdot 10^{-5}\,,
\end{equation}
where the left equality defines the CKM as the overlap of the matrices $U_{u,d}$ diagonalizing the up / down Yukawa couplings.  The translation of these bounds to the $\theta_{ij}^q$ and $\delta_q$ basis is given in Table \ref{tab:mixingangles}.

Leptonic mass and mixing constraints are of course deeply sensitive to ongoing neutrino oscillation, cosmology, and $\beta$-decay experiments.  The authors of \cite{Esteban:2020cvm} have compiled a global fit to the available oscillation data, finding (e.g.)
\renewcommand\arraystretch{1.8}
\begin{equation}
\label{eq:Nufit}
\vert V_{\rm PMNS} \vert \equiv \vert U_l^\dagger U_\nu \vert \in
\left(
\begin{array}{ccc}
\left(\stackanchor{0.801}{0.845}\right)  & \left(\stackanchor{0.513}{0.579}\right) & \left(\stackanchor{0.144}{0.156}\right) \\
\left(\stackanchor{0.244}{0.499}\right)   & \left(\stackanchor{0.505}{0.693}\right) & \left(\stackanchor{0.631}{0.768}\right)\\
\left(\stackanchor{0.272}{0.518}\right)  & \left(\stackanchor{0.471}{0.669}\right) & \left(\stackanchor{0.623}{0.761}\right)
\end{array}
\right)\,,
\end{equation}
where the LHS again gives the standard definition of the PMNS matrix as it appears in the charged-current interactions in terms of constituent charged-lepton and neutrino mixing matrices $U_{l,\nu}$, and the $3\sigma$ confidence bounds on the RHS further assume a unitary $V_{\rm PMNS}$ and include Super-Kamiokande atmospheric data --- see \cite{Esteban:2020cvm} for details.  As seen in \cite{Esteban:2020cvm} and also in Table \ref{tab:mixingangles}, current $3 \sigma$ oscillation constraints do not yet fully determine the quadrant of the atmospheric mixing angle $\theta_{23}^l$ and, at least in the normal ordering scenario, have only excluded $\sim 43\%$ of the available domain of the leptonic CP-violating phase $\delta^l$, i.e. $\delta^l$ is only constrained within a $\sim 200^{\degree}$ arc. This is reduced to an exclusion of only $\sim 20\%$ of the phase domain when not including SK data. 

\begin{table}[tp]
\centering
\renewcommand{\arraystretch}{1.75}
\begin{tabular}{|c||c|c|c|c|}
\hline
\multicolumn{5}{|c|}{Uncertainties on Fermionic Mixing Parameters}\\
\hline
\hline
\text{CKM Parameters}  & $\sin \theta_{12}^{q}$ & $\sin \theta_{23}^{q}$ &$\sin \theta_{13}^{q}$ & $\sin \delta_{CP}^{q}$  \\
\hline
($\mu = M_{IR}$) & $\left(\stackanchor{0.226}{0.224}\right)$ & $\left(\stackanchor{0.427}{0.411}\right)\cdot 10^{-1}$  & $\left(\stackanchor{0.380}{0.358}\right)\cdot 10^{-2}$ & $\left(\stackanchor{0.921}{0.899}\right)$ \\ 
\hline
($\mu = M_{UV}$) & $\left(\stackanchor{0.226}{0.224}\right)$ & $\left(\stackanchor{0.463}{0.219}\right)\cdot 10^{-1}$ & $\left(\stackanchor{0.409}{0.184}\right)\cdot 10^{-2}$ & $\left(\stackanchor{1.000}{0.194}\right)$ \\
\hline

\hline
\hline
\text{PMNS Parameters}   & $\sin \theta_{12}^{l}$ & $\sin \theta_{23}^{l}$ &$\sin \theta_{13}^{l}$ & $\sin \delta_{CP}^{l}$  \\
\hline
($\mu = M_{{IR,UV}}$) & $\left(\stackanchor{0.586}{0.519}\right)$ &   $\left(\stackanchor{0.776}{0.639}\right)$ & $\left(\stackanchor{0.156}{0.144}\right)$ & $\left(\stackanchor{0.588}{-1.000}\right)$ \\ 
\hline
\end{tabular}
\caption{Uncertainty estimates for fermionic mixing parameters.  In the CKM sector, the (experimental) IR bounds are given in \cite{Workman:2022ynf}, while the UV bounds are estimated by considering various input RGE/threshold correction parameter choices from \cite{Ross:2007az}, and accounting for the propagated IR experimental uncertainties.  In the PMNS sector we take 3$\sigma$ global bounds from NuFit, in the normal ordering scenario and incorporating Super-Kamiokande atmospheric data.}
\label{tab:mixingangles}
\end{table}
\begin{table}[t]
\centering
\renewcommand{\arraystretch}{1.75}
\begin{tabular}{|c||c|c||c|c|}
\hline
\multicolumn{5}{|c|}{Uncertainties on Fermionic Masses}\\
\hline
\hline
\text{Quarks}  & $m_{u}/m_{t}$ &  $m_{c}/m_{t}$ & $m_{d}/m_{b}$ & $m_{s}/m_{b}$ \\
\hline 
($\mu = M_{IR}$)  & $\left(\stackanchor{1.543}{1.097}\right)\cdot 10^{-5}$ & $\left(\stackanchor{7.509}{7.217}\right)\cdot 10^{-3}$  & $\left(\stackanchor{1.238}{1.069}\right)\cdot 10^{-3}$ & $\left(\stackanchor{2.452}{2.138}\right)\cdot 10^{-2}$  \\
\hline
($\mu = M_{UV}$) & $\left(\stackanchor{5.807}{1.592}\right)\cdot 10^{-6}$  & $\left(\stackanchor{2.576}{0.911}\right)\cdot 10^{-3}$ & $\left(\stackanchor{8.303}{3.570}\right)\cdot 10^{-4}$ & $\left(\stackanchor{1.729}{0.880}\right)\cdot 10^{-2}$ \\
\hline
\hline
\text{Leptons} & $m_{e}/m_{\tau}$ & $m_{\mu}/m_{\tau}$ & \multicolumn{2}{c|}{$\Delta m_{sol}^2 / \Delta m^2_{atm}$}   \\
\hline 
($\mu = M_{IR}$)  & $\left(\stackanchor{2.876}{2.876}\right)\cdot 10^{-4}$ & $\left(\stackanchor{5.947}{5.946}\right)\cdot 10^{-2}$ & \multicolumn{2}{c|}{$\left(\stackanchor{3.31}{2.63}\right)\cdot 10^{-2}$}   \\
\hline
($\mu = M_{UV}$)  & $\left(\stackanchor{2.875}{2.343}\right)\cdot 10^{-4}$ & $\left(\stackanchor{5.753}{5.096}\right) \cdot 10^{-2}$ & $\left(\stackanchor{3.31}{2.63}\right)\cdot 10^{-2}$ & $\left(\stackanchor{3.421}{0.124}\right)\cdot 10^{-1}$ \\
\hline
\hline
\text{Neutrinos} & $m_\beta$ [GeV] & $\langle m_{\beta\beta} \rangle$  [GeV]& $m_\Sigma$  [GeV]& \multirow{3}{*}{\shortstack{See text for \\ UV Estimates} }  \\
\cline{1-4} 
($\mu = M_{IR}$)  & $8\cdot10^{-10}$ & $6\cdot10^{-12}$ & $1.2\cdot10^{-10}$ &     \\
\cline{1-4}
($\mu = M_{UV}$)  & $1.12\cdot10^{-9}$ & $8.4\cdot10^{-12}$ & $1.68\cdot10^{-10}$ &  \\
\hline
\end{tabular}
\caption{The same as Table \ref{tab:mixingangles}, but for fermion masses.  We estimate the neutrino mass squared difference in the UV from \cite{Antusch:2003kp}, and recall that the $\xi$ ratio only differs from the IR when $\tan \beta$ is large and/or the neutrino mass spectrum is partially degenerate (see text), hence the two UV bounds for $\xi$ in the last row, with the left (right) cell corresponding to the low (high) $\tan \beta$ scenario. IR bounds for $m_{\beta(\beta)}$ and $m_\Sigma$ are taken from \cite{Planck:2018vyg,KamLAND-Zen:2016pfg,KATRIN:2021uub}, and their corresponding UV bounds are given by conservatively setting $\mathbb{s} = 1.4$ (see text).  
}
\label{tab:massfits}
\end{table}

The authors of \cite{Esteban:2020cvm} have also obtained global constraints on the differences of squared neutrino mass eigenvalues, finding at the $3\sigma$ confidence level
\begin{align}
\nonumber
\Delta m^2_{sol} &\equiv m_2^2 - m_1^2 \in \lbrace 6.82, 8.04 \rbrace^{3\sigma}\cdot10^{-5}\, \text{eV}^2 \,,\\
\label{eq:NuFitNuMassConstraint}
\Delta m^2_{atm} &\equiv m_3^2 - m_2^2 \in \lbrace 2.430, 2.593\rbrace^{3\sigma} \cdot10^{-3}\, \text{eV}^2\,,
\end{align}
in the normal-mass-ordering scenario relevant to the UTZ construction, and again including Super-Kamiokande atmospheric data.\footnote{Note that, by definition, the mass eigenvalues are labeled according to their relative magnitudes, i.e. $m_3 > m_2 > m_1$ in the normal-ordering scenario.}
We have translated this to a bound on the ratio $\xi \equiv \Delta m^2_{sol}/ \Delta m^2_{atm}$ in Table \ref{tab:massfits}.

A second class of neutrino mass constraints comes directly from cosmological probes.  For example, assuming the $\Lambda$CDM model and using data from the Cosmic Microwave Background's (CMB) angular spectra, the Planck experiment has put an upper bound on the sum of cosmologically stable neutrino masses $m_\Sigma$ of \cite{Planck:2018vyg}
\begin{equation}
\label{eq:plancksumconstraint}
    m_\Sigma \equiv \sum_i \, m_{\nu_i} < 0.26 \,\,\,\,\text{eV}
\end{equation}
at the 95 $\%$ confidence level.  When also including data from Baryon Acoustic Oscillations this bound is reduced to $m_\Sigma < 0.12$ eV \cite{Planck:2018vyg}, which can be yet further reduced to $m_\Sigma < 0.09$ eV when including Type Ia supernova luminosity distances and growth rate parameter determinations \cite{DiValentino:2021hoh} --- see \cite{DiValentino:2021imh} for a recent review.  

Finally, additional constraints in the neutrino mass and mixing sector originate in the effective mass terms controlling electron-neutrino 0$\nu\beta\beta$ decay and single $\beta$ decays,
\begin{align}
\label{eq:effectivemassconstraints1}
\langle m_{\beta \beta} \rangle &\equiv \big\vert \sum_{i} V_{e i}^2 \, m_i \big\vert \,<\, (61-165)\cdot 10^{-3}\,\,\text{eV}\,, \\
\label{eq:effectivemassconstraints2}
m_\beta &\equiv \sqrt{\sum_i \vert V_{e i} \vert^2 \, m_i^2} \,<\, 0.8 \,\,\text{eV}\,,
\end{align}
where $V_{ei}$ is the matrix element of the first row and $i$-th column of the PMNS matrix defined in \eqref{eq:Nufit}, and $m_i$ is the corresponding neutrino mass eigenvalue. Robust bounds for these quantities are provided by dedicated terrestrial experiments.  In \eqref{eq:effectivemassconstraints1} we have cited the KamLAND-Zen collaboration \cite{KamLAND-Zen:2016pfg}, while the limit in \eqref{eq:effectivemassconstraints2} is the 90\% confidence-level bound from KATRIN \cite{KATRIN:2021uub}.  KATRIN's future sensitivity is expected to reach $m_\beta < 0.2$ eV \cite{KATRIN:2019yun}.\footnote{See e.g. \cite{Agostini:2022zub,Cirigliano:2022oqy} and references therein for a recent summary review of these independent flavour constraints and the ability to use them to probe neutrino mass models.}

At this point we should clarify that, while in practice we must specify a numerical value for the scale $M_\theta$ when applying the seesaw formula \eqref{eq:seesaw},\footnote{...which amounts to specifying \emph{both} the Dirac and Majorana neutrino mass scales and hence their relative hierarchy, as we have no way of differentiating $M_\theta$ from $ \tilde{M}_\theta \equiv s_\nu^2/M_\theta$ in the seesaw formula...} and must therefore count this as a relevant IR model parameter in our theory (unlike the charged fermion case), this choice completely determines the overall neutrino mass scale.  As this is a free UTZ model parameter, we can vary it between sensible UV seesaw scales to accommodate (e.g) \eqref{eq:NuFitNuMassConstraint}-\eqref{eq:effectivemassconstraints2}, and probe said variation's effect on the relative-scale-setting parameters $\lbrace a_\nu, b_\nu, x, y \rbrace$.  We have done so between $M_\theta \in 10^{10-12}$ GeV, and observe in Figure \ref{fig:MCMCparamplots} that, at least qualitatively, larger values of $M_\theta$ are preferred. Regardless, as with the charged fermion spectrum, we only consider UTZ predictions for \emph{ratios} of observables that depend on the overall neutrino mass scale, where $M_\theta$ cancels, as truly meaningful.  For this reason we will present $\xi$, $\langle m_{\beta(\beta)}\rangle / m_\Sigma$, and $\langle m_{\beta\beta}\rangle /m_\beta$ as predictions in Section \ref{sec:MCMC}, but not their individual constituent parameters $m_{\nu_i}$ or $m_{\beta(\beta)}$, although we can report these based on the various $M_\theta$'s identified in the MCMC evolution, and indeed $m_{\beta(\beta)}$, $\Delta m^2_{sol,atm}$, and $m_\Sigma$ should be seen as giving reliable \emph{constraints} on the MCMC algorithm, in that once a value for $M_\theta$ has been settled on, their associated values must still be consistent with observation.
\subsection{Renormalization Group Evolution Uncertainties}
\label{sec:RGE}
 A critical uncertainty for any prediction of the UTZ model comes from the fact that \eqref{eq:MDIR}-\eqref{eq:MMAJ} are the textures associated to the \emph{UV} theory.  Any comparison with data must be made at the scale where said experimental constraints are obtained, which in the case of the UTZ model is orders of magnitude below where \eqref{eq:MDIR}-\eqref{eq:MMAJ} hold.  Thankfully the Renormalization Group (RG) evolution required to account for this scale separation is well-studied in the context of a background SM or minimally-SUSY SM (MSSM) spectrum (consistent with our vacuum alignment mechanism discussed above), for both the charged fermion \cite{Ross:2007az,Olechowski:1990bh,Chiu:2016qra} and neutrino sectors \cite{Antusch:2005gp,Antusch:2003kp,Chankowski:1999xc,Casas:1999tg,Casas:1999ac,Gupta:2014lwa}. For example, assuming a large flavour-breaking scale and a background SUSY spectrum allowing for high-scale gauge-coupling unification, the phenomenological charged-fermion structures discussed between \eqref{eq:mm1}-\eqref{eq:GST} are already consistent with RGE and threshold corrections from the IR to UV --- see e.g. \cite{Ross:2007az} --- up to uncertainties regarding (e.g.) the underlying SUSY breaking scale and parameter spaces (in particular the ratio of Higgses, $\tan \beta$), which can drive some mass and mixing-angle splittings.  As we do not specify $\tan \beta$ or other parameters and/or fields beyond those of the Yukawa sector of the EFT presented in Section \ref{sec:LOLAGRANGIAN}, we have used  \cite{Ross:2007az} to estimate the overall uncertainty associated to UV quark mass and mixing parameters, accounting for the broad range of possible theory parameters studied therein, and of course propagating updated IR experimental uncertainties from \cite{Workman:2022ynf} to the UV. These estimates are reported in Table \ref{tab:mixingangles}-\ref{tab:massfits}.

Moving to the neutrino sector and assuming a Type-I seesaw mechanism,  detailed RGE and threshold correction analyses can be found in \cite{Antusch:2003kp,Antusch:2005gp} and references therein. There one concludes that, in the absence of a conspiracy between special alignments of phases, large $\tan \beta$, and/or a (partially-)degenerate\footnote{Here a `degenerate' spectrum corresponds to $ \Delta m^2_{atm} \ll m^2_3 \sim m^2_2 \sim m^2_1$, while a `partially degenerate' spectrum corresponds to $\Delta m^2_{sol} \ll m_1^2 \lesssim \Delta m^2_{atm}$.  Hierarchical neutrinos satisfy $m_1^2 \ll \Delta m^2_{sol}$ in the normal-ordering scenario.} light neutrino spectrum, radiative corrections to PMNS mixing angles and phases between disparate scales is generally minimal,
\begin{align}
    \Delta \theta_{ij}^l \equiv \theta_{ij}^l (\Lambda_{M_1}) - \theta_{ij}^l (\Lambda_{M_Z}) &\sim \frac{1}{2} \cdot \ln \frac{M_1}{M_Z}\cdot 10^{-6}\cdot \left(1+\tan^2 \beta \right)\cdot \Gamma_{\text{enh}} \,,
\end{align}
for the lowest seesaw scale of $M_1$, and with $\Gamma_{\text{enh}} = \lbrace 1, \sqrt{\xi}, 1, \sqrt{\xi}/\theta_{13}^l, \sqrt{\xi} \rbrace$ for $\lbrace \theta_{12}^l, \theta_{13}^l, \theta_{23}^l, \delta^l, \phi_i  \rbrace$, respectively.  Taking rough order-of-magnitude estimates for  $\Gamma_{\text{enh}}$ and allowing for $M_1$ as large as $M_{\text{GUT}} \sim 10^{16}$ GeV and $\tan \beta$ as large as 50, one sees that typically $\Delta \theta_{ij}^l \lesssim \mathcal{O}(10^{-2})$, which is largely insignificant in comparison to the experimental uncertainties on mixing parameters given in Table \ref{tab:mixingangles}, except for possible corrections to $\theta_{13}^l$.  Given that we predict a normally-ordered, hierarchical mass spectrum as a result of the sequential dominance condition of \eqref{eq:sequentialdominance},\footnote{Note that the presence of a dominant scale $M_3$ in the RH neutrino mass matrix also minimizes the \emph{inter-}threshold radiative effects between $M_i$ \cite{Antusch:2005gp}.} we can take the $3\sigma$ bounds of \eqref{eq:Nufit} to hold in the UV, implying that the experimental uncertainties are large in comparison to radiative effects.  

On the other hand, the light neutrino mass eigenvalues are far more sensitive to RGE  than are the PMNS parameters, even in a hierarchical system.  Assuming small $\tan \beta$, neutrino mass eigenvalues generally evolve with a common scaling, $m_{\nu,i}(\mu) \approx \mathbb{s}(\mu,\mu_0)\, m_{\nu,i}(\mu_0)$  with (e.g.) $\mathbb{s} \approx 1.1-1.2$ for $\tan \beta \approx 10$ or $\mathbb{s} \approx 1.35-1.4$ for SM-like running.  This obviously leads to a UV enhancement of the neutrino mass differences in \eqref{eq:NuFitNuMassConstraint} $\propto \mathbb{s}^2$, but this effect cancels in the ratio $\xi$.  On the other hand, large $\tan \beta$ can drive UV flavour splittings amongst the neutrino mass eigenvalues, evolving both $\Delta m_{sol,atm}^2$ and $\xi$, an effect which is especially enhanced in the case of a (partially-)degenerate spectrum, and which is considerably uncertain when allowing for generic phase configurations.  We have used \cite{Antusch:2003kp} to estimate the effect on $\xi$ in this regime in Table \ref{tab:mixingangles}, where one sees that an uncertainty greater than an order of magnitude in principle exists, although this is quite conservative given the neutrino mass domain considered in \cite{Antusch:2003kp}, and the fact that we can constrain our MCMC scan to prefer a hierarchical mass spectrum, i.e. $\Delta m^2_{sol}/m_{\nu_1}^2 \gg 1$.\footnote{While sequential dominance \eqref{eq:sequentialdominance} naturally generates a hierarchical spectrum, variations of the relative-scale-setting neutrino coefficients $\lbrace a_\nu, b_\nu, x, y\rbrace$ can in principle edge the spectrum towards partial degeneracy.  We have applied a likelihood of 1 to any value of $\Delta m^2_{sol}/m_{\nu_1}^2 > 10$ found in our MCMC scans, and have applied a smoothing, Gaussian-like corrective factor to assign likelihoods for values close to this threshold.}  
Finally, we note that RGE discussed above also impacts the UV values of  \eqref{eq:plancksumconstraint}-\eqref{eq:effectivemassconstraints2}, which serve as constraints on the MCMC system. In Table \ref{tab:massfits} we have estimated these in the (conservative) SM-like scenario, with $\mathbb{s}=1.4$ for all neutrino species.

In summary, we will apply the UV bounds in Tables \ref{tab:mixingangles}-\ref{tab:massfits} to account for a rather generic class of RGE and threshold corrections to fermionic mass and mixing in the UTZ.  They will allow us to robustly explore the UTZ's predictions without introducing unnecessary assumptions about the background field content and/or non-flavour parameter spaces that are irrelevant to the EFT construction at hand, which is designed to be as model-independent as possible.

\section{An MCMC Scan of Parameter Space}
\label{sec:MCMC}

A proof-in-principle numerical analysis of the UTZ predictions derived from \eqref{eq:MDIR}-\eqref{eq:MMAJ} was originally performed in \cite{UTZ}, in order to show that the model was consistent with available mass and mixing data at the time.  This semi-analytic study, while successful, relied on a largely heuristic contour analysis to identify a viable region of the UTZ parameter space. However, the analysis was incomplete in many ways, in that it did not
\begin{enumerate}
    \item \label{item1} exhaustively explore the available UTZ model space, robustly accounting for all theory correlations amongst its Lagrangian parameters and therefore conclusively determine whether the LO UTZ effective Lagrangian adequately describes nature;
    \item \label{item2} explore the complete set of corrections coming from NLO effective operators as discussed in Section \ref{sec:HO}.  Only the largest corrections identified in the Dirac Lagrangian were briefly considered in \cite{UTZ}, and only in the down-quark sector (the corrections parameterized by $d_d$ and $\psi_d$).
    \item \label{item3} identify sufficiently generic \emph{predictions} for (e.g.) the  CP-violating phases $\delta^l$ and $\phi_{1,2}$ or PMNS atmospheric angle $\theta_{23}^l$, when all other (well-measured) flavour parameters were simultaneously resolved by the UTZ;
    \item \label{item4} consider in any way the experimental constraints from, nor predictions for, neutrino-sector observables like $0\nu\beta\beta$, single $\beta$-decay rates, or the sum of neutrino masses $m_\Sigma$.
\end{enumerate}
Furthermore, the experimental datasets available for theory comparison have been updated since the original publication of \cite{UTZ}. All of these considerations motivate us to revisit our phenomenological analysis of the UTZ in order to better determine its viability and identify means of falsifying it.  However, given the number of free parameters introduced by \eqref{eq:DiracUTZLO}, \eqref{eq:MajoranaUTZLO}, and even \eqref{eq:DiracUTZHO}, numerical techniques more sophisticated than those applied in \cite{UTZ} will be necessary to achieve \ref{item1}-\ref{item4}. To that end, in this work we consider a Markov Chain Monte Carlo (MCMC) algorithm for exploring the UTZ.

\subsection{The Generic MCMC Algorithm}
\label{sec:ALGORITHM}

Our numerical analysis will rely on a Metropolis-Hasting MCMC algorithm. The purpose of this approach is to find the posterior distribution of the model after applying relevant experimental constraints, thereby obtaining viable, high-likelihood UTZ parameter regions. The MCMC technique has proven to be very powerful when applied to the exploration of high-dimensional parameter spaces, with physics applications originating in phenomenological studies of SUSY extensions of the SM \cite{RuizdeAustri:2006iwb,Baer:2008jn, DeCausmaecker:2015yca}, cosmology \cite{Trotta:2008qt}, and the determination of parton distribution functions \cite{Mangin-Brinet:2017vkg}. More recently, two publications have used this approach to study the viability of a flavoured SUSY SU(5) model \cite{Bernigaud:2021kpw} and a scotogenic model for loop-induced neutrino masses \cite{Sarazin:2021nwo}, from which we will follow most of the methodology.

The algorithm is based on an iterative process where every new proposed parameter point is selected in an area near to the previous one, and its estimated viability drives its acceptance in the chain. To be more explicit, every Markov chain starts on a randomly selected point within the parameter interval ranges. Then, on every iteration, a new point with parameters $\Vec{\theta}^{n+1}$ is proposed in the vicinity of the previous point with parameters $\Vec{\theta}^n$. In our study, the new proposed parameter value is computed according to
\begin{equation}
    \theta_{i}^{n+1} = \Pi\left\{ \theta_{i}^{n},\, \kappa \left( \theta_{i}^{\text{max}} - \theta_{i}^{\text{min}}\right) \right\} \,,
\end{equation}
where $\Pi \left\{ a,b \right\}$ is a Gaussian distribution with mean value $a$ and standard deviation $b$. The parameter $\kappa$ parametrizes the allowed jump length between two iterations, and its value is chosen empirically in order to maximize the efficiency of the algorithm. If the calculated value exceeds the limits of the defined intervals for the model parameters, the point is rejected.

We then compute the global likelihood associated with the proposal point $\mathcal{L}^{n+1}$, which is accepted with a probability
\begin{equation}
    p = \text{min}\left( 1,\, \mathcal{L}^{n+1}/\mathcal{L}^n\right)\,,
\end{equation}
which enforces the acceptance of points with higher likelihood and conditions the acceptance to the viability of the proposal point with respect to the previously accepted one. Since our objective is to evaluate the global posterior distribution for the parameters, accepting points with lower likelihoods and using a large number of chains is important both in attempting to avoid an MCMC evolution where the chain gets trapped in local maxima that might be fine-tuned, and also in helping to enforce a better distribution across the full parameter space. 

For simplicity, we assume here that our experimental constraints are not correlated and that the global likelihood is simply the products of the individual likelihoods, i.e
\begin{equation}
    \mathcal{L}^n ~\equiv~ \mathcal{L}(\Vec{\theta}^n, \Vec{O}) 
        ~=~ \prod_{i} \mathcal{L}_{i}^n(\Vec{\theta}^n,O^i) \,,
    \label{Eq:Likelihood}
\end{equation}
where $\Vec{O}$ is the set of experimental observables used as constraints. Furthermore, we assume a Gaussian shape for all the constraint likelihoods where uncertainties are given in Table \ref{tab:mixingangles}-\ref{tab:massfits}, except for constraints that only correspond to upper or lower bounds.  In these latter cases we apply a step function whose likelihood is assigned to 1 if the bound is satisfied, and which otherwise employs a Gaussian `corrective' factor that diminishes the likelihood assigned to the phase-space point as a function of the extent to which the bound is violated.   Within this numerical setup, the chain will converge to high-likelihood domains whose area represents the viability of the models according to the applied uncertainties on the constraints.

Additionally, in order to speed up the convergence process, we modify our jump parameter $\kappa$ to include a memory of proposal tries
\begin{equation}
\label{eq:adaptiveproposal}
    \kappa(t) = (1-\epsilon)^t \kappa_0 \,,
\end{equation}
where $t$ is the number of tries before accepting a new point in the chain. This becomes extremely helpful as the chain converges since some parameters might have a very thin data-compatible range. As soon as one point is accepted, $t$ is set to 0 again, maintaining the chain's ability to jump to another parameter region.\footnote{Note that, by setting $\epsilon = 0$, we have checked that the additional proposal in \eqref{eq:adaptiveproposal} does not impact our conclusions; upon running our scripts with this modification (albeit with slightly lower statistics) we find the same qualitative results as presented in Figure \ref{fig:MCMCparamplots} for the model parameters and subsequent Figures for associated physical parameters, which are generated with \eqref{eq:hyperranges}.}  

Finally, we focus solely on points within a chain for which the convergence already occurred (i.e. where maximums of likelihoods are reached). Therefore, we set up a `burning-length' parameter which automatically removes the first $N_{\text{burn}}$ points of each chains. This parameter is once again chosen empirically during the pre-runs by studying multiple likelihood evolution plots.

For clarification and to summarize, we emphasize that the MCMC strategy presented above is associated to Bayesian statistical methods. Our goal is to obtain the posterior distribution for UTZ model parameters while taking into account available experimental constraints. During this process, our algorithm will successively compare parameter points while keeping the best ones, within a probability associated to their respective likelihood ratios. In this approach, prior knowledge is represented by the parameter distributions in the absence of any constraints.  We have performed such `constraint-free' scans and found that every parameter in the scan follows a uniform prior distribution, which demonstrates that the final posterior distributions we present in the following Sections are solely explained by the use of experimental constraints (also see Footnote 12).

Given the above discussion, we present the different MCMC hyper-parameters that we use in our setup : $N_{\text{burn}}$, $\kappa_0$, $\epsilon$, the number of chains launched $N_\text{chains}$, and the length of the chains $L_\text{chain}$ which determine the target number of accepted points for every chain. All these parameters are chosen empirically depending on the model and the final statistics desired for the distributions. As a final comment, we note that it is usually better to allow for more chains, rather than longer chains, as this ensures a more reliable parameter exploration.

\subsection{UTZ Specifics}
\label{sec:UTZMCMC}

\begin{table}[t]
\centering
\renewcommand{\arraystretch}{1.45}
\begin{tabular}{|c||c|c|c|c|}
\hline
\multicolumn{5}{|c|}{LO UTZ Model Parameter MCMC Ranges \& Global Best Fits}\\
\hline
\hline
 & $(a,b)_{d} \cdot10^{3}$ & $(a,b)_{u} \cdot10^{5}$  & $(a,b)_{\nu} \cdot 10^{1}$ & $(x,y)\cdot10^{3}$ \\
\hline
\text{Range} & $\left(\left[2,6\right],\left[10,20\right]\right)$  &$\left(\mp30, \mp800\right)$ & $\mp 5$ & $\mp 5$ \\
\hline
\text{LO} & $\left(3.579, 15.924\right)$  &  $\left(6.720,-192.922\right)$ &  $\left(-1.166,1.818\right)$ & $\left(-0.146,-4.641\right)$ \\
\hline
\hline
\text{HO} & $\left(3.415, 15.416\right)$ & $\left(7.604,-200.279\right)$ & $\left(-1.819,2.440\right)$ & $\left(3.728,3.501\right)$\\
\hline
\hline
& $(\gamma,\delta)_{d}$ & $(\gamma,\delta)_{\nu}$ & $(\rho,\phi)$ & $M_\theta \cdot10^{-11}$ [GeV]  \\
\hline
\text{Range} &  $\left[0, 2\pi\right]$   &  $\left[0, 2\pi\right]$ &  $\left[0, 2\pi\right]$  & $\left[0.1,10\right]$  \\
\hline
\text{LO} & $\left(3.910, 5.782\right)$  &  $\left(3.163,4.553\right)$ & $\left(2.964,4.784\right)$ & $3.084$  \\
\hline
\text{HO} & $\left(4.228, 6.134\right)$ &  $\left(0.464,2.293\right)$ &  $\left(3.636,3.976\right)$ & $9.918$\\
\hline
\end{tabular}
\\[0.5cm]
\begin{tabular}{|c||c|c|c|}
\hline
\multicolumn{4}{|c|}{HO UTZ Model Parameter MCMC Ranges \& Global Best Fits}\\
\hline
\hline
& $(c,d)_{d} \cdot10^{5}$ & $(c,d)_{u} \cdot10^{6}$  & $(c,d)_{\nu} \cdot10^{3}$  \\
\hline
\text{Range} & $\left(\mp5,\mp50\right)$  & $\left(\mp5,\mp50\right)$ & $\left(\mp5,\mp50\right)$ \\
\hline
\text{HO} &  $\left(0.640, 10.811\right)$ & $\left(0.916,-37.298\right)$ & $\left(-0.896,-1.565\right)$\\
\hline
\end{tabular}
\caption{The scan ranges of UTZ model parameters, along with the 
value of the model parameter in the global best-fit dataset, for both LO and HO fits.  Recall that only two charged-fermion phase parameters are non-redundant at LO \cite{Roberts:2001zy}, and so we have chosen $\lbrace \gamma_d,  \delta_d \rbrace$, as in \cite{UTZ}.  Graphical representations of the MCMC evolution of these parameters are given in Figure \ref{fig:MCMCparamplots}. Also recall that there are no relevant HO Majorana corrections, that we have kept the HO corrections real, and that the global best-fit values identified for the phases are not terribly meaningful, as we do not observe very strong MCMC preferences for any phase values in our scans (they are all relatively evenly distributed across $\left[0,2\pi\right]$).}
\label{tab:MCMCranges}
\end{table}
\begin{figure}[tp]
\centering
\includegraphics[scale=0.23]{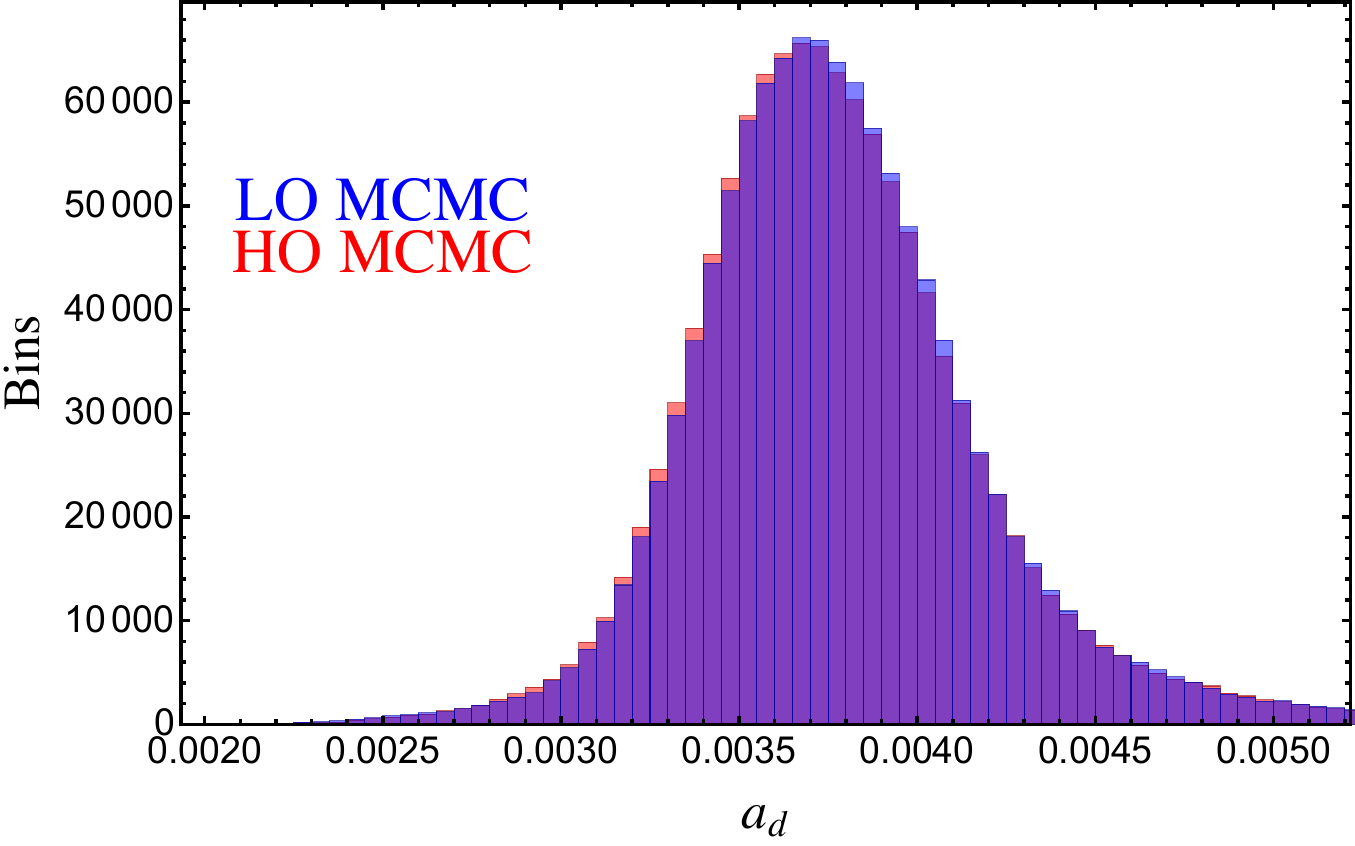}
\includegraphics[scale=0.225]{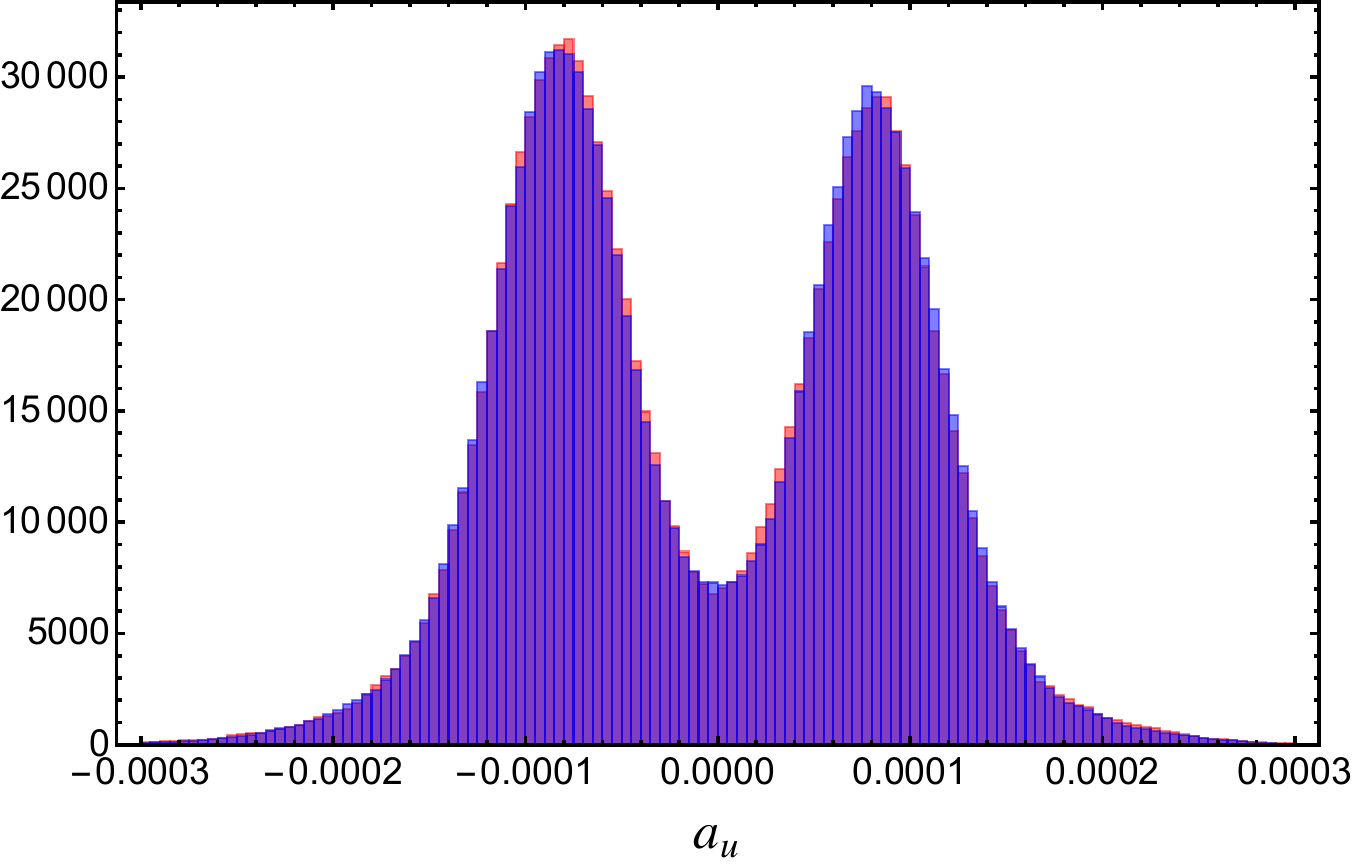}
\includegraphics[scale=0.225]{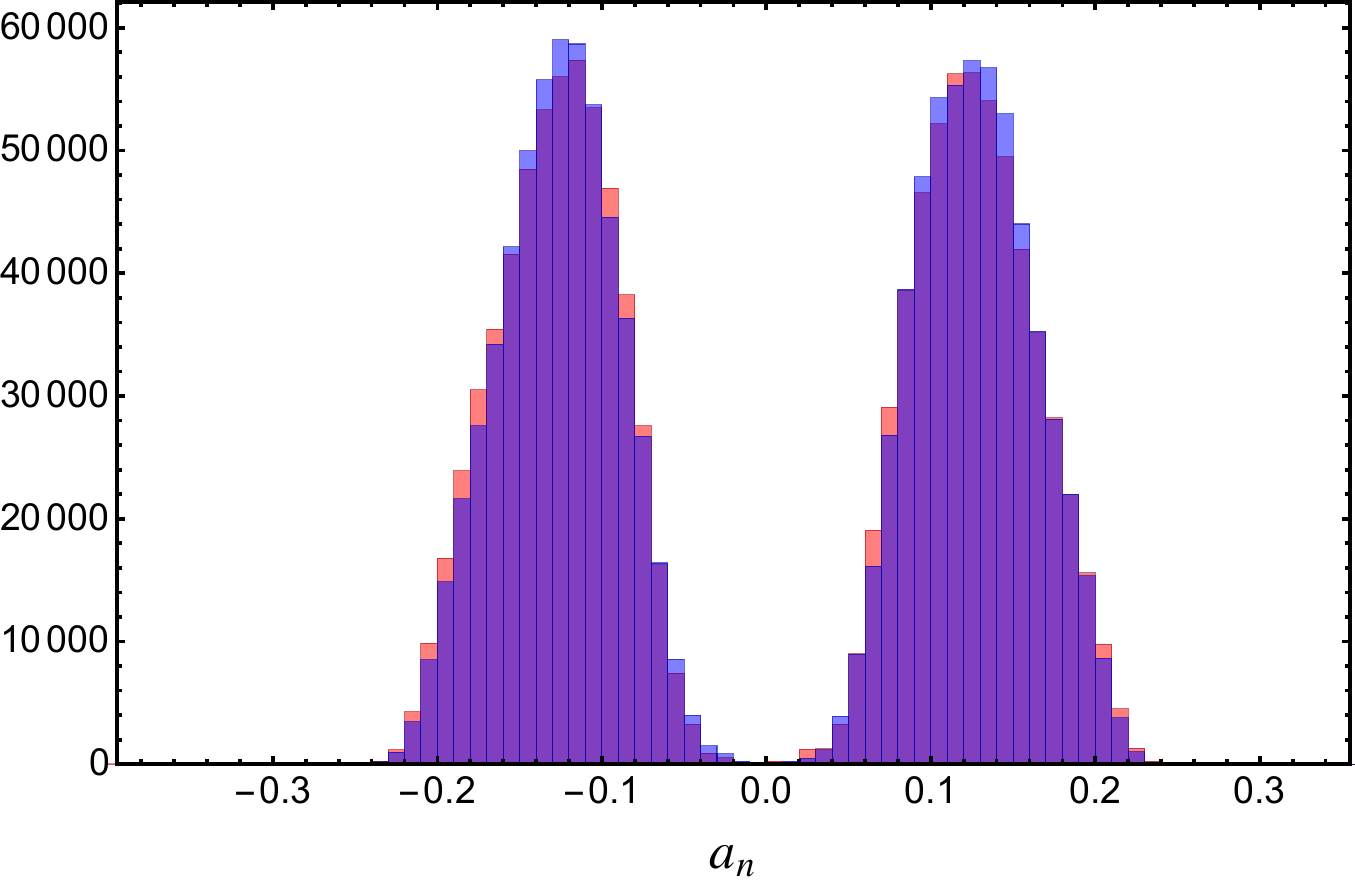}\\
\includegraphics[scale=0.225]{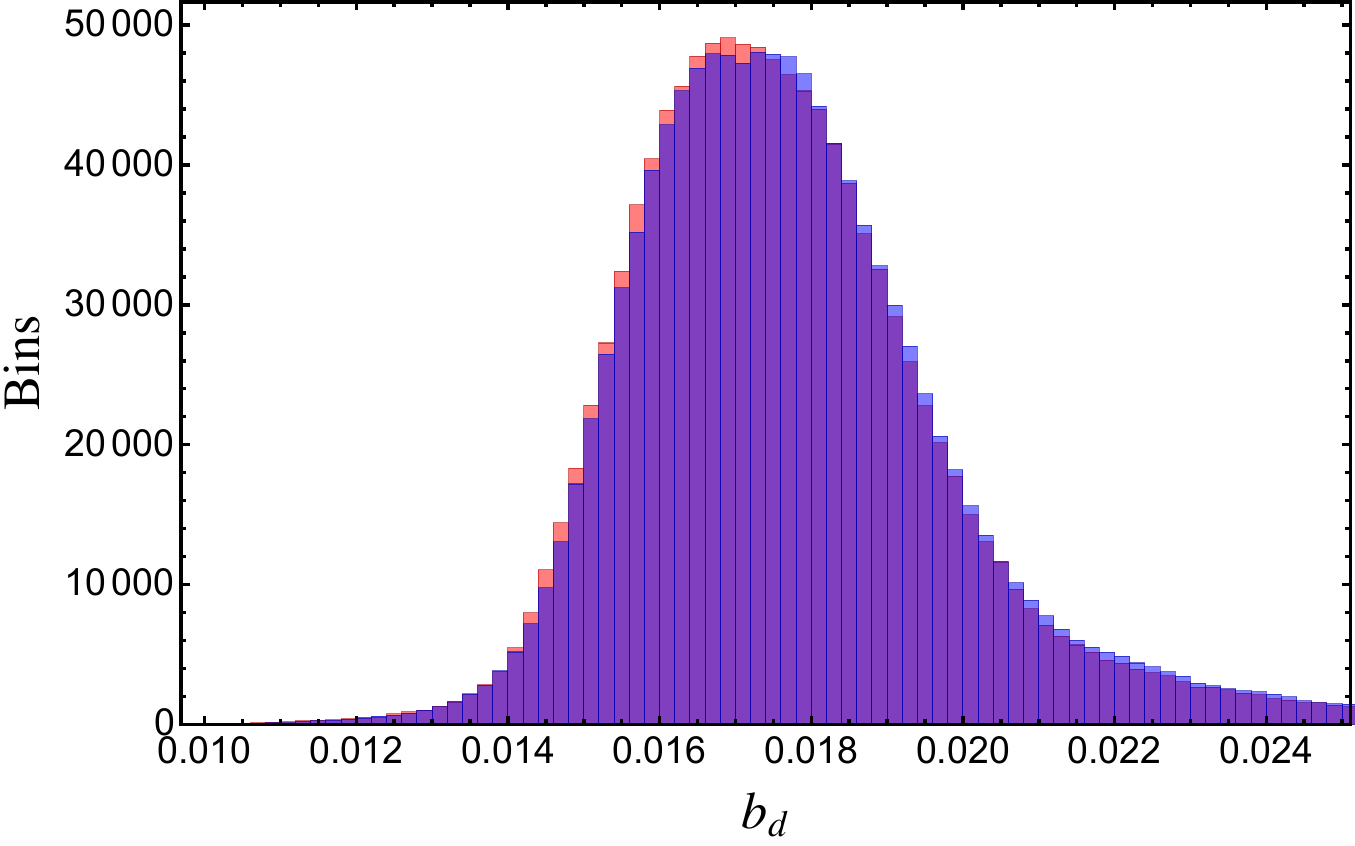}
\includegraphics[scale=0.225]{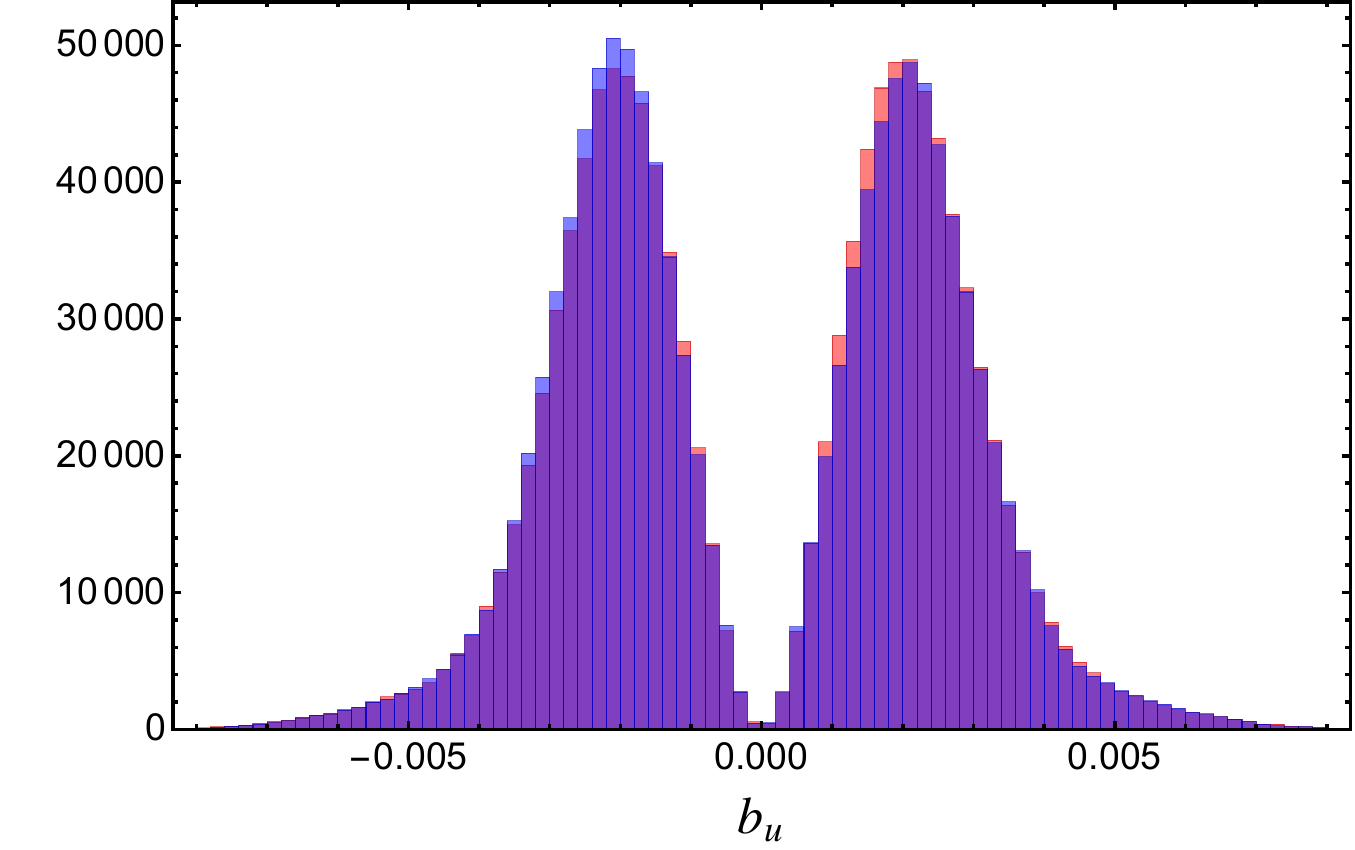}
\includegraphics[scale=0.225]{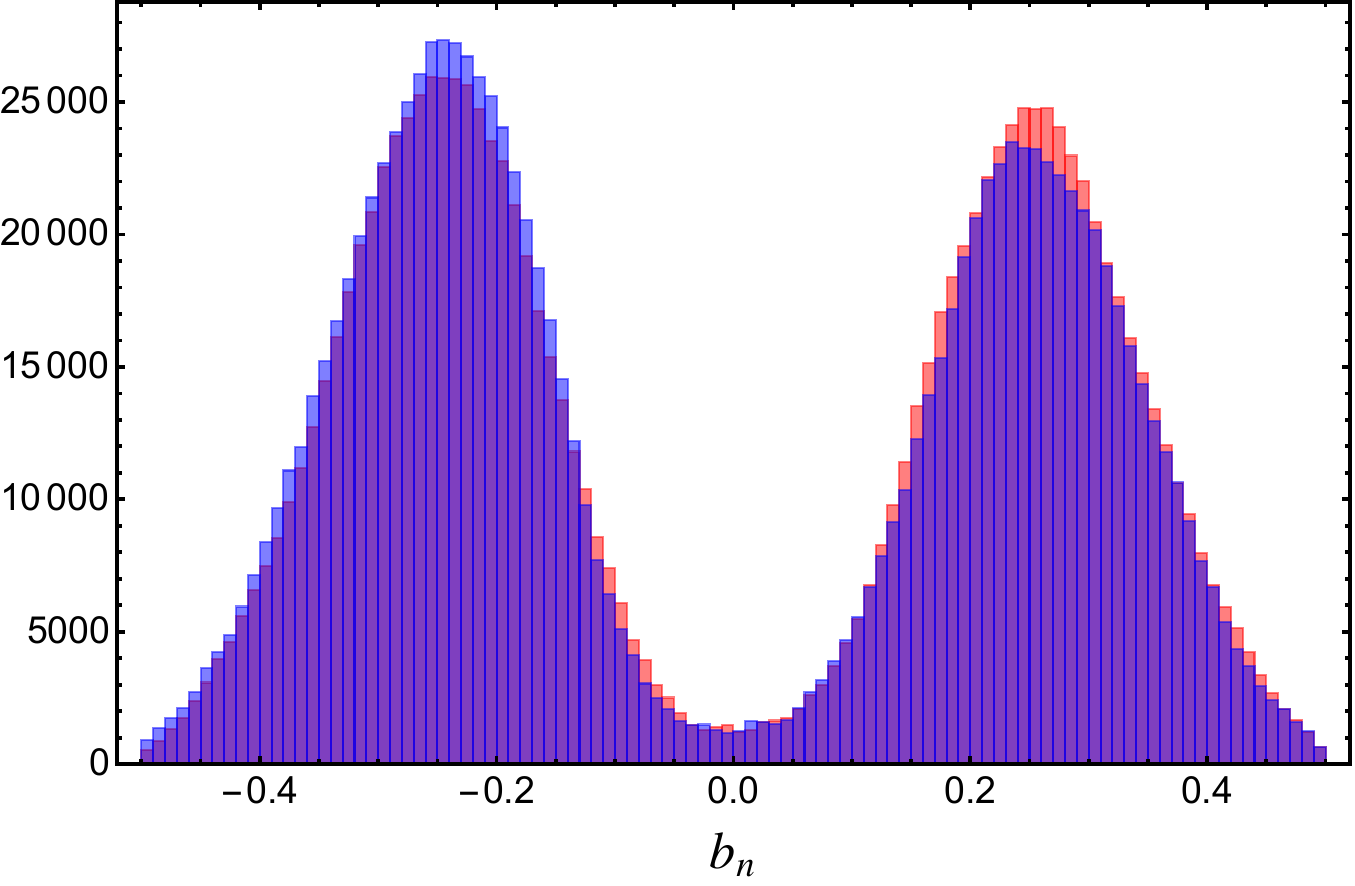}\\
\includegraphics[scale=0.225]{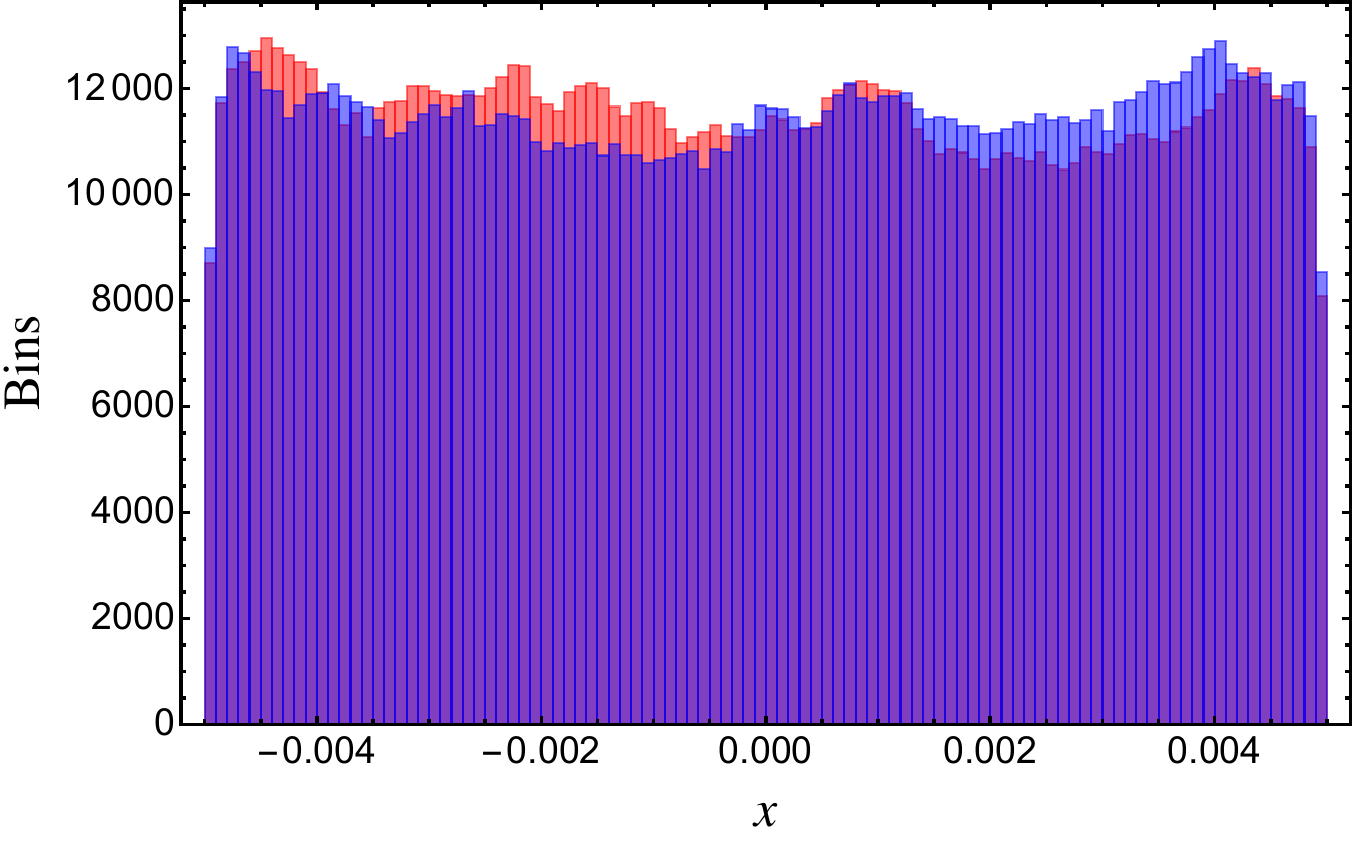}
\includegraphics[scale=0.225]{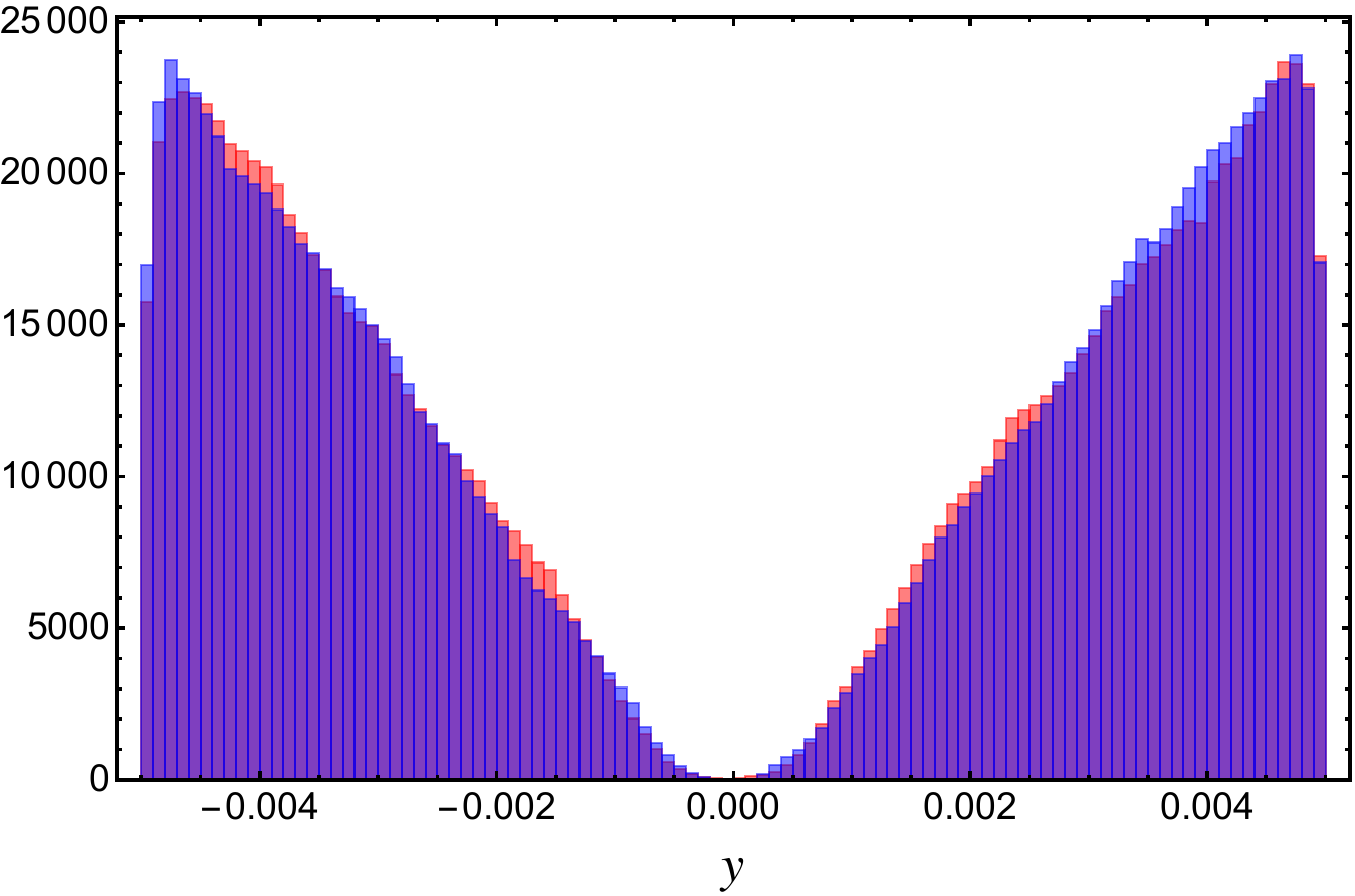}
\includegraphics[scale=0.225]{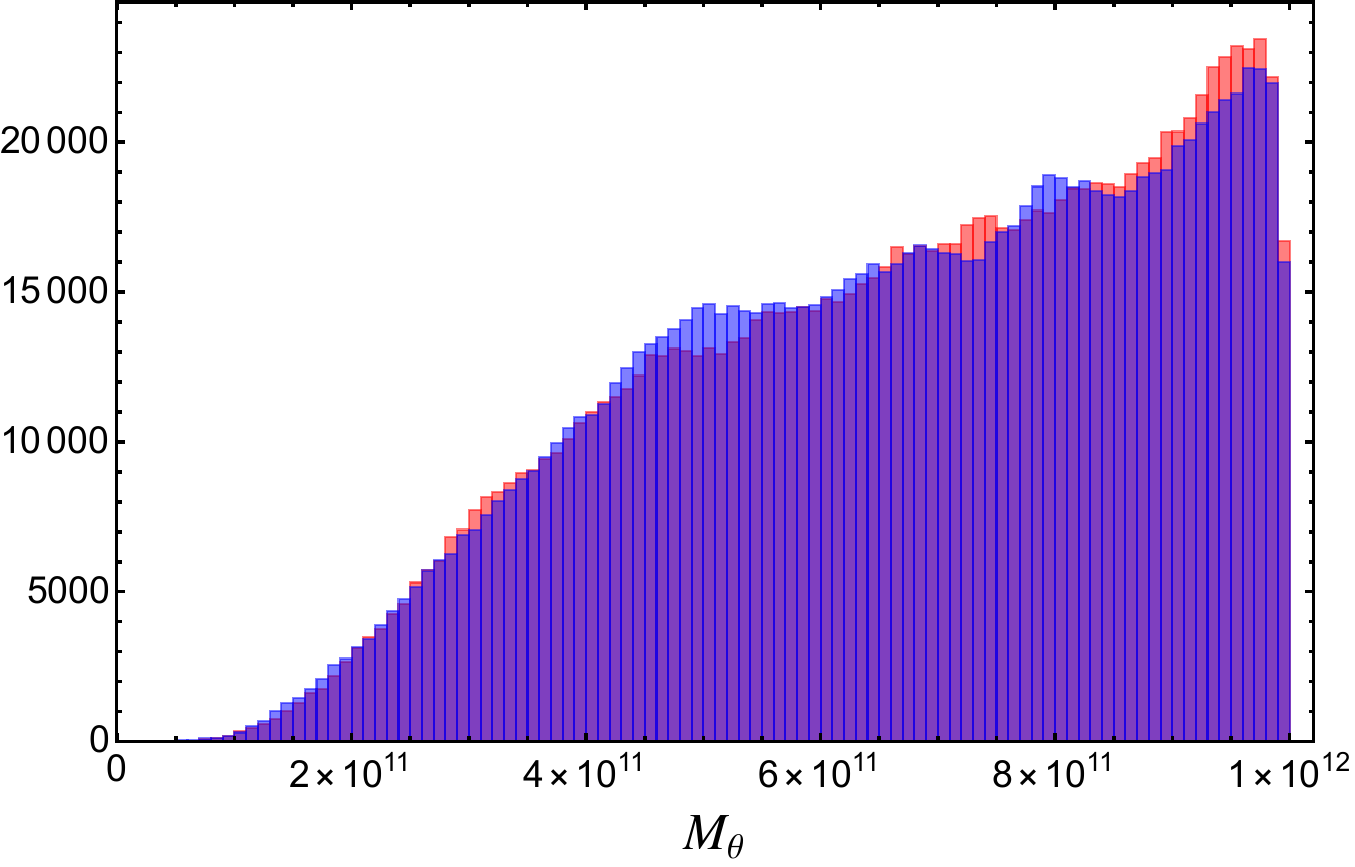}\\
\includegraphics[scale=0.225]{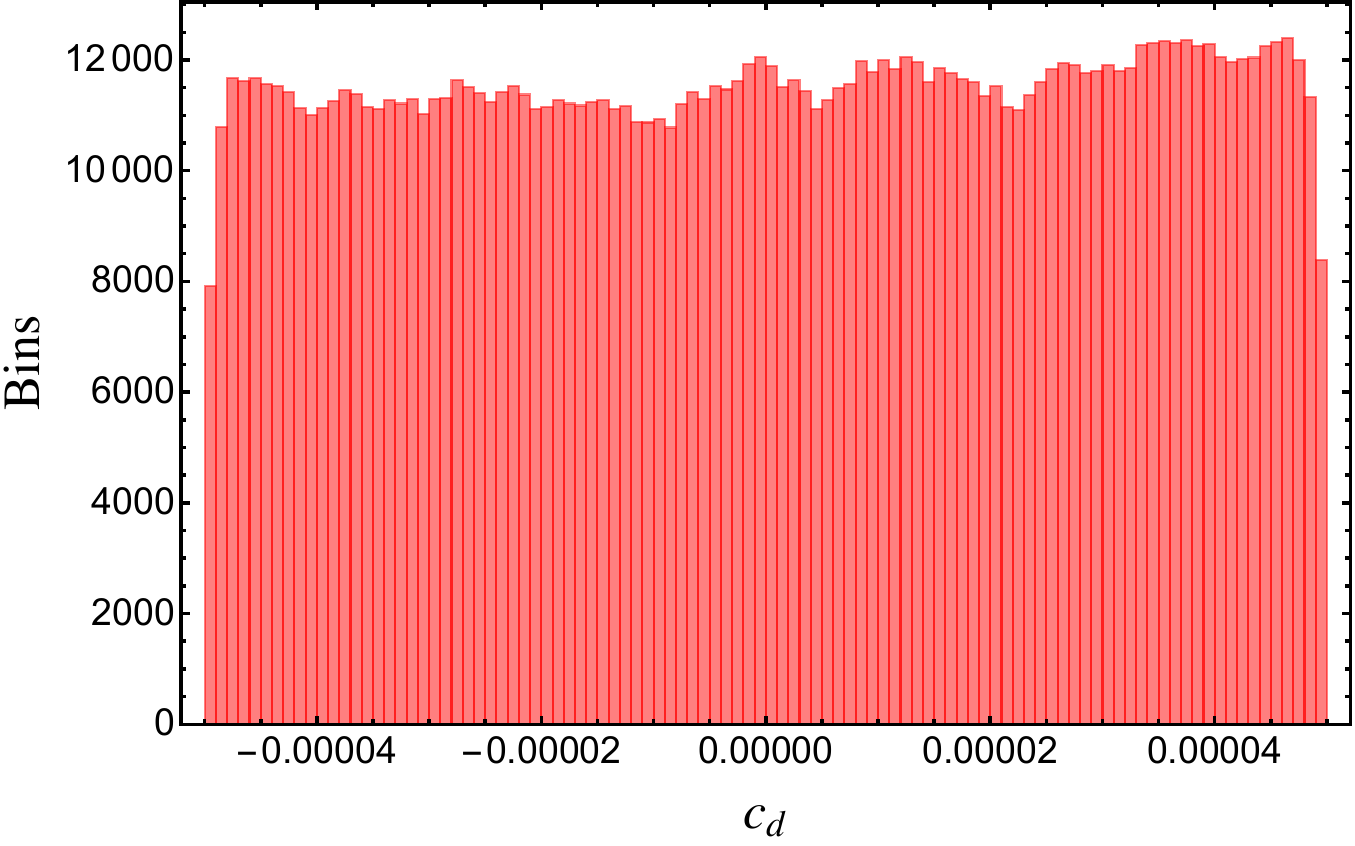}
\includegraphics[scale=0.225]{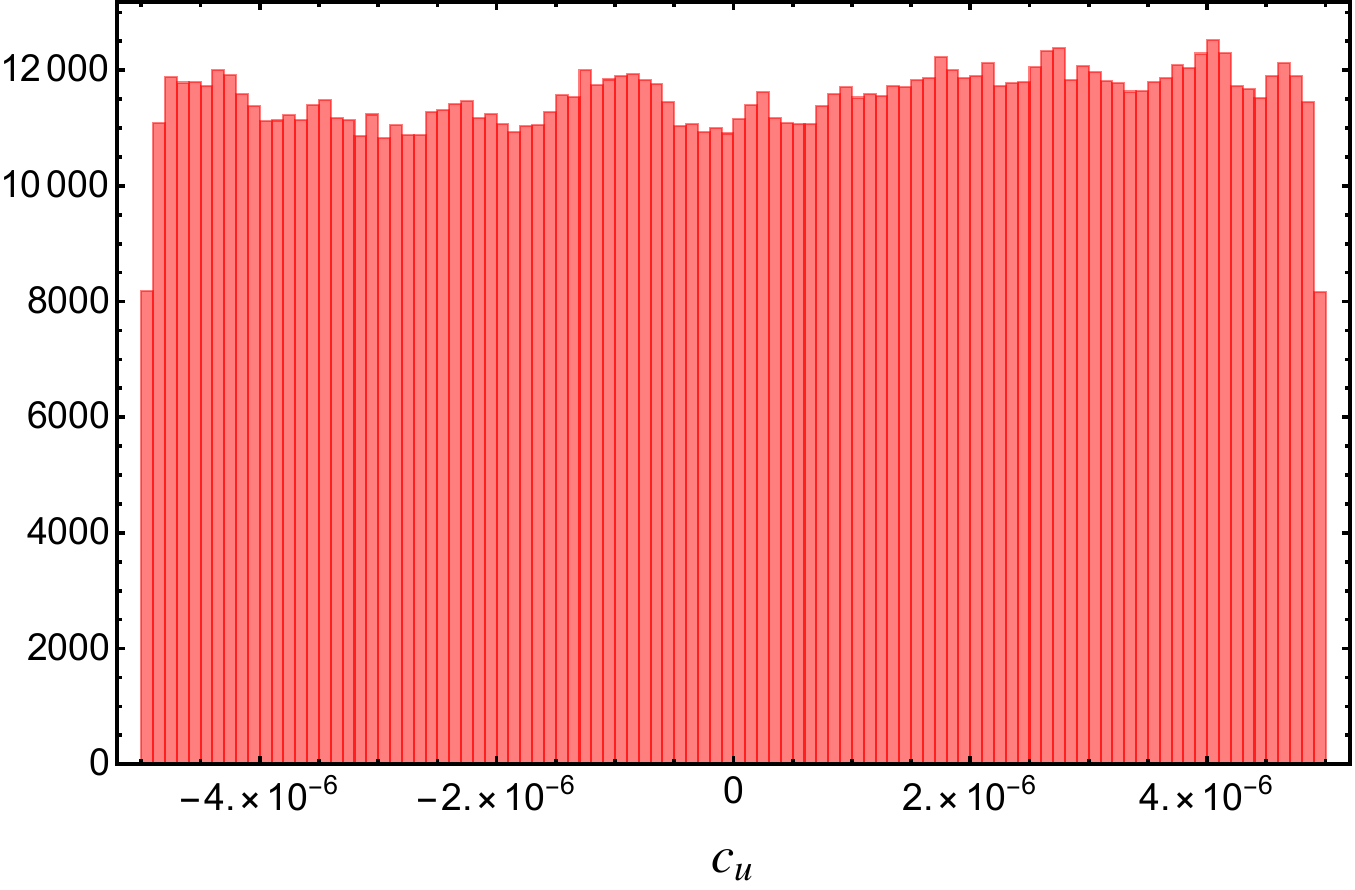}
\includegraphics[scale=0.225]{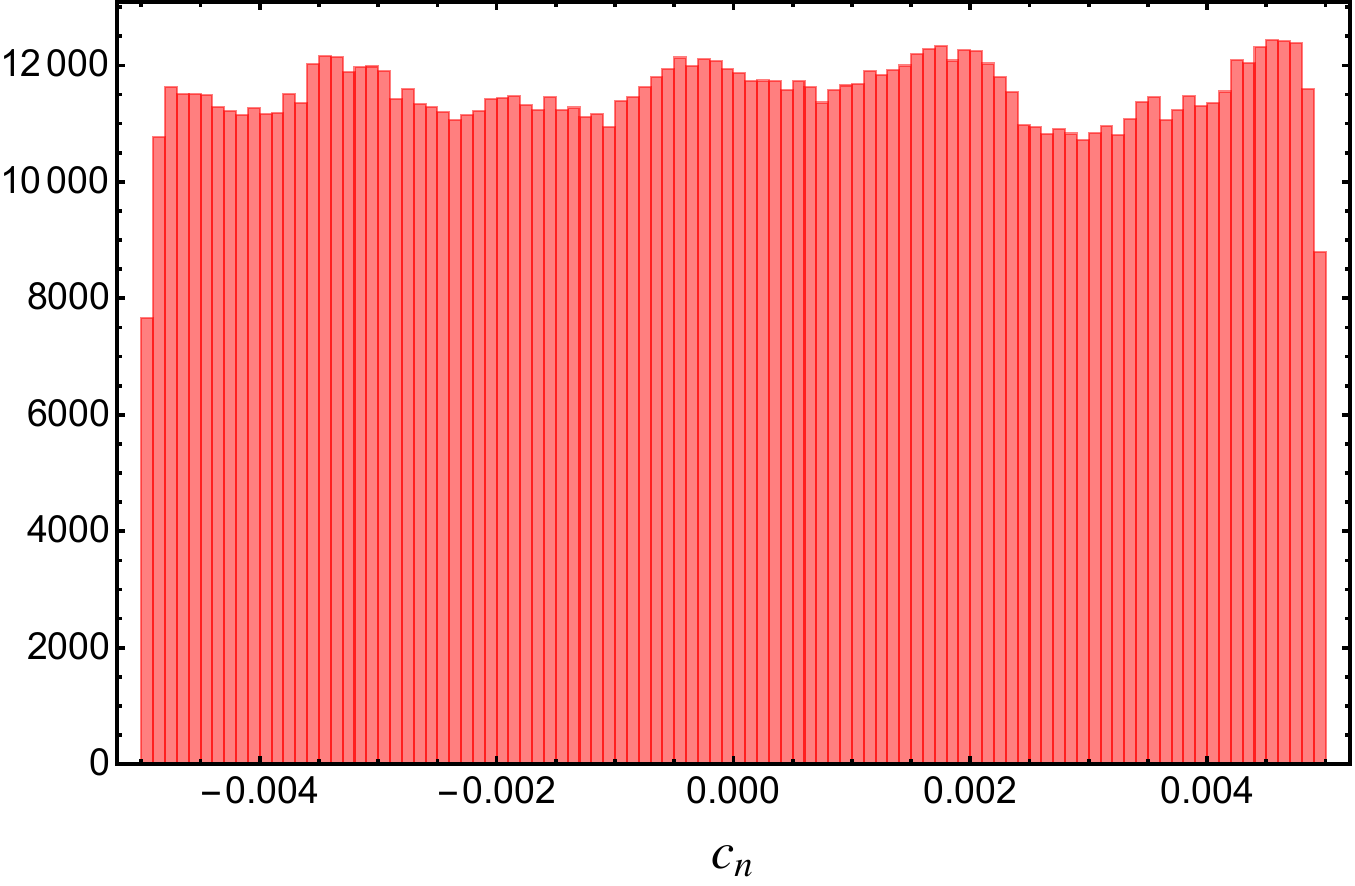}\\
\includegraphics[scale=0.225]{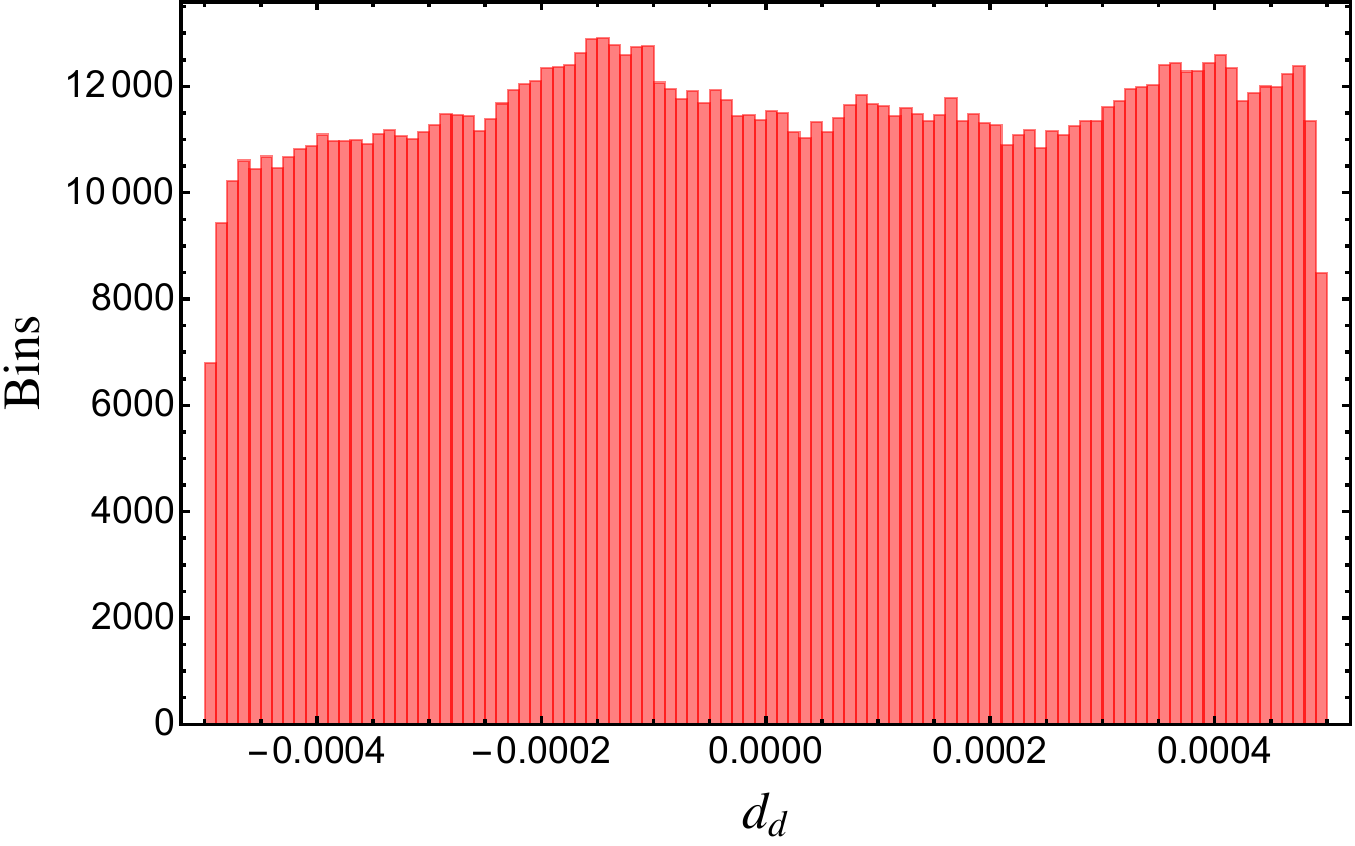}
\includegraphics[scale=0.225]{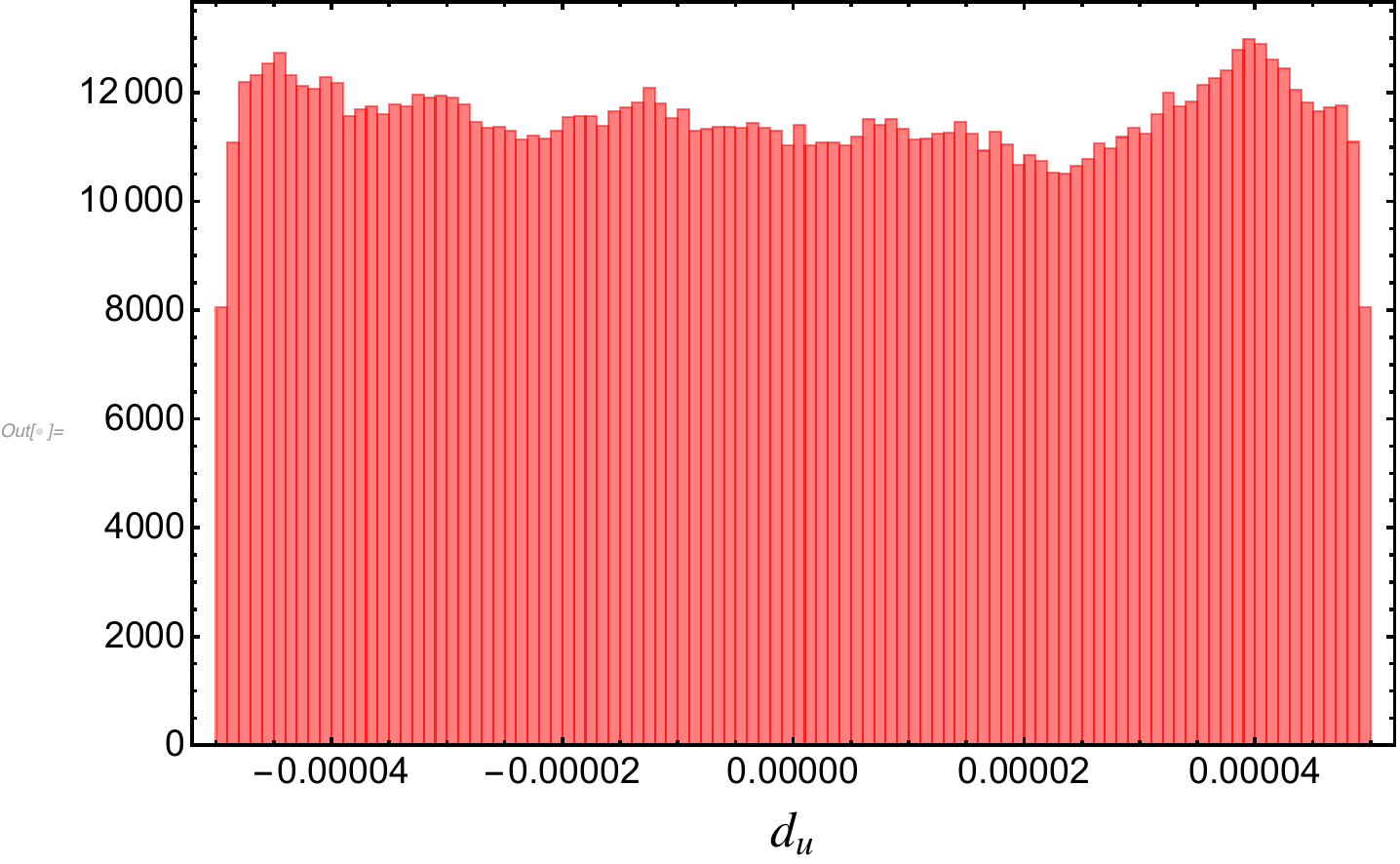}
\includegraphics[scale=0.225]{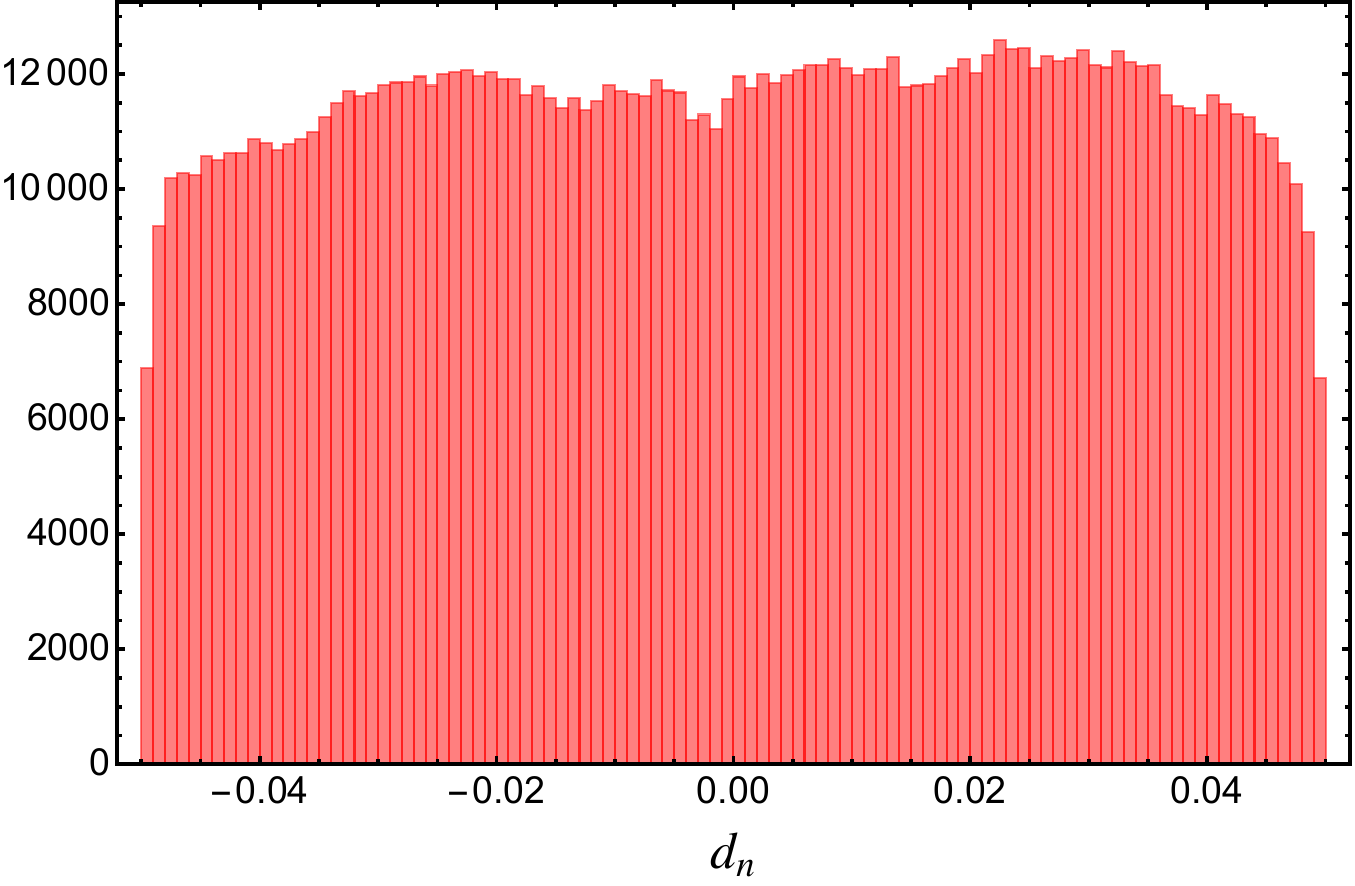}\\
\caption{
Histograms demonstrating the distribution of MCMC iterations for the Dirac (Majorana) scale-setting UTZ parameters $\lbrace a,b,c,d \rbrace_{d,u,\nu}$ ($\lbrace x, y, M\rbrace$), in the LO (blue) and HO (red) scans. We have distributed our results across 100 horizontal bins, while the sum of all vertical histogram values in a given plot is equal to $N_\Theta$. By and large, phases, like the $\lbrace c,d \rbrace_f$ shown, do not exhibit strong preferential values.
}
\label{fig:MCMCparamplots}
\end{figure}

Following the algorithm above, we now specify the constraints that will guide our likelihood evolution in the MCMC, and also the hyper- and model-parameter choices that control our statistics.  Regarding the former, we have identified / implemented the following set of MCMC \emph{constraints} and \emph{predictions}:
\begin{align}
\nonumber
    \text{{\bf{Constraints}}}:&\,\,\,\,\,  \lbrace R_{f_i f_3} \,(f \in u,d,e),\, \sin \theta_{ij}^{q,l}\,,\sin \delta^{q,l}\,,\Delta m^2_{sol,atm},\, m_{\beta(\beta)},\, m_\Sigma,\, \xi, \,\text{n.h.} \rbrace \\
\nonumber
    \text{{\bf{Predictions}}}:& \,\,\,\,\, \lbrace R_{\nu_i \nu_3} , \, m_{\beta\beta} / m_\Sigma,\, m_{\beta} / m_\Sigma,\, m_{\beta\beta} / m_\beta ,\,\sin\phi_1, \,\sin\phi_2 \rbrace \\
\label{eq:Constraint&ObservableList}
    \text{{\bf{Quasi-Predictions}}}:& \,\,\,\,\, \lbrace \sin \delta^{q,l}\,,\,\xi \rbrace
\end{align}
where $R_{f_i f_3}$ corresponds to the ratio of the $i$th generation over third-generation mass ($R_{i3} \equiv m_{f_i}/m_{f3}$) for the corresponding family $f$, and where `n.h.' corresponds to the constraint $\Delta m^2_{sol}/m_{\nu_1}^2 \gg 1$, which enforces a strictly-hierarchical normal-ordered light neutrino spectrum.  The associated numerical constraints correspond to the UV bounds from Tables \ref{tab:mixingangles}-\ref{tab:massfits}. Hence there are $N_{\text{cons}} = 21$ constraints to guide the MCMC likelihood evolution, and $N_{\text{obs}} = 7$ additional predictions that depend on correlated theory parameters, but which do not impact MCMC likelihoods. Observe that $\sin \delta^{q,l}$ and $\xi$ are listed as \emph{quasi-predictions} because, as discussed above, the UV bounds associated to them are extremely large, either due to IR experimental uncertainties ($\sin \delta^l$) or due to theory uncertainties associated to radiative corrections ($\sin \delta^q$, $\xi$).  We will therefore use Tables \ref{tab:mixingangles}-\ref{tab:massfits} as (weak) MCMC constraints, but will also present these results as novel predictions of the UTZ framework, along with those already listed as such in \eqref{eq:Constraint&ObservableList}. 

Given \eqref{eq:Constraint&ObservableList}, we then set the values of the MCMC hyper-parameters we have employed to\footnote{We have explored the stability of our results under variations of each of these parameters. We already mentioned in Footnote 11 that setting $\epsilon \rightarrow 0$ does not qualitatively impact our conclusions, a fact that we have also confirmed with respect to $\kappa_0$, by varying it by a factor of three.  Furthermore, $N_{\text{burn}}$ was determined empirically by conservatively analyzing the likelihood evolution of numerous individual chains; decreasing it will only serve to include lower-likelihood regions of parameter space.  Finally, preliminary scans of the UTZ implementing \eqref{eq:Constraint&ObservableList} were done using much lower statistics than implied by $\lbrace N, L \rbrace_{\text{chains}}$, finding results in qualitative agreement with the high-statistics run generated by \eqref{eq:hyperranges}. Hence we believe these choices are quite conservative.\\
\\
Finally, as mentioned above, we performed a preliminary scan with the same statistics as in \eqref{eq:hyperranges}, but without applying the constraints of \eqref{eq:Constraint&ObservableList} (all model parameter configurations generate a likelihood of 1), in order to confirm that the UTZ does not exhibit built-in preferred regions. Taking the generic scan range $-5\cdot 10^{-1} \le \lbrace a_f, b_f, x, y \rbrace \le 5\cdot 10^{-1}$, we find that all parameter distributions are flat.  In other words, the shapes of the distributions presented in Fig. \ref{fig:MCMCparamplots} are truly driven by \eqref{eq:Constraint&ObservableList}}
\begin{equation}
\label{eq:hyperranges}
    N_{\text{chains}} = 2500,\,\,\,\,\,\,\,\, L_{\text{chains}} = 500,\,\,\,\,\,\,\,\,  N_{\text{burn}} = 40,\,\,\,\,\,\,\,\,\kappa_0 = 0.01,\,\,\,\,\,\,\,\,\epsilon = 0.00005\,,
\end{equation}
 while Table \ref{tab:MCMCranges} gives the ranges scanned over for the actual UTZ model parameters outlined in Section \ref{sec:REVIEW}. The ranges listed for both were identified from successful preliminary MCMC runs with broader model-parameter ranges, coarser hyper-parameter specifications and, most importantly, general physics considerations from Section \ref{sec:REVIEW}, which we now discuss.  

Considering the Majorana sector, we heuristically observe that establishing the sequential dominance condition in \eqref{eq:sequentialdominance} with $M_3/M_2 \sim 10^{n}$ GeV requires $\text{max}(x,y) \sim \mathcal{O}(10^{-n})$, and this is largely independent of the scale $M_\theta$.  We have required $n \ge 3$, to truly establish the third-family Majorana dominance implied by \eqref{eq:MajoranaUTZLO}.  Requiring $M_2 \gg M_1$ of course requires further suppression between $x$ and $y$, such that $M_2/M_1 \sim 10^{n}$ GeV (roughly) corresponds to $\text{min}(x,y) \sim \text{max}(x,y) \cdot 10^{-n/2}$, and we recall that the qualitative physics leading to \eqref{eq:PMNSsumrules} does indeed imply said additional hierarchy.  However, we also notice from \eqref{eq:MajoranaUTZLO} that the two coefficients are sourced from Lagrangian terms that enter at the same power counting (suppressed by $M^4$).  While the combinations of vevs and coefficients can lead to additional suppression, in Table \ref{tab:MCMCranges} we have kept the scan range for $y$ on the same generic order of magnitude as $x$, which of course allows for the additional suppression, but does not enforce it.  

Scan ranges for the LO Dirac parameters $\lbrace a,b \rbrace_f$ are determined by observing that the LO Dirac Lagrangian in \eqref{eq:DiracUTZLO} only exhibits one order of messenger mass suppression w.r.t. the leading third-generation scale-setting terms.  Allowing for a broad range of Wilson coefficients and flavon vevs, we consider $ -5 \cdot 10^{-1} < \lbrace a,b \rbrace_f < 5 \cdot 10^{-1} $ as a reasonable first constraint on preliminary MCMC scans, which we then iteratively refine given observed preferential domains.  Following this procedure, we have noticed that the up-family parameters prefer to be (roughly) symmetrically distributed around zero, and extend to $\pm \mathcal{O}(10^{-4})$ ($\mathcal{O}(10^{-3})$) for the $a_u$ ($b_u$) terms. The down-quark and charged-lepton parameters are also symmetric about zero, but with centers around $\mathcal{O}(10^{-3})$ ($\mathcal{O}(10^{-2})$) for $a_d$ ($b_d$).  In Table \ref{tab:MCMCranges} and Figure \ref{fig:MCMCparamplots} we have only considered the positive branch of these parameters.  Finally, the Dirac neutrino parameters are also distributed in a roughly symmetric way about zero, with both $a_\nu$ and $b_\nu$ peaked around $\mathcal{O}(10^{-1})$. 

Upon identifying the final LO scan ranges as above, we then consider the HO Dirac parameters $\lbrace c, d \rbrace_f$, which we recall from \eqref{eq:DiracUTZHO} contribute at $1/M_{i,f}^4$ in the UTZ OPE, i.e. at one order higher than the leading terms.  Consistent with our power-counting philosophy at LO, we require these terms be at least one order of magnitude smaller than their LO counterparts.  We then consider the analytic hierarchy between the HO operators identified in \eqref{eq:HOcomp} suggesting the $c_f$ correction $\propto S^2$ be yet further suppressed w.r.t. $d_f$.  To this end, if we have identified a scan range of $ \vert \text{min}\lbrace a, b \rbrace_f \vert < \mathcal{O}(10^{-n})$, we require $\vert d_f \vert < \mathcal{O}(10^{-n-1})$ and $\vert c_f \vert < \mathcal{O}(10^{-n-2})$.  While this of course does not forbid the possibility that $\vert d_f \vert \sim \vert c_f \vert$ (as is also in principle allowed given slightly non-universal messenger masses and/or hierarchical Wilson coefficients), it is sufficiently generic for our purposes and, in any event, we observe that these HO corrections do not converge on highly-preferential domains regardless, which is clearly visible in the last two rows of Figure \ref{fig:MCMCparamplots}. Note that, for simplicity, we have kept these HO corrections real.  

\begin{table}[t]
\centering
\renewcommand{\arraystretch}{1.35}
\begin{tabular}{|c||c|c|c|c|c|c|c|}
\hline
Phase Combos & $\lbrace \gamma_d, \delta_d \rbrace$ & $\lbrace \gamma_d, \gamma_u \rbrace$ & $\lbrace \gamma_d, \delta_u \rbrace$ & $\lbrace \gamma_u, \delta_d \rbrace$ &$\lbrace \delta_u, \delta_d \rbrace$  & $\lbrace \gamma_u, \delta_u \rbrace$ \\
\hline
\hline
LO Max Likelihood $\times 100$ & 3.22 & 3.57 & 4.35 & 1.17 & 1.30 & 2.35 $\cdot 10^{-3}$ \\
\hline
\end{tabular}
\caption{The maximum likelihoods returned from preliminary LO MCMC scans with $N_\text{chains} = 500$, $L_\text{chain} = 250$, and $L_\text{burn} = 20$, upon choosing different charged fermion phase configurations.  Note that we have shown the results for the $\lbrace \gamma_d, \delta_d \rbrace$ configuration in Figures \ref{fig:massplots}-\ref{fig:betaplots}, to readily compare with \cite{UTZ}.}
\label{tab:phaselikelihoods}
\end{table}

Finally, we note that in all of the above considerations we have allowed for generic LO phase configurations in the neutrino sector,\footnote{Although recall that in the sequentially-dominant IR limit only one neutrino phase, formed from a combination of said UV phases, dominates the phenomenology.} but have chosen the two non-redundant LO phases in the charged fermion sector as in \cite{UTZ}, i.e. we allow for non-zero $\gamma_d$ and $\delta_d$.  This allows us to readily compare the physical conclusions of our analysis with those of \cite{UTZ}.  In Table \ref{tab:phaselikelihoods} we show that this choice is amongst the higher-likelihood configurations given the six possible pairings, having considered other configurations in preliminary MCMC scans with limited statistics.  Up to this choice, we have otherwise allowed for arbitrary phases in our MCMC scans; all are constrained to $\left[0, 2\pi \right]$, and we observe that there is typically no strong MCMC preference for said phase domains. For this reason we do not show their MCMC histogram distributions in Fig. \ref{fig:MCMCparamplots}.

\begin{table}[tp]
\centering
\renewcommand{\arraystretch}{1.75}
\begin{tabular}{|c||c|c|c|c|}
\hline
\multicolumn{5}{|c|}{Global Best-Fit UV Fermionic Mass Parameters}\\
\hline
\hline
\text{Quark Masses}  & $m_u/m_t$ & $m_c/m_t$ &$m_d/m_b$ & $m_s/m_b$  \\
\hline
\hline
LO Fit & $2.518\cdot 10^{-6}$ & $1.805\cdot 10^{-3}$  & $7.738\cdot 10^{-4}$ & $1.563\cdot 10^{-2}$ \\ 
\hline
HO Fit & $3.126\cdot 10^{-6}$ & $1.862\cdot 10^{-3}$ & $7.397\cdot 10^{-4}$ & $1.490\cdot 10^{-2}$ \\
\hline
\hline
\text{Lepton Masses}   & $m_e/m_\tau$ & $m_\mu/m_\tau$ &$m_{\nu_1} / m_{\nu_3}$ & $m_{\nu_2} / m_{\nu_3}$  \\
\hline
LO Fit & $2.621\cdot 10^{-4}$ &   $5.423\cdot 10^{-2}$ & $8.450\cdot 10^{-3}$ & $1.149\cdot 10^{-1}$ \\ 
\hline
HO Fit & $2.475\cdot 10^{-4}$ &   $5.341\cdot 10^{-2}$ & $2.540\cdot 10^{-3}$ & $1.298\cdot 10^{-1}$ \\ 
\hline
\end{tabular}
\\[0.5cm]
\begin{tabular}{|c||c|c|c|c|}
\hline
\multicolumn{5}{|c|}{Global Best-Fit UV Fermionic Mixing Parameters}\\
\hline
\hline
\text{CKM Parameters}  & $\sin \theta_{12}^{q}$ & $\sin \theta_{23}^{q}$ &$\sin \theta_{13}^{q}$ & $\sin \delta_{CP}^{q}$  \\
\hline
LO Fit & $0.225$ & $1.762\cdot 10^{-2}$  & $3.429\cdot 10^{-3}$ & $0.485$ \\ 
\hline
HO Fit & $0.225$ & $1.752\cdot 10^{-2}$ & $3.247\cdot 10^{-3}$ & $0.446$ \\
\hline
\hline
\hline
\text{PMNS Parameters}   & $\sin \theta_{12}^{l}$ & $\sin \theta_{23}^{l}$ &$\sin \theta_{13}^{l}$ & $\sin \delta_{CP}^{l}$  \\
\hline
LO Fit & $0.550$ &   $0.704$ & $0.146$ & $-0.845$ \\ 
\hline
HO Fit & $0.547$ &   $0.714$ & $0.150$ & $-0.975$ \\ 
\hline
\end{tabular}
\caption{The UTZ global best-fits for fermionic mass (top) and mixing (bottom) parameters in the UV, as determined from the MCMC scan described in Section \ref{sec:MCMC}. The upper lines correspond to the fit allowing for only LO UTZ Lagrangian parameters, while the lower lines also account for HO parameters, both of whose MCMC distributions are given in Table \ref{tab:MCMCranges}. Figures \ref{fig:massplots}-\ref{fig:mixplots} show the total spread of MCMC predictions in this sector, and also highlight the LO global fits presented in this table with a black target marker.}
\label{tab:mass&mixFIT}
\end{table}

\subsection{Results and Analysis}
\label{sec:RESULTS}

We implement the MCMC scan as described above on a computing cluster.  The output is a data-set composed of UTZ model parameters, associated values for the constraints and predictions from \eqref{eq:Constraint&ObservableList}, and the corresponding likelihood of said predictions for \emph{each} saved MCMC iteration.  We denote the corresponding data-set $\Theta_i^j$, where our notation implies that the $i$th data-set has $j$ entries corresponding to the model / physical / likelihood parameter(s).  Hence $i \in \lbrace 1,2,..., N_{\Theta} \rbrace$, where
\begin{equation}
\label{eq:NTheta}
N_{\Theta} = N_{\text{chains}} \cdot \left(L_{\text{chains}} - N_{\text{burn}} \right)\,,
\end{equation}
i.e the overall number of data-sets, each of which has $j \in \lbrace 1, 2, ..., L_{\Theta} \rbrace$ constituents, where
\begin{equation}
\label{eq:LTheta}
L_\Theta = N_{\text{model}} + N_{\text{cons}} + N_{\text{obs}} + 1 \,,
\end{equation}
with the additional unit in $L_{\Theta}$'s counting coming from the standard likelihood function $\mathcal{L}$ for a set of model predictions compared to experiment.  Given \eqref{eq:hyperranges}, $N_\Theta = 1.15 \cdot 10^6$, and we now examine the physical and model parameters embedded therein.

\subsubsection*{Fermion Flavour Mass Ratios}
\label{sec:MASSRESULTS}
\begin{figure}[t]
\centering
\includegraphics[scale=0.31]{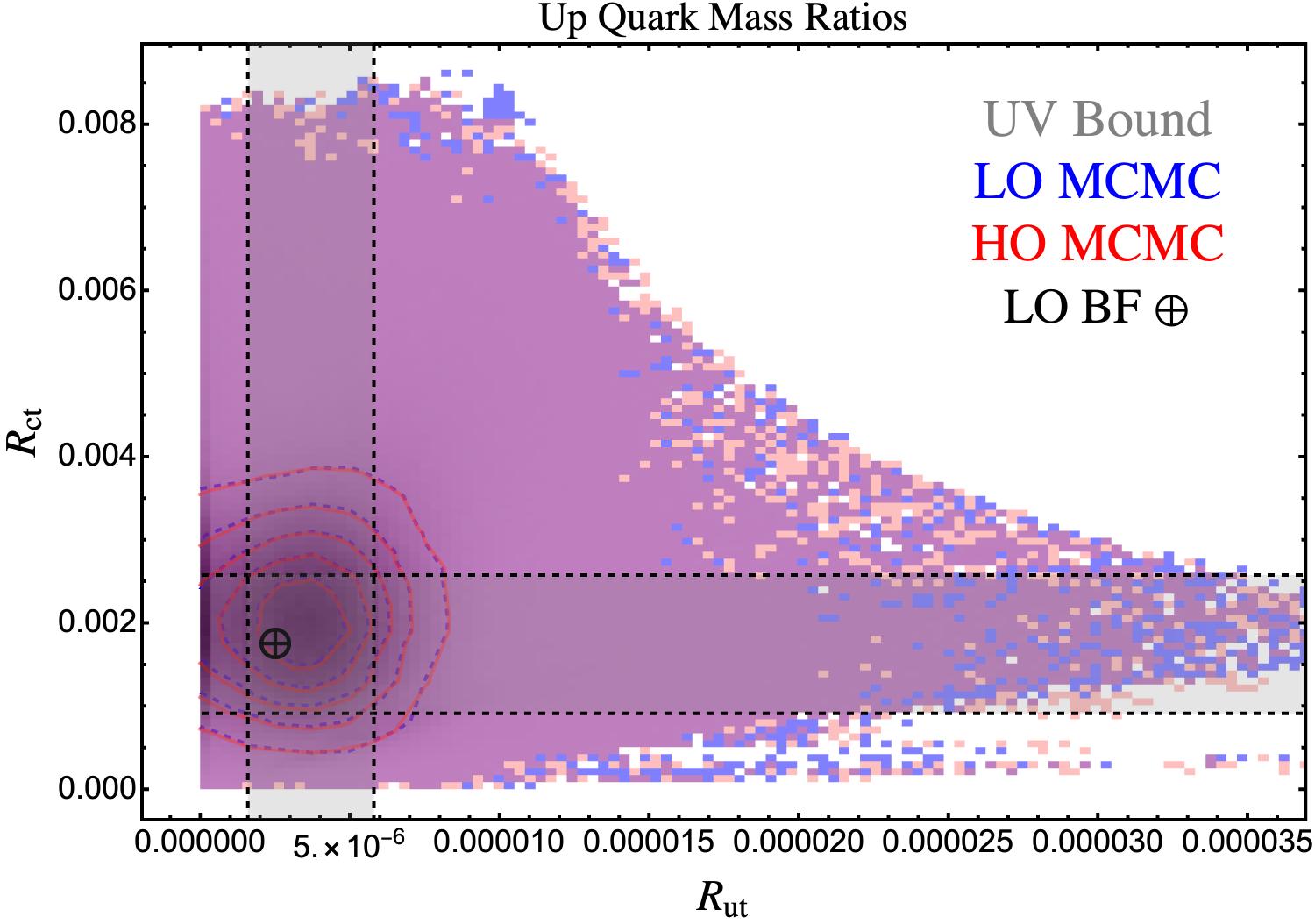}
\includegraphics[scale=0.35]{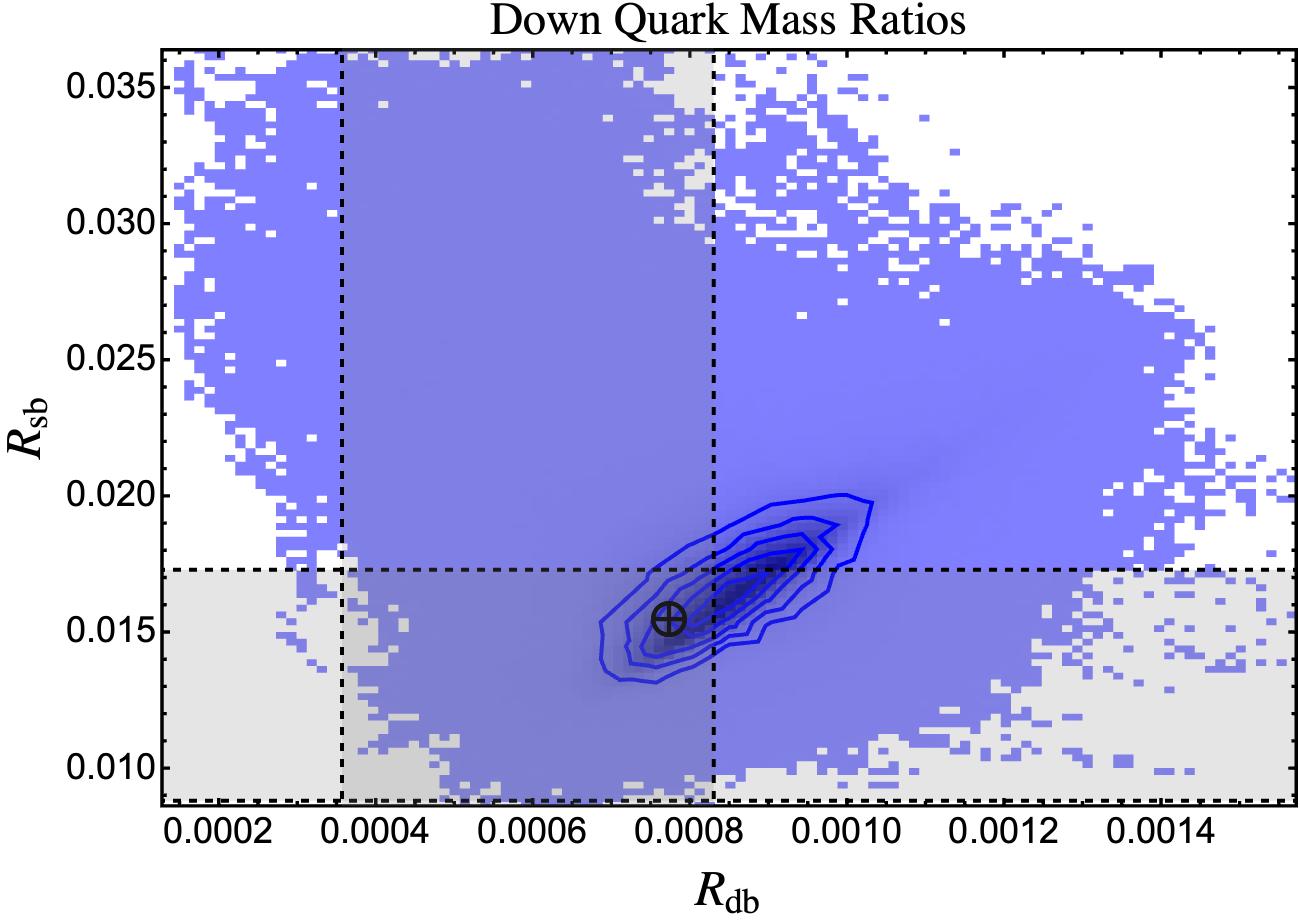}\\
\,\,\includegraphics[scale=0.35]{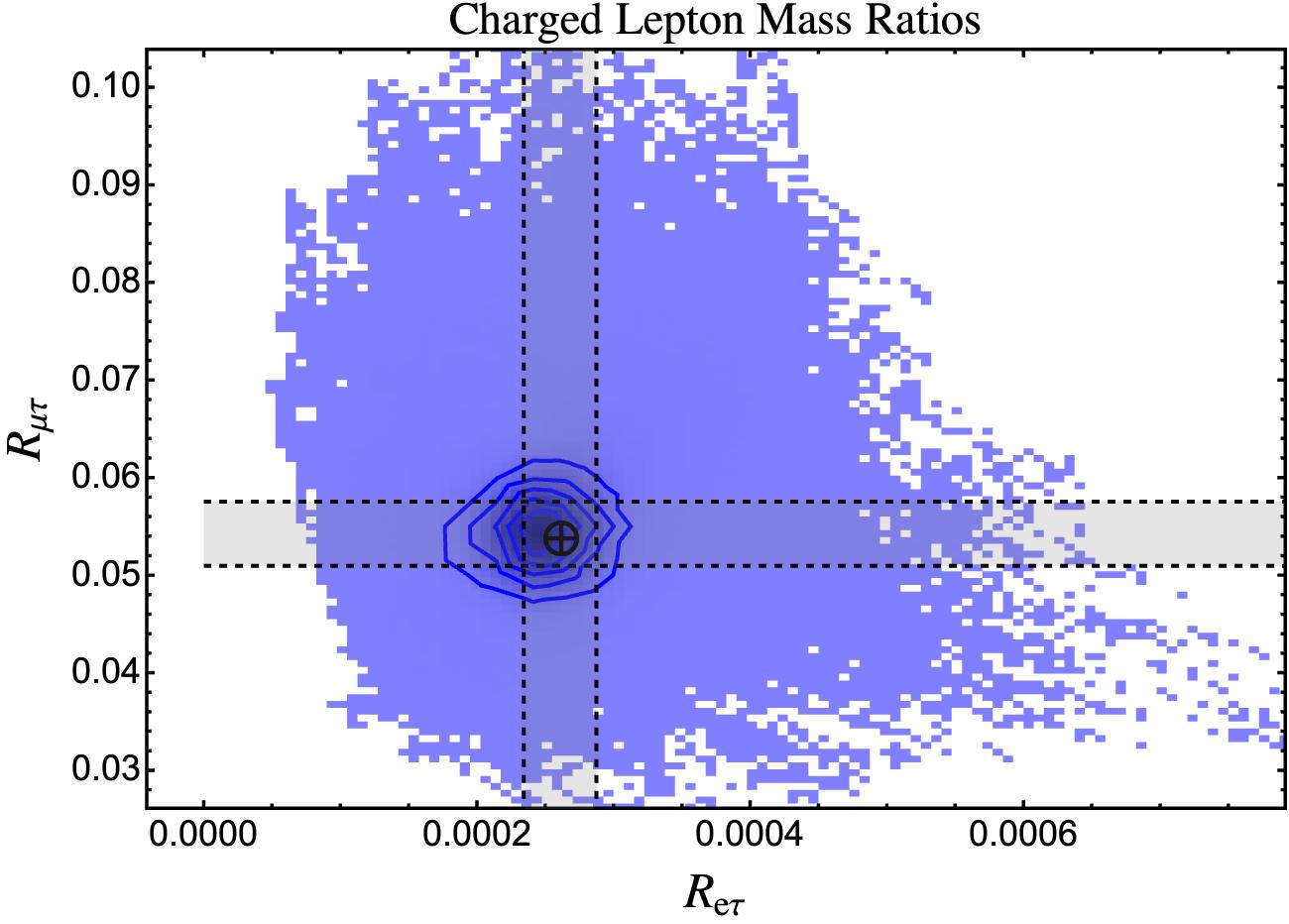}
\,\,\,\,\,\includegraphics[scale=0.35]{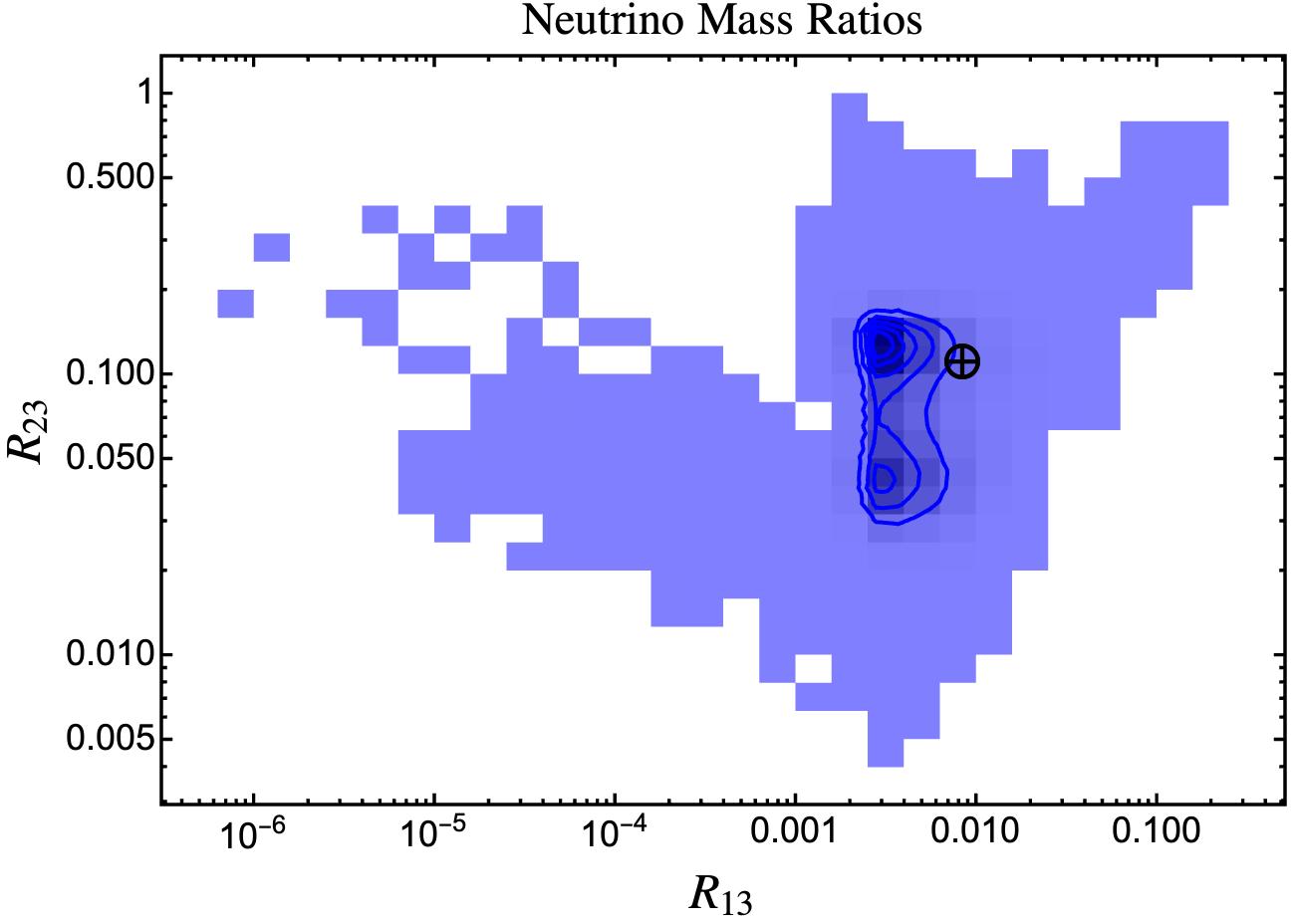}
\caption{
MCMC density plots for UTZ quark and lepton flavoured mass ratio predictions.  Plots are generated with the hyper-parameter choices in \eqref{eq:hyperranges} with model-parameter variations as given in Table \ref{tab:MCMCranges}.  The blue (red) regions correspond to the 
LO (HO) MCMC scan results, with darker regions corresponding to places of higher density.  Gray regions represent the UV bounds for the mass ratios as presented in Table \ref{tab:massfits}, and the black target markers correspond to the global best-fit values shown numerically in Table \ref{tab:mass&mixFIT}. 
}
\label{fig:massplots}
\end{figure}
\begin{figure}[t]
\centering
\includegraphics[scale=0.35]{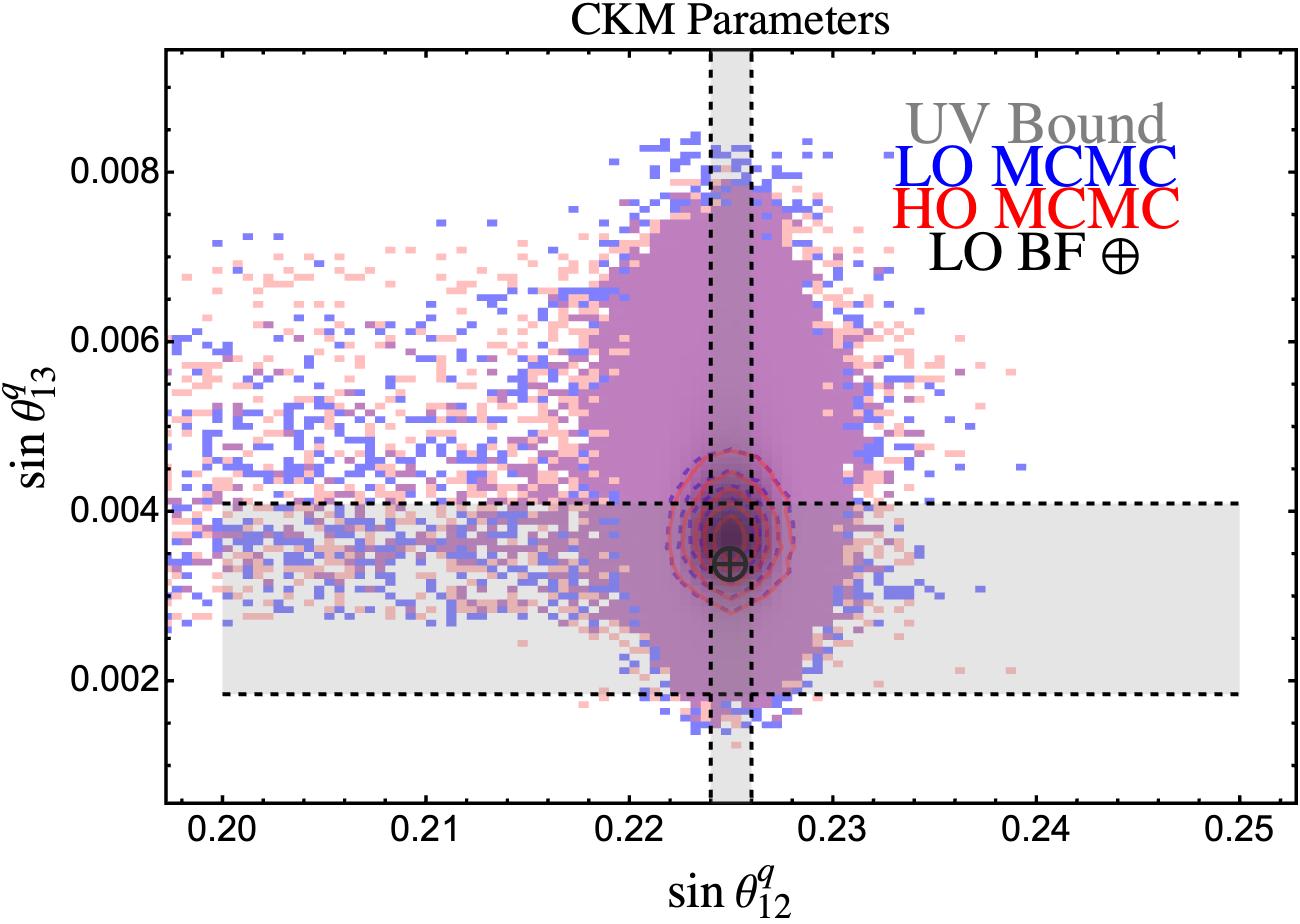}
\includegraphics[scale=0.35]{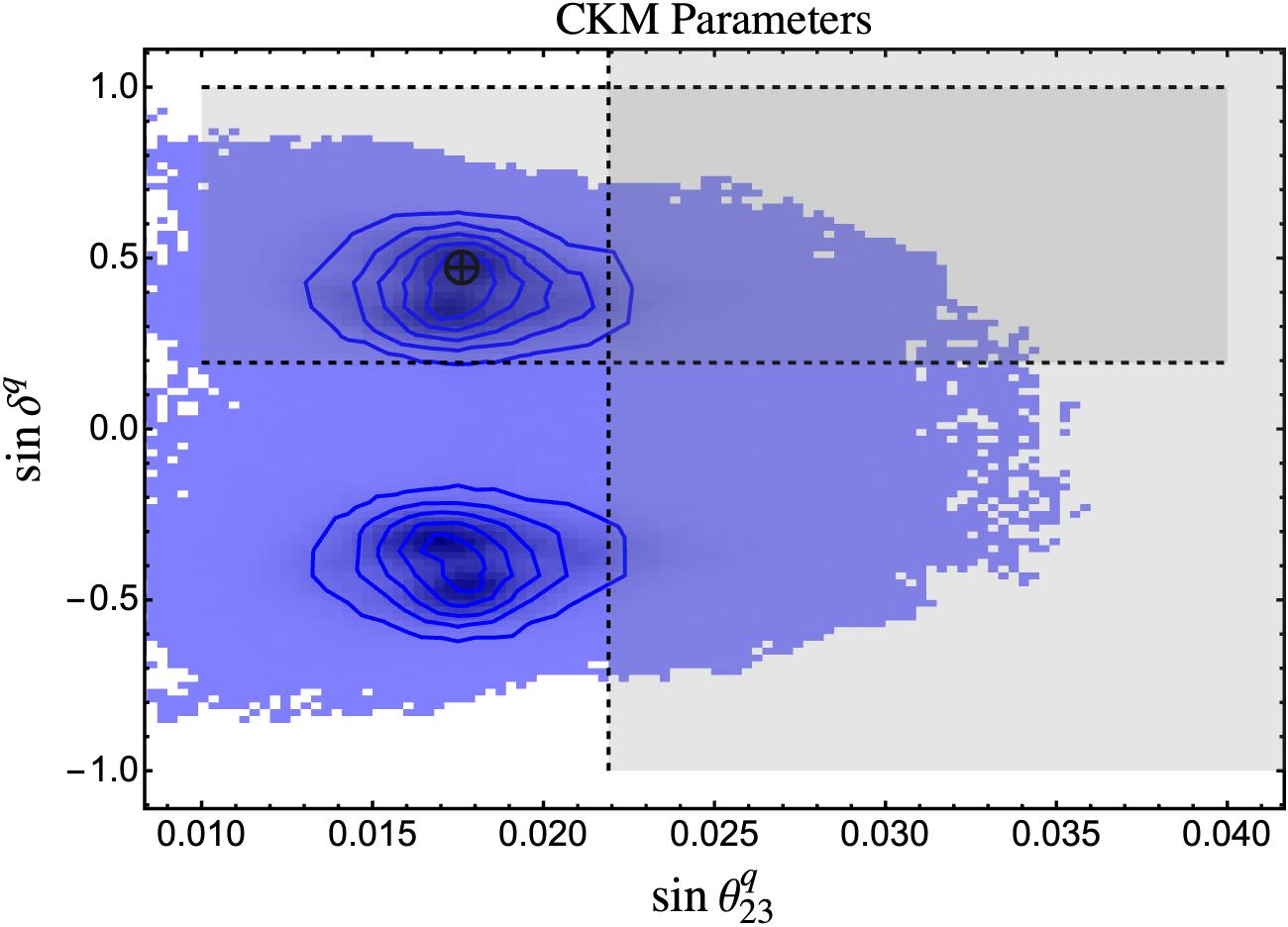}\\
\includegraphics[scale=0.35]{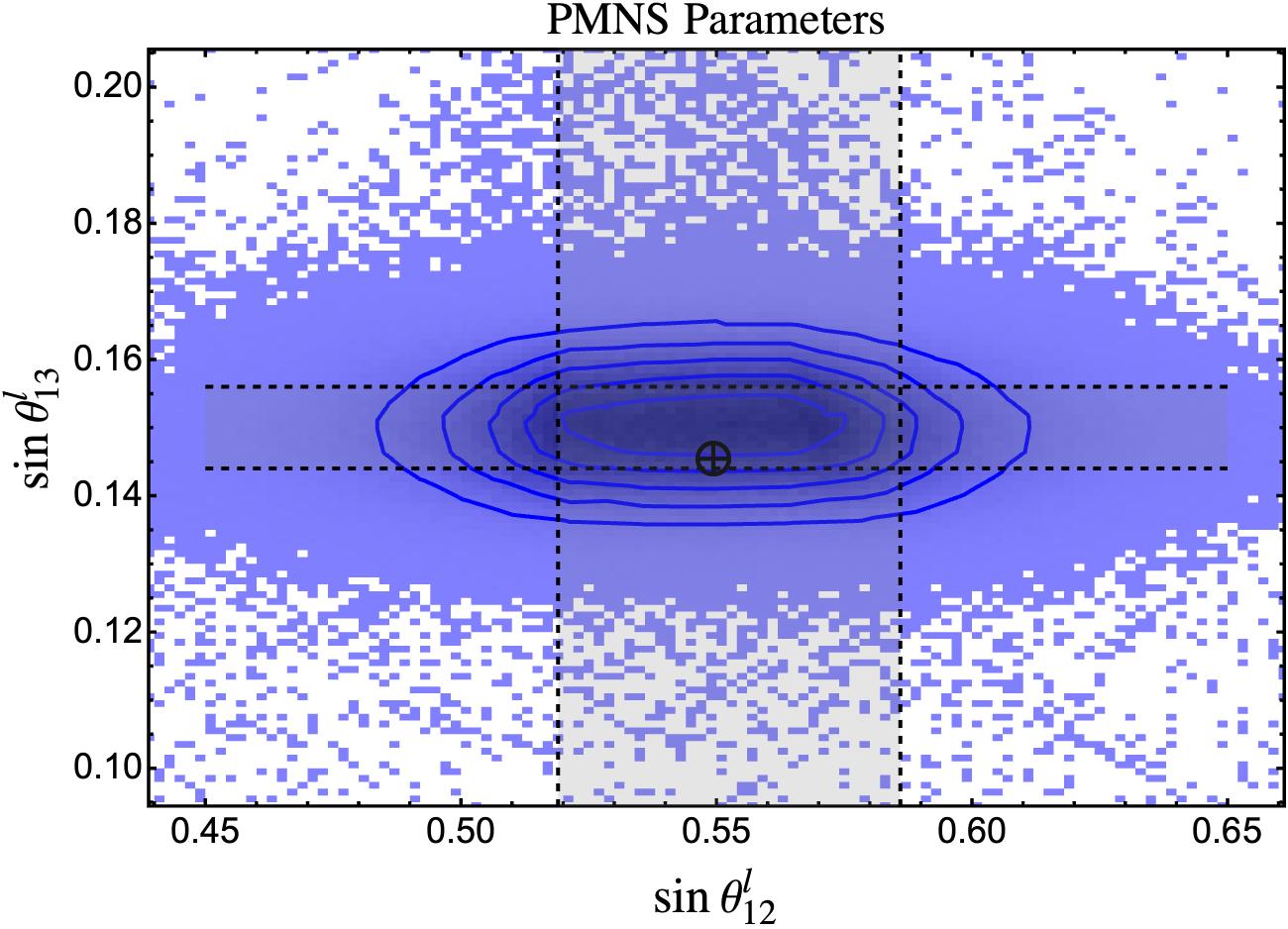}
\includegraphics[scale=0.35]{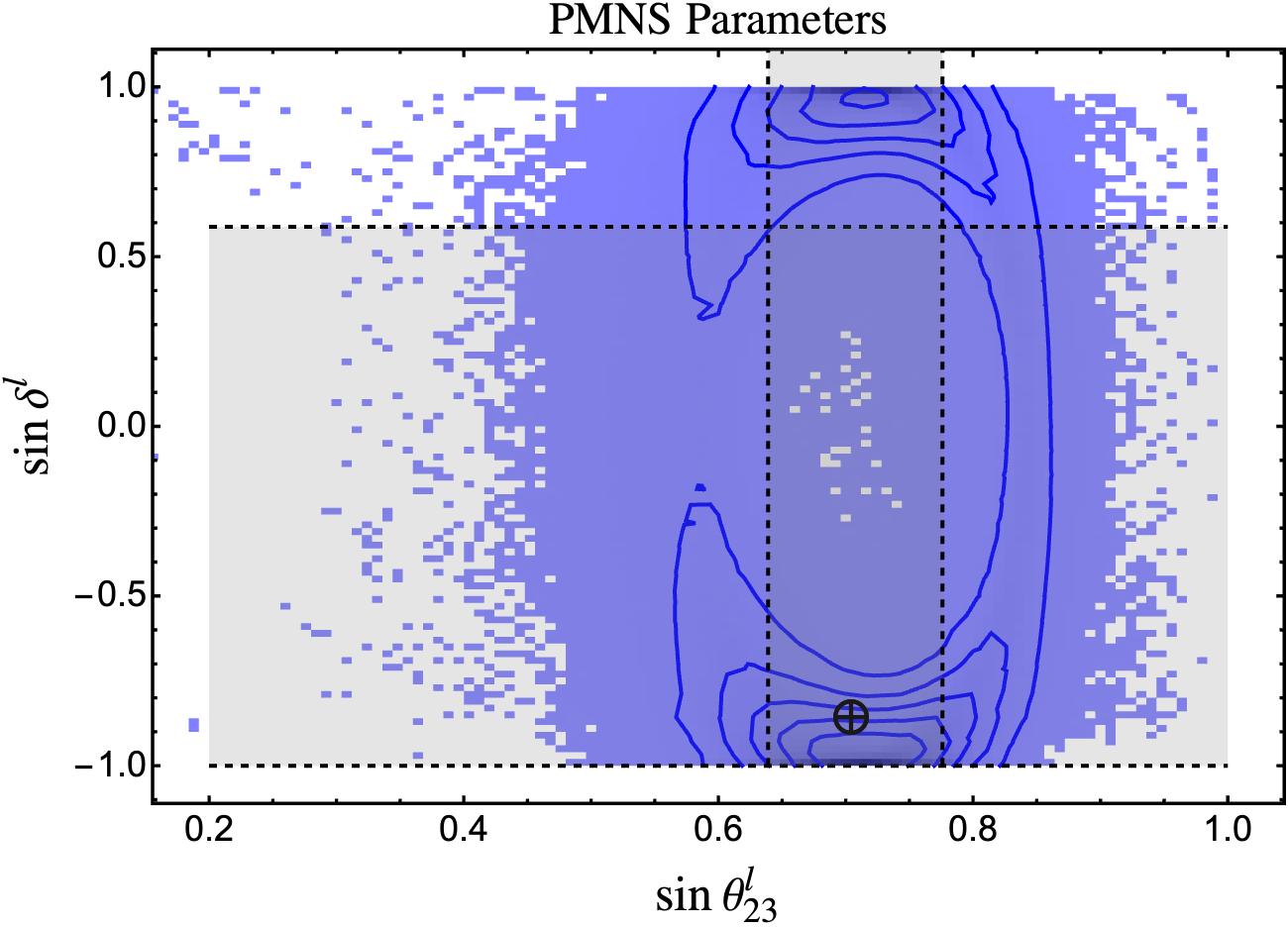}
\caption{
The same as Figure \ref{fig:massplots}, but for the CKM and PMNS mixing angles and Dirac CP-violating phases. 
}
\label{fig:mixplots}
\end{figure}
\begin{figure}[tp]
\centering
\includegraphics[scale=0.35]{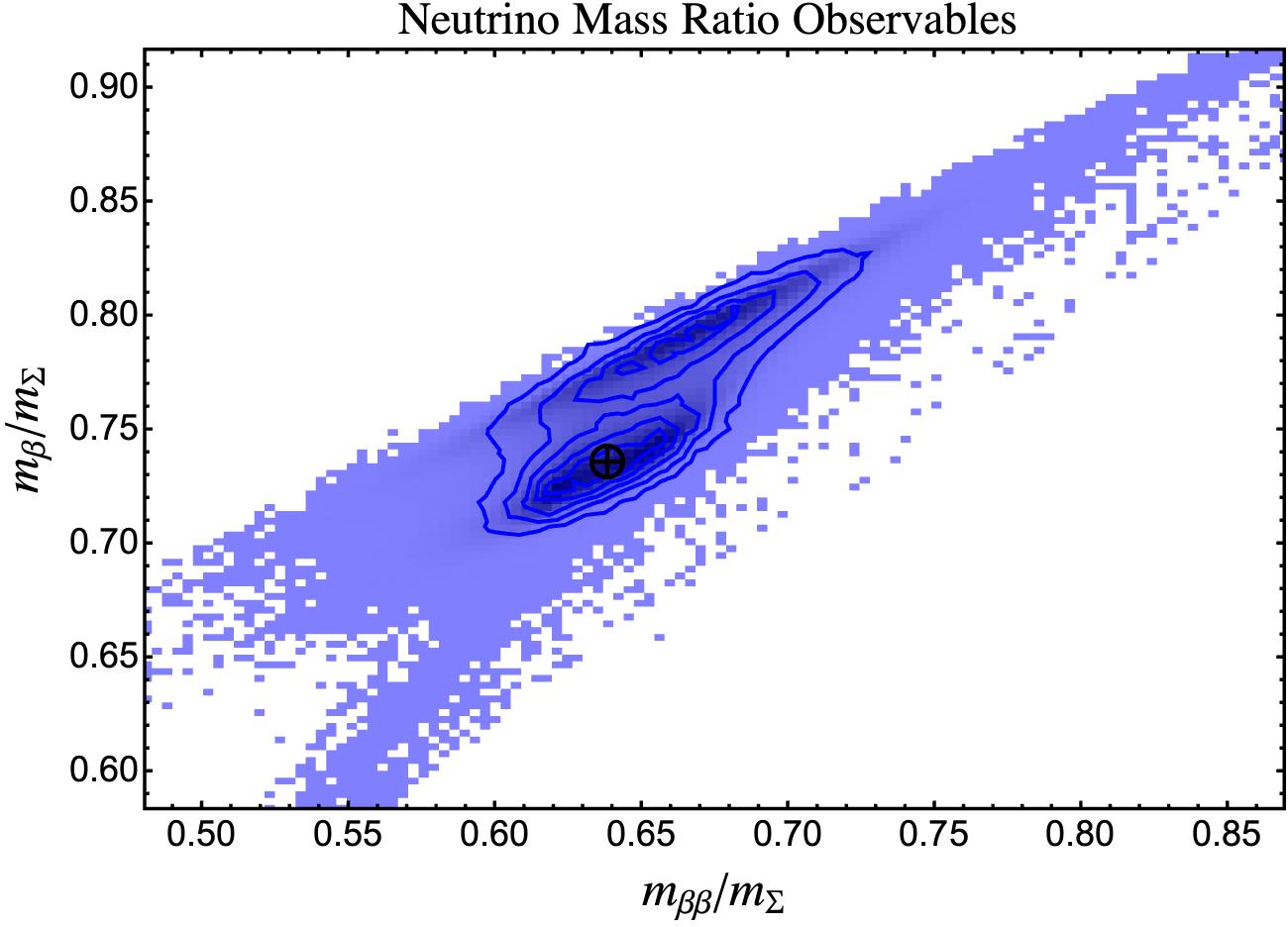}
\includegraphics[scale=0.335]{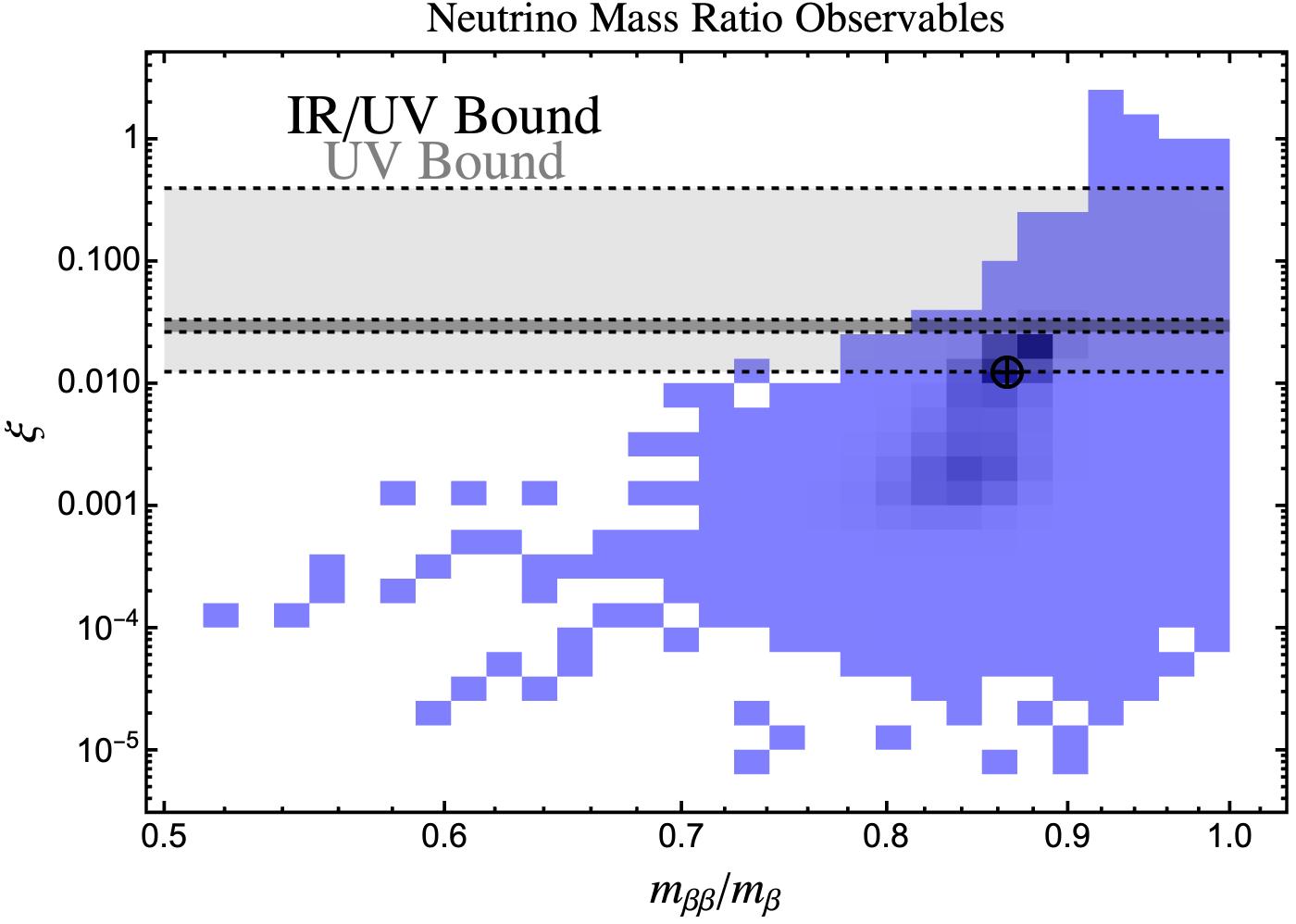}
\caption{
The same as Figure \ref{fig:massplots}, but now comparing the neutrino mass ratio observables predicted by the UTZ.
}
\label{fig:betaratioplots}
\end{figure}
\begin{figure}[t]
\centering
\includegraphics[scale=0.375]{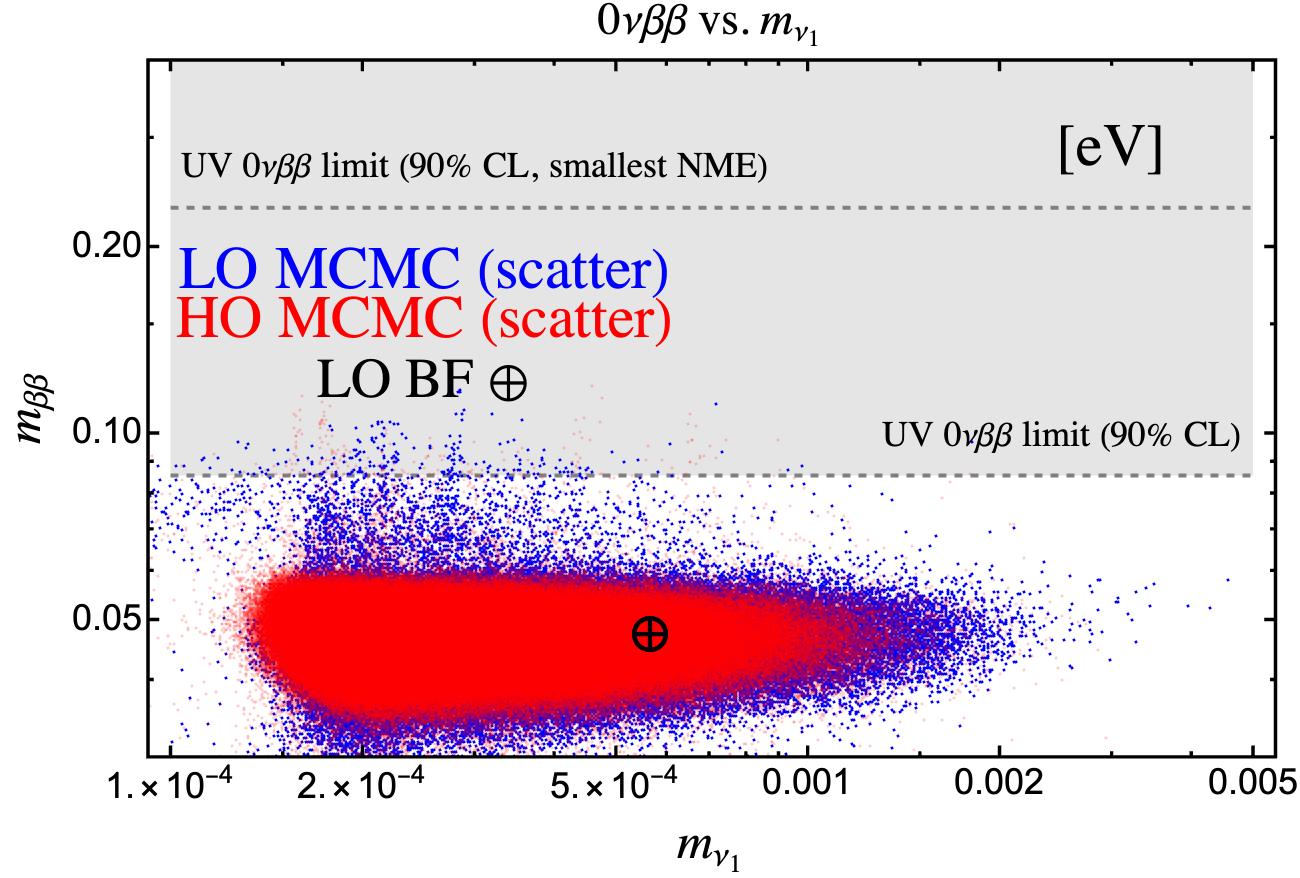} \\
\includegraphics[scale=0.35]{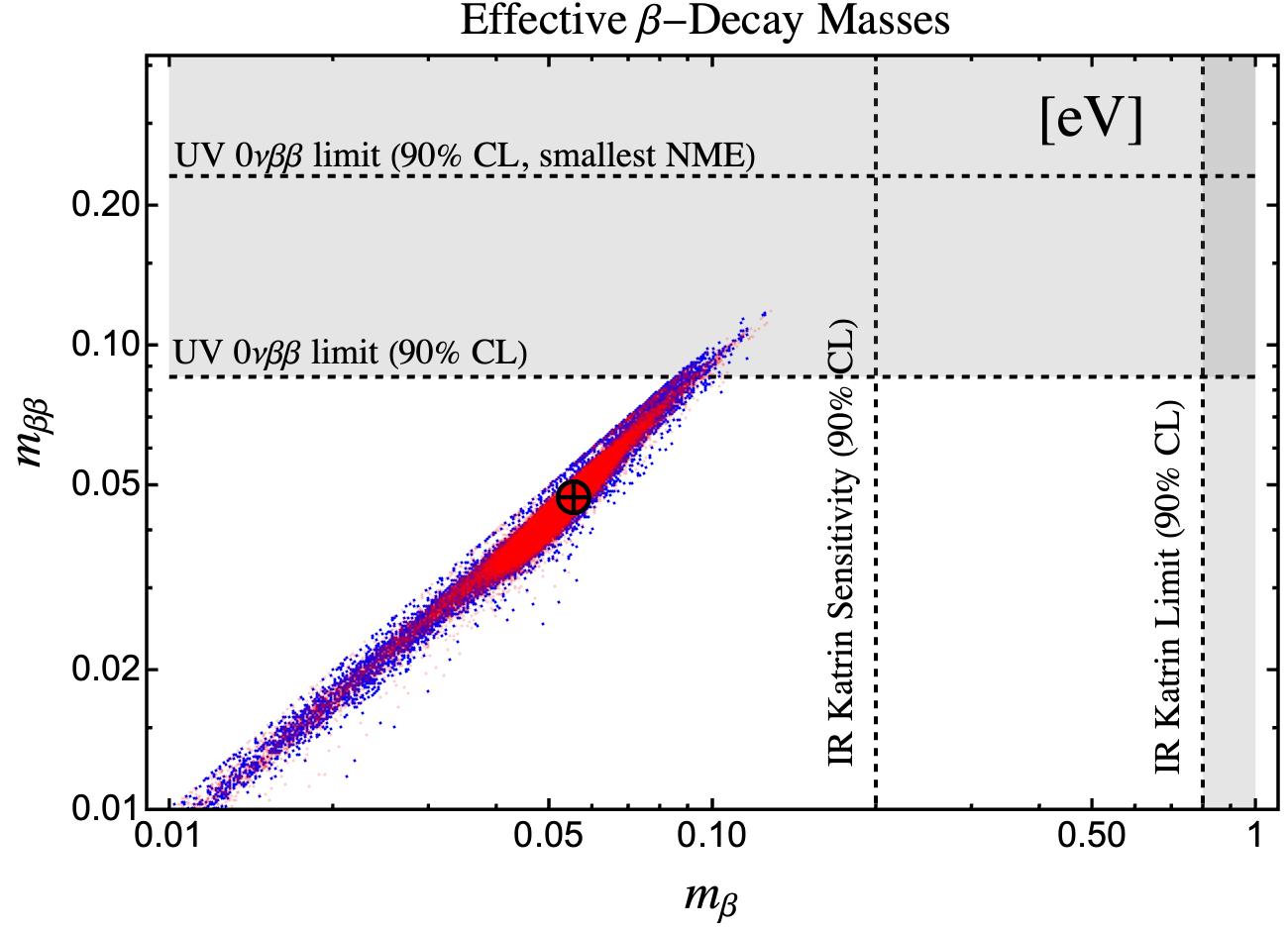}
\includegraphics[scale=0.375]{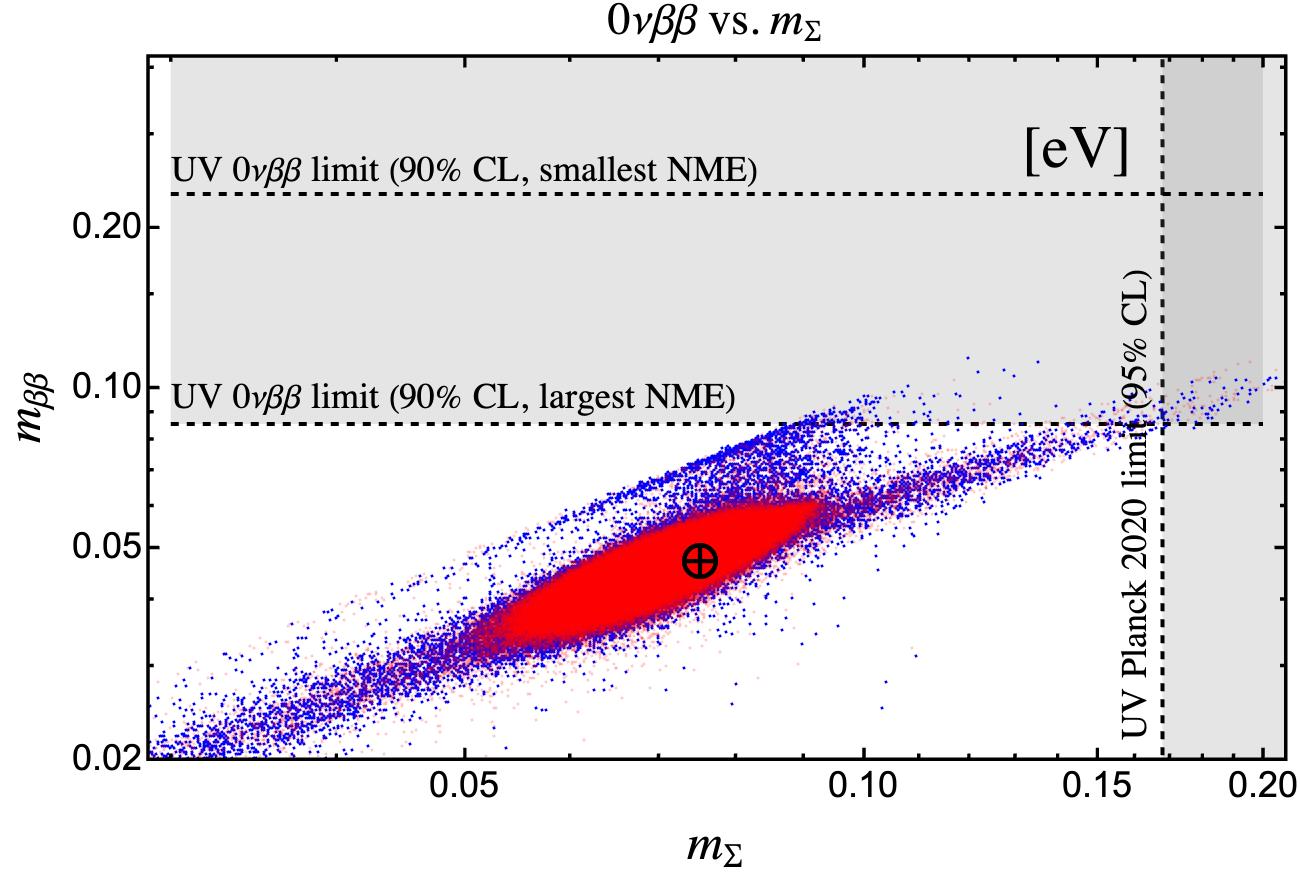}
\caption{
MCMC scatter results for $\beta$-decay and absolute neutrino mass scale observables. Gray regions correspond to IR bounds from KamLAND-Zen ($m_{\beta\beta}$), KATRIN ($m_\beta$), and Planck ($m_\Sigma$), corrected to the UV (if stated) with the conservative $\mathbb{s}$ evolution factor for the heaviest generation. As noted in the text, these results are presented as consistency tests of our approach.
}
\label{fig:betaplots}
\end{figure}

We first examine the MCMC results for the UTZ's predictions in the fermion mass sector. Figure \ref{fig:massplots} illustrates these for mass ratios in the up quark, down quark, charged lepton, and neutrino families.  Both Figure \ref{fig:massplots} (and upcoming figures) and Table \ref{tab:mass&mixFIT} give results for the LO and, when indicated, HO MCMC scans, with the former given in blue and the latter in red. Note that these figures represent density plots, in that darker regions correspond to parameter domains where more Markov chains evolved.  Also, the black `target' marker in Figures \ref{fig:massplots}-\ref{fig:betaplots} corresponds to the location of the \emph{overall} (global) best-fit data-set $\Theta_i$, 
which is also given numerically in Table \ref{tab:mass&mixFIT}. 

The gray bands correspond to the global data available from the PDG (NuFit) collaborations for the charged fermion (neutrino) masses, corrected to the UV according to the discussion in Section \ref{sec:RGE}.  Comparing these to the blue and red regions, we see that the UTZ is capable of successfully resolving the entire charged fermion mass spectrum, for both quarks and leptons, up to the RGE and threshold correction uncertainties. Furthermore, the UTZ predictions for (currently unmeasured) neutrino mass ratios are shown in the bottom-right panel; given the model parameter ranges explored in Table \ref{tab:MCMCranges}, the ratio $R_{i3} \equiv m_{\nu_i}/m_{\nu_3}$ is densely populated within $2.5\cdot10^{-2} < R_{23} < 2\cdot10^{-1}$ for the heavier generations while the smaller mass ratio is densely populated between $1.5\cdot10^{-3} < R_{13} < 2\cdot10^{-2}$.  However, we see that, albeit less frequent, much larger neutrino mass hierarchies are also resolved, with $R_{13} \,(R_{23})$ falling below $10^{-6} \,(5\cdot10^{-3})$.  

Finally, we notice from the up-family plot that the inclusion of the red HO corrections sourced from \eqref{eq:DiracUTZHO} do not qualitatively change the physics conclusions of the blue LO regions.  We have in fact observed this quite generically across family and observable sectors, and hence for visual clarity we only display the dominant LO results in what follows, unless otherwise specified.

\subsubsection*{Fermion Mixings and CP-Violation}
\label{sec:MIXINGRESULTS}
In Figure \ref{fig:mixplots} we have presented the MCMC UTZ predictions for the CKM (top two plots) and PMNS (bottom two plots) mixing angles $\theta_{ij}^{q,l}$ as well as the associated Dirac CP-violating phases $\delta^{q,l}$.  Here we again compare to (radiatively corrected) data from the PDG and {\tt{NuFit}} given in gray, and notice that the blue (red) LO (HO) UTZ Lagrangian is again highly successful at resolving these parameters.  Indeed, while we observe that the overlap with PMNS uncertainty bands is perhaps qualitatively more successful than that of the quarks, the  regions overlap with the UV bounds for all fermion families.   

\begin{figure}[tp]
\centering
\includegraphics[scale=0.35]{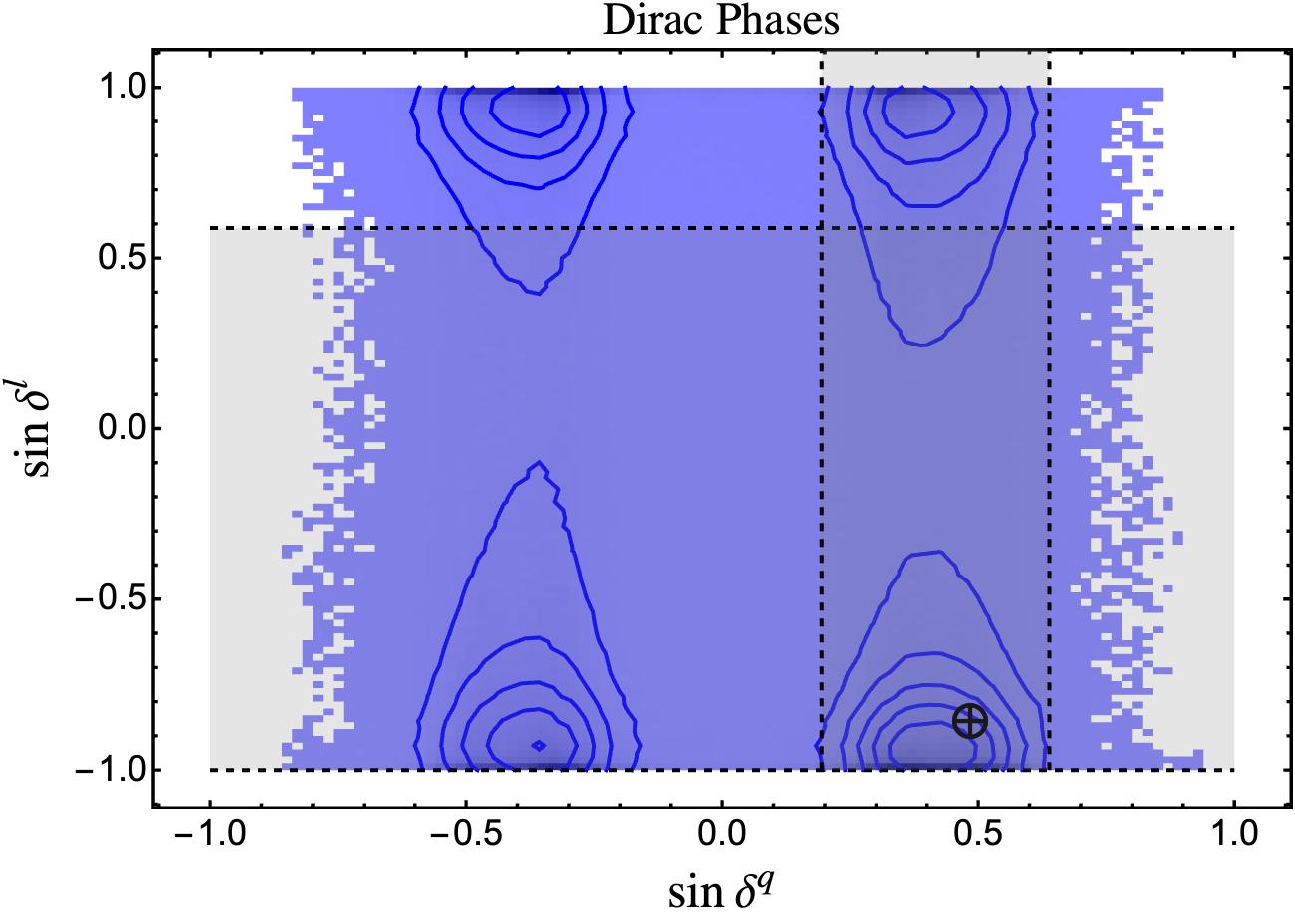}
\includegraphics[scale=0.35]{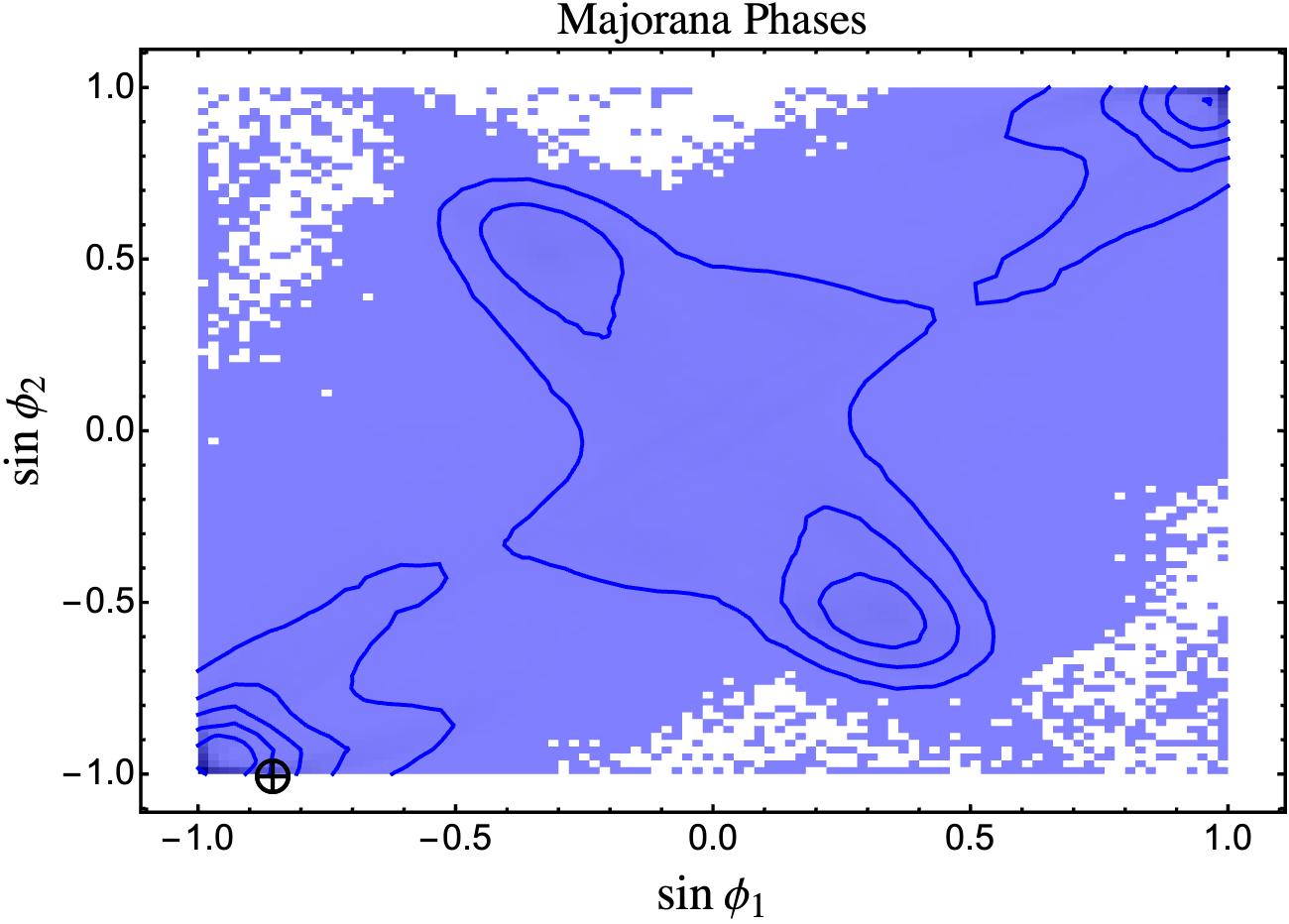}\\
\includegraphics[scale=0.352]{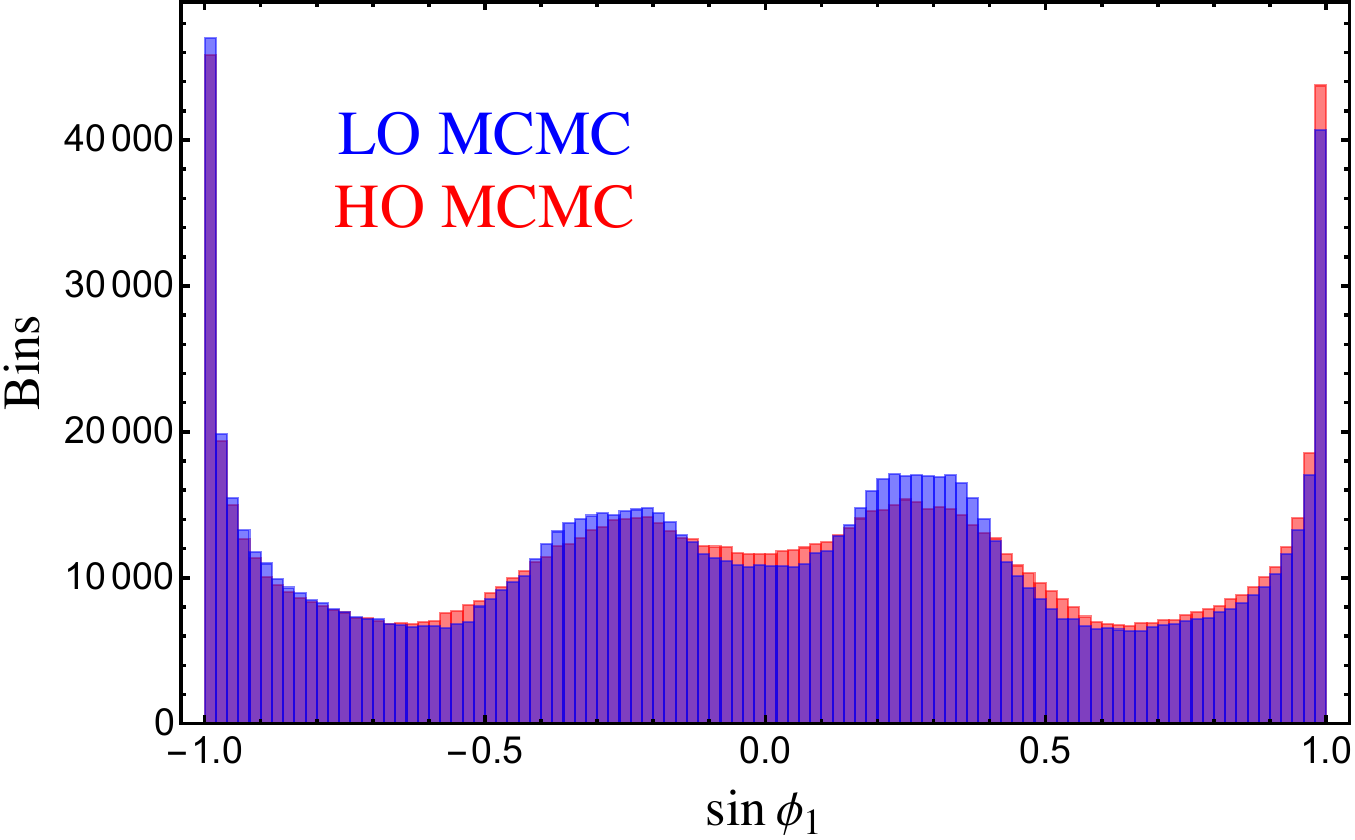}
\includegraphics[scale=0.34]{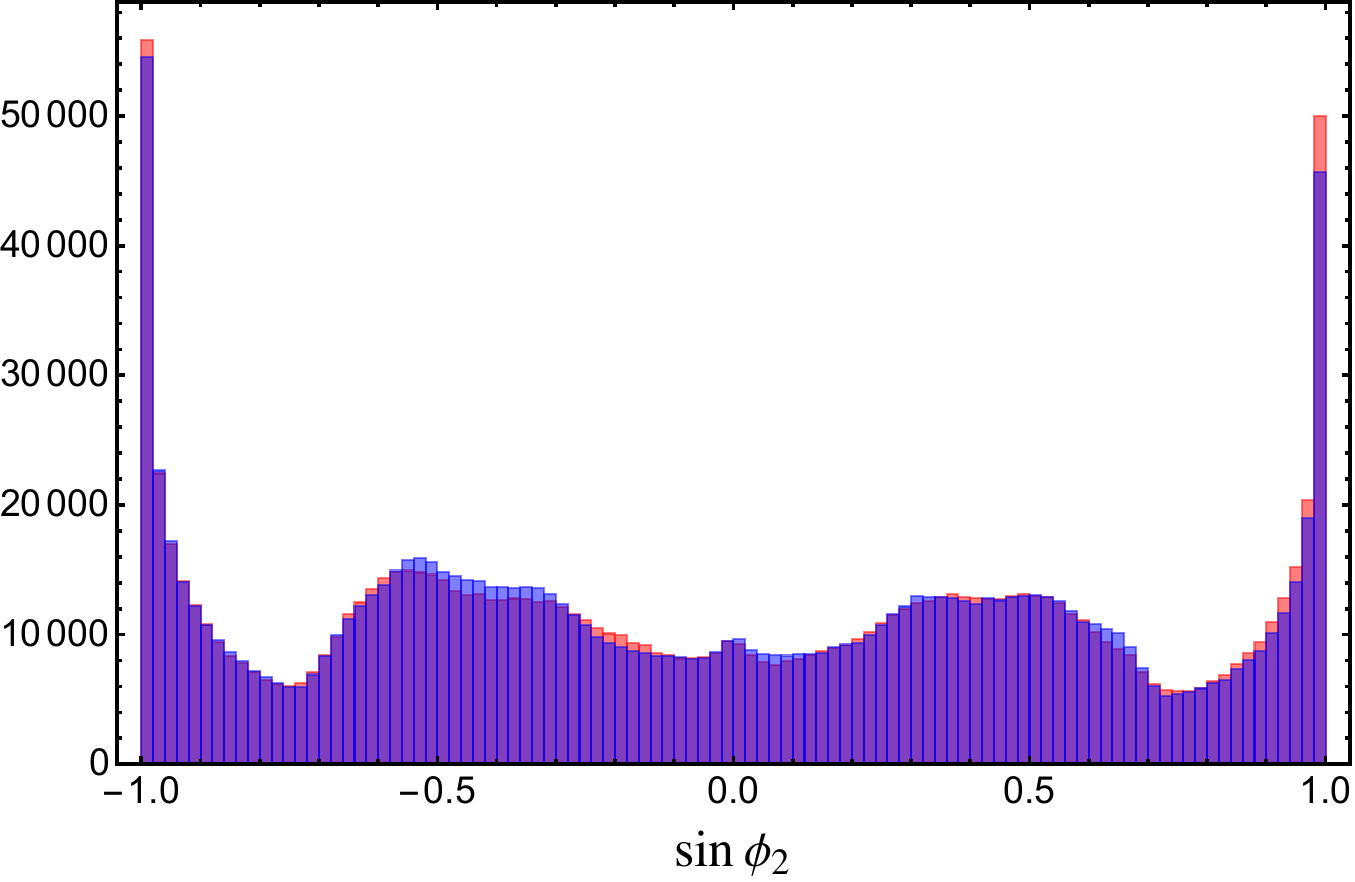}
\caption{
The same as Figure \ref{fig:mixplots}, but now comparing the quark and lepton Dirac CP-violating phases, and presenting the novel predictions for the Majorana CP-violating phases $\phi_{1,2}$. 
}
\label{fig:phaseplots}
\end{figure}
Note that this conclusion differs from the naive analysis in \cite{UTZ}, which found elements in the third row and column of the CKM to be outside of the UV uncertainty bands considering only the LO UTZ Lagrangian, a deviation sourced by the $\theta_{23}^q$ mixing angle.  While we observe that the bulk of the MCMC sample points for $\sin \theta_{23}^q$ are indeed lower than the allowed uncertainty region, a significant number of LO points do overlap successfully.  Studying \cite{Ross:2007az}, one concludes that lower values of $\theta_{23}^q$ tend to correspond to higher $\tan\beta$ RGE scenarios. Hence independent evidence that a background spectrum imitating this UV MSSM structure\footnote{...assuming a certain threshold correction structure and SUSY breaking scale, of course...} is not physical would in principle also disfavor the UTZ theory of flavor, up to the extent the bounds on $\sin \theta_{23}^q$ drive our current MCMC likelihoods.   

In addition to $\theta_{23}^q$, the bottom-right panel of Figure \ref{fig:mixplots} suggests that resolving values of $\sin \delta^l \sim 0$ simultaneously with $\theta_{23}^l$ in the UTZ is disfavored compared to larger $\vert \sin \delta^l \vert$ in the UV. Hence the $\lbrace \theta_{23}^l, \delta^l \rbrace$ sector of the PMNS represents an exceptional opportunity to constrain significant portions of the UTZ parameter space, as information on $\delta^l$ from neutrino oscillations continues to improve.  

To fully present the CP-violating sector of the UTZ, we have also presented our MCMC results for $\sin \delta^{q,l}$ side-by-side in Figure \ref{fig:phaseplots}, along with results for the Majorana CP-violating phases $\phi_{1,2}$ of the PMNS matrix.  Reliable experimental constraints on $\phi_{1,2}$ are presently non-existent, and so they also represent opportunities to falsify / further constrain our model space.  However, one observes that a broad range of Majorana phases are predicted in the UTZ.  We have contextualized this observation by including the MCMC histograms for these phases, analogous to the model parameters presented in Fig. \ref{fig:MCMCparamplots}, in the bottom two panels of Fig. \ref{fig:phaseplots}. These histograms reveal that, while it is true that virtually all values of $\sin \phi_{1,2}$ are acceptable, a huge number of Markov chains evolved to $\vert \sin \phi_{1,2}\vert \sim 1$. Hence improving data (and therefore more rigid constraints in \eqref{eq:Constraint&ObservableList}) could allow us to resolve more precise predictions for these Majorana phases.

Finally, as with the masses presented in Figure \ref{fig:massplots}, we have noticed that HO corrections (as shown in red the top left panel of Fig. \ref{fig:mixplots}, and in the histograms of \ref{fig:phaseplots}) do not qualitatively alter the LO physical conclusions we present above.  This is because we have been conservative in Table \ref{tab:MCMCranges} regarding the relative magnitude of said HO corrections w.r.t. LO parameters, which is of course motivated by the relative suppression of these Lagrangian terms.\footnote{We did not enforce this suppression on $d_d$ in \cite{UTZ} and hence allowed it to be comparatively large.}

 \subsubsection*{$\beta$-Decay and Cosmological Probes}
\label{sec:BETARESULTS}

 We now focus on the sector of observables sensitive to the absolute neutrino mass scale and the Majorana (vs. Dirac) nature of the neutrino field, i.e. $m_{\Sigma}$ and $m_{\beta(\beta)}$. As mentioned above, we only consider ratios of these observables as highly meaningful predictions in the UTZ, and we present their regions in Fig. \ref{fig:betaratioplots}, where it is clear that, at least for the parameter-space we have explored, the UTZ largely prefers values for the ratios $m_{\beta(\beta)}/m_{\Sigma}$ and $m_{\beta\beta}/m_\beta$ given by 
 \begin{equation}
 \label{eq:UTZbetabounds}
 5.9 \cdot 10^{-1} \lesssim \frac{m_{\beta\beta}}{m_\Sigma} \lesssim 7.3 \cdot 10^{-1}\,,\,\,\,\,\,\,\,\,\,\, 7.0 \cdot 10^{-1} \lesssim \frac{m_{\beta}}{m_\Sigma} \lesssim 8.4 \cdot 10^{-1}\,,\,\,\,\,\,\,\,\,\,\, 7.8 \cdot 10^{-1} \lesssim \frac{m_{\beta\beta}}{m_\beta} \lesssim 9.2\cdot 10^{-1}\,.
 \end{equation}
 In the event that a positive signal for $m_{\beta(\beta)}$ is ever observed, \eqref{eq:UTZbetabounds} will serve as an excellent probe of the UTZ construction.  One also notices in the right panel of Fig. \ref{fig:betaratioplots} that the MCMC has evolved such that relatively small values of the neutrino-mass-squared difference ratio $\xi$ are preferred, with respect to the possible UV upper bound in Table \ref{tab:massfits}.  However, the observed region is still consistent with both the low-$\tan \beta$ / SM-like RGE and high-$\tan \beta$-like RGE scenarios discussed in Section \ref{sec:RGE}.\footnote{For consistency with the quark sector we have trained our MCMC on the more uncertain UV scenario for $\xi$, which allows for the possibility of high-$\tan\beta$-like RGE.  If we instead train on the low-$\tan \beta$ / SM-like RGE scenario, the preferred regions in Fig. \ref{fig:betaratioplots} shift upwards to center on the darker, smaller UV/IR uncertainty band.}
 
 Of course, as mentioned, we can also report the actual values of the constituent functions $m_{\Sigma}$ and $m_{\beta(\beta)}$, despite them being less meaningful due to their sensitivity to $M_\theta$.  For completeness we do so in Figure \ref{fig:betaplots}.  As expected due to their use as constraint in \eqref{eq:Constraint&ObservableList}, we observe that the UTZ readily evades available bounds from (e.g.) KATRIN, Planck, and KamLAND-Zen.  However we emphasize that this statement effectively amounts to a consistency check on the MCMC framework implemented.

\subsection{Summary Comments}
\label{sec:PHENOSUMMARY}
Before concluding, we summarize the results in the above Sections:
\begin{itemize}
    \item The LO UTZ Lagrangian in \eqref{eq:DiracUTZLO} and \eqref{eq:MajoranaUTZLO} is sufficient to describe all available data on fermionic mass and mixing, as well as data constraining the overall scale of neutrino masses.  This result is novel, and represents a substantial improvement on the phenomenological findings of \cite{UTZ}, which found that HO corrections were necessary to describe the third row and column of the CKM matrix (due to $\theta_{23}^q$).  This illustrates the power of our MCMC algorithm to robustly explore viable UTZ parameter spaces, in comparison to less sophisticated methods.  However,  $\theta_{23}^q$ still represents an excellent parameter to exclude UTZ parameter spaces in the future.
    \item The MCMC algorithm also allows us to present robust predictions for observables that are not well constrained by data --- e.g. leptonic CP-violating phases $\sin \lbrace \delta^l, \phi_1, \phi_2 \rbrace$ and neutrino mass ratios $m_{\nu_i}/m_{\nu_3}$, $\xi$, $m_{\beta(\beta)}/m_\Sigma$, and $m_{\beta\beta}/m_\beta$ --- despite the fact that said observables depend sensitively on theory parameters that are highly-correlated to other, well-constrained observable sectors.  These findings provide excellent opportunities for the falsification / exclusion of UTZ parameter spaces.  We have presented these predictions using the (N)LO UTZ Lagrangian in Figs. \ref{fig:massplots}-\ref{fig:betaratioplots}.
    \item The HO corrections generated by the operators in \eqref{eq:DiracUTZHO} do not qualitatively change the physics conclusions driven by the dominant operators in \eqref{eq:DiracUTZLO}-\eqref{eq:MajoranaUTZLO}.  This is due to our (natural) assumption that said HO parameters are suppressed with respect to LO parameters, a constraint that we did not impose as rigorously in \cite{UTZ}.  As a result, the UTZ's predictions are dominated by as few as nine IR theory parameters.  Hence the UTZ is realized as a \emph{well-defined}, \emph{stable}, and \emph{predictive} effective theory of flavour.
    \item The results we have presented are of course sensitive to the hyper- and model-parameter ranges we have explored, which are presented in \eqref{eq:hyperranges} and Table \ref{tab:MCMCranges}.  While we have taken care in identifying these ranges, and have indeed demonstrated that they are successful, they are not necessarily unique. Exploring alternative parameter spaces, possibly with even more statistics than implied by \eqref{eq:hyperranges}, will be especially motivated in the event data fully excludes the predictions presented in Figures \ref{fig:massplots}-\ref{fig:betaratioplots}, and/or a specific renormalizable completion (with exact RGE / threshold behavior) of the UTZ is identified.      
\end{itemize}

\section{Summary and Outlook}
\label{sec:CONCLUDE}

We have re-examined the Universal Texture Zero (UTZ) model of flavour presented in \cite{UTZ} in light of updated experimental constraints and in the context of a novel Markov Chain Monte Carlo (MCMC) analysis routine.  We have considered the UTZ's predictions at both leading- and next-to-leading orders in its effective theory operator product expansion, and the associated phenomenological pre- and post-dictions are given in Figures \ref{fig:massplots}-\ref{fig:betaratioplots}.  There we observe that the UTZ is capable of fully resolving the fermionic mass and mixing spectrum as constrained by global data sets, for both the quark and lepton sectors, up to uncertainties regarding radiative corrections to/from the ultraviolet.  We have also presented a host of novel, robust predictions for poorly-constrained leptonic observables, in particular the PMNS CP-violating phases $\delta^l$, $\phi_{1,2}$ and neutrino mass-sector ratio observables $m_{\beta(\beta)}/m_\Sigma$ and $m_{\beta\beta}/m_\beta$.  These latter results offer a route to UTZ falsification and/or parameter-space exclusions.

Our analysis therefore greatly improves on the proof-in-principle phenomenology pursued in the original UTZ publication \cite{UTZ}, which was incapable of yielding a robust prediction for even (e.g.) $\delta^l$, and which did not consider observables like $m_{\beta(\beta)}$ and $m_\Sigma$.  Indeed, our updated MCMC analysis revises the claim from \cite{UTZ} that the leading UTZ Lagrangian is insufficient to account for all fermionic data, before considering next-to-leading corrections.  However, as discussed at the end of Section \ref{sec:MCMC}, there is still room for improvement, as a yet more exhaustive scan of the UTZ parameter space is in principle possible.  We also note that, while our MCMC algorithm fully accounts for theory correlations amongst UTZ model parameters, we have not accounted for \emph{experimental} correlations, beyond what is already accounted for in the global fits presented in Section \ref{sec:CONSTRAINTS}. While we do not expect such correlations to qualitatively change our conclusions, pursuing such an analysis in the future could be interesting. 
 
Besides these future technical/phenomenological improvements, we also note that significant progress has recently been made in rigorously connecting theories of flavour controlled by non-Abelian discrete symmetries (and additional shaping symmetries) to string theories with toroidal orbifold compactifications --- see e.g. \cite{Baur:2019kwi,Baur:2019iai,Baur:2021bly,Baur:2022hma}. It would be interesting to determine whether the UTZ (or a close cousin) could be formally embedded into one of these structures, thereby providing a UV origin for the field and symmetry content of Table \ref{tab:Zcharges}, and an unambiguous background spectrum that would minimize the radiative uncertainties that we have considered agnostically in our effective field theory setup. After all, the absence of $\Delta(27)$ contractions with non-trivial singlets in \eqref{eq:DiracUTZLO}, \eqref{eq:MajoranaUTZLO}, and \eqref{eq:DiracUTZHO} is already consistent with the stringy models examined in \cite{Nilles:2012cy}. 

We leave these questions to future work, and simply conclude that Figures \ref{fig:massplots}-\ref{fig:betaplots} indicate that the UTZ represents an appealing, minimal, and phenomenologically viable model of flavour physics, and therefore provides some support for the idea that observed flavour patterns are the result of yet-discovered Beyond-the-Standard Model dynamics, rather than (e.g.) random chance.

\section*{Acknowledgements}
 JT and IdMV are indebted to Prof. Graham Garland Ross (1944-2021), whose supervision and collaboration greatly inspired our research interests, including the Universal Texture Zero theory we originally co-authored in \cite{UTZ} and further studied in this paper.  We thank him for his mentorship, both personal and professional, and for the time we shared together over the years. 
 
 Our collaboration also thanks Ben Risebrow for his helpful contributions over the course of his summer research project, and Andrew Fowlie for comments on our MCMC approach.
 
 JB is supported by the Deutsche Forschungsgemeinschaft (DFG, German Research Foundation) under grant  396021762 -- TRR 257.
 IdMV acknowledges funding from Funda\c{c}\~{a}o para a Ci\^{e}ncia e a Tecnologia (FCT) through the contract UID/FIS/00777/2020 and was supported in part by FCT through projects CFTP-FCT Unit 777 (UID/FIS/00777/2019), PTDC/FIS-PAR/29436/2017, CERN/FIS-PAR/0004/2019 and CERN/FIS-PAR/0008/2019 which are partially funded through POCTI (FEDER), COMPETE, QREN and EU.
The work of ML is funded by
FCT Grant
No.PD/BD/150488/2019, in the framework of the Doctoral Programme
IDPASC-PT.  
 JT gratefully acknowledges funding from the European Union's Horizon 2020 research and innovation programme under the Marie Sk\l{}odowska-Curie grant agreement No. 101022203. 
 
\bibliographystyle{unsrt}
\bibliography{biblioUTZpheno}
\end{document}